\documentclass[11pt,letterpaper]{article}
\usepackage[margin=1in]{geometry}
\usepackage[T1]{fontenc}
\usepackage{lmodern,amsmath,amssymb,amsthm,mathtools,microtype}
\usepackage{aliascnt,enumitem,xspace,booktabs,array,longtable,needspace,tabularx}
\usepackage{placeins,float}
\usepackage{caption}
\usepackage{etoolbox}
\patchcmd{\abstract}{\small}{\normalsize}{}{}
\newsavebox{\TableBody}
\usepackage{xcolor}
\definecolor{RefLinkBorder}{RGB}{37,99,166}
\definecolor{CiteLinkBorder}{RGB}{33,126,84}
\definecolor{URLLinkBorder}{RGB}{132,78,158}
\usepackage{hyperref}
\usepackage{tikz}
\usetikzlibrary{arrows.meta,positioning,calc,shapes.geometric}
\usepackage[capitalise,nameinlink,noabbrev]{cleveref}
\usepackage[backend=biber,style=alphabetic,sorting=nyt,maxbibnames=99,maxalphanames=5,minalphanames=3,giveninits=true,isbn=false]{biblatex}
\AtEveryBibitem{%
  \interlinepenalty=10000
  \ifentrytype{misc}{\clearfield{howpublished}}{}%
  \iffieldequalstr{entrykey}{CKK21}{\clearfield{eprint}}{}%
  \iffieldequalstr{entrykey}{FHHH24}{\clearfield{eprint}}{}%
}

\theoremstyle{plain}
\newtheorem{theorem}{Theorem}[section]
\newaliascnt{lemma}{theorem}
\newtheorem{lemma}[lemma]{Lemma}
\aliascntresetthe{lemma}
\Crefname{lemma}{Lemma}{Lemmas}
\newaliascnt{proposition}{theorem}
\newtheorem{proposition}[proposition]{Proposition}
\aliascntresetthe{proposition}
\Crefname{proposition}{Proposition}{Propositions}
\newaliascnt{corollary}{theorem}
\newtheorem{corollary}[corollary]{Corollary}
\aliascntresetthe{corollary}
\Crefname{corollary}{Corollary}{Corollaries}
\newaliascnt{claim}{theorem}

\aliascntresetthe{claim}
\Crefname{claim}{Claim}{Claims}
\theoremstyle{definition}
\newaliascnt{definition}{theorem}
\newtheorem{definition}[definition]{Definition}
\aliascntresetthe{definition}
\Crefname{definition}{Definition}{Definitions}
\newaliascnt{remark}{theorem}

\aliascntresetthe{remark}
\Crefname{remark}{Remark}{Remarks}

\Crefname{theorem}{Theorem}{Theorems}
\Crefname{equation}{Equation}{Equations}
\Crefname{section}{Section}{Sections}
\Crefname{figure}{Figure}{Figures}
\Crefname{table}{Table}{Tables}
\Crefname{appendix}{Appendix}{Appendices}
\numberwithin{equation}{section}

\newcommand{\FIXP}{\mathsf{FIXP}}
\newcommand{\FIXPa}{\mathsf{FIXP}_{a}}
\newcommand{\PPAD}{\mathsf{PPAD}}
\newcommand{\SL}{\mathrm{SL}}

\newcommand{\clip}{\operatorname{clip}_{[0,1]}}

\newcommand{\Sq}{\operatorname{Sq}}

\providecommand{\Description}[1]{}
\allowdisplaybreaks
\hypersetup{
  colorlinks=true,
  pdfborder={0 0 0},
  linkcolor=RefLinkBorder,
  citecolor=CiteLinkBorder,
  urlcolor=URLLinkBorder,
  filecolor=URLLinkBorder,
  pdfauthor={}
}
\title{On the Complexity of Exact Autobidding Equilibria}

\author{%
  Yiyang Huang\thanks{\normalsize Beijing Institute of Technology, \texttt{huangyiyang@bit.edu.cn}}
  \and Zhengyang Liu\thanks{\normalsize Beijing Institute of Technology, \texttt{zhengyang@bit.edu.cn}}
  \and Yonglei Yan\thanks{\normalsize Beijing Institute of Technology, \texttt{yanyonglei@bit.edu.cn}}
}

\date{}

\hypersetup{
  pdftitle={On the Complexity of Exact Autobidding Equilibria},
  pdfauthor={Yiyang Huang, Zhengyang Liu, Yonglei Yan}
}
\hypersetup{pdftitle={On the Complexity of Exact Autobidding Equilibria}}
\begin{document}
\pagenumbering{roman}
\maketitle
\thispagestyle{empty}

\begin{abstract}
Autobidding equilibria link advertisers' value multipliers or participation probabilities through aggregate budget and return-on-investment (ROI) constraints. We establish the exact computational complexity of finding these equilibria across several canonical settings: exact equilibrium computation is $\FIXP$-complete for second-price ROI pacing, budget- or ROI-constrained generalized second-price (GSP) sponsored shopping, and second-price throttling. Furthermore, approximating the equilibrium control vector in distance to a true equilibrium is $\FIXPa$-complete when the target accuracy is provided as input.
\end{abstract}

\clearpage
\setcounter{tocdepth}{2}
\tableofcontents
\clearpage
\pagenumbering{arabic}

\section{Introduction}
Modern online advertising platforms rely on repeated auctions to allocate impressions among competing advertisers. To manage the scale and frequency of these auctions, platforms increasingly depend on autobidding: advertisers specify high-level campaign constraints, such as budgets or target return-on-investment (ROI), and automated agents optimize bids on their behalf~\cite{ABM19,Survey24,LPSZ24}. A bidding agent must therefore manage global campaign objectives across a vast inventory rather than optimize individual auctions in isolation.

Platforms typically operationalize campaign management via two widespread market mechanisms: \emph{multiplicative pacing} and \emph{throttling}. Under pacing, an agent scales its underlying valuations by a uniform multiplier; under throttling, the agent adjusts its auction participation probability. In standard equilibrium formulations, each bidder exercises a single scalar parameter across all eligible queries. Although each agent optimizes locally through this low-dimensional control, market outcomes are deeply interdependent: one bidder's multiplier determines the market-clearing price for its rivals, while its participation probability dictates how frequently rivals face competition and incur payments.

These interactions give rise to natural complementarity conditions. Consider an advertiser with an ROI target normalized to one: aggregate payment $P_i$ cannot exceed aggregate acquired value $V_i$. If the constraint is slack ($P_i < V_i$), the scaling multiplier hits an upper bound $A$; conversely, if the multiplier lies strictly below $A$, the target must bind tightly ($P_i = V_i$). Budget constraints follow an analogous complementarity rule with a fixed spending cap replacing acquired value. Because local bid adjustments shift prices and allocated values for others, these aggregate conditions induce rich feedback loops-propagating dependencies across the market even when individual bidders share only a sparse set of common auctions.

The distinction between budget and ROI constraints is already visible in ordinary second-price auctions: exact budget-pacing equilibrium computation is $\PPAD$-complete~\cite{CKK24,FHHH24}, whereas exact ROI-pacing equilibria can be irrational and are known to lie in $\FIXP$~\cite{FHHH24}. For GSP sponsored shopping and second-price throttling, prior work likewise establishes $\FIXP$ membership and hardness of weak approximation, while leaving exact completeness open~\cite{HLN26,CKK21}. The class $\FIXP$ captures fixed points of algebraic circuits and includes exact three-player Nash equilibrium computation~\cite{EY10}. Proving completeness requires showing that auction equilibrium conditions can implement general algebraic fixed points, beyond the threshold relations used for $\PPAD$-hardness.

We prove that exact equilibrium computation is $\FIXP$-complete and approximation in distance is $\FIXPa$-complete for second-price ROI pacing, budget- and ROI-constrained GSP, and second-price throttling. Ordinary ROI pacing and throttling remain hard with four goods per bidder and fixed finite rational data; GSP remains hard in a single full-display, unit-CTR auction with equilibrium distributions over legal tied orders.

\subsection{Main Results}

Our hardness reductions map a fixed-point problem to an auction instance and decode each source variable from a single bidder's control parameter $y_j$ via $x_i = a_i y_j + b_i$. This constitutes a \emph{separable linear reduction} for $\FIXP$ \cite{EY10} (see \Cref{sec:complexity}). Every auction equilibrium decodes to a source solution, and every source solution extends to an equilibrium, giving the solution-set correspondence in \Cref{eq:global-correspondence}. The following theorem presents the exact complexity classifications along with the associated instance restrictions.

\begin{theorem}[Exact autobidding equilibria]\label{thm:main}
The following equilibrium-computation problems are $\FIXP$-complete under separable linear reductions. In (ii) and (iii), hardness holds in one no-self-pricing GSP auction with full display, unit CTRs, positive rational values, and a distribution over legal tied orders as part of the equilibrium.

\begin{enumerate}[label=(\roman*),itemsep=2pt]
\item \textbf{Second-price ROI pacing}, for every fixed rational cap $A>1$, with at most two interested bidders per good, at most four goods per bidder, and all positive valuations in a fixed finite rational set depending only on $A$.
\item \textbf{Budget-constrained GSP}, with positive rational budgets and a universal constant bound on goods per bidder.
\item \textbf{ROI-constrained GSP}, with a universal constant bound on goods per bidder for the canonical instance-defined cap, and a constant bound depending only on $A$ for every prescribed fixed rational cap $A>1$.
\item \textbf{Second-price throttling}, with at most three positive bids per good, at most four goods per bidder, a common fixed tie priority, and fixed finite rational bid and budget sets.
\end{enumerate}
\end{theorem}

We next consider approximating equilibria in distance to an exact equilibrium. Affine decoding transfers distance guarantees after a rational rescaling of the accuracy parameter. The algorithmic task requires computing an approximate control vector, omitting allocations and tie-breaking probabilities, given the target accuracy $\varepsilon$ as part of the input. The following corollary establishes $\FIXPa$-completeness for this task under the same instance restrictions.

\begin{corollary}[Approximation in Distance]\label{cor:strong}
For the four models in \Cref{thm:main}, finding a rational control vector within a supplied rational $\varepsilon>0$ in $\ell_\infty$ of some exact equilibrium is $\FIXPa$-complete, with the same hardness restrictions.
\end{corollary}

Parts~(i)--(iv) of Theorem~\ref{thm:main} and the corresponding distance-approximation classifications in Corollary~\ref{cor:strong} follow, respectively, from Theorems~\ref{thm:spa-arbitrary-cap}, \ref{thm:budget-gsp}, \ref{thm:roi-gsp-rewritten}, and~\ref{thm:throttle}.

These results directly resolve several open questions in the literature. For second-price ROI pacing, Filos-Ratsikas, Hansen, H{\o}gh, and Hollender~\cite{FHHH24} established $\FIXP$ membership while explicitly leaving its completeness open. For second-price throttling, while Chen, Kroer, and Kumar~\cite{CKK21} established hardness under approximate equilibrium conditions, the complexity of finding an exact equilibrium remained open, a question also highlighted by Liu~\cite{Liu26}. Finally, in sponsored shopping, D{\"u}tting, Li, Paes Leme, Spendlove, and Teng~\cite{DLPST26} posed the computational complexity of GSP equilibria as an open problem, after which Hansen, Li, and Nie~\cite{HLN26} proved $\FIXP$ membership for both budget- and ROI-constrained variants. Our work settles these questions by establishing $\FIXP$-completeness across all these settings.

\subsection{Related Work}\label{sec:related-work}

\paragraph{Pacing and throttling.}
First-price budget pacing admits a convex program~\cite{CKP22}. Suitable conditions on the mechanism extend existence, uniqueness, and convex-program characterizations to broader pacing settings~\cite{AK26}. Early work studies second-price pacing structure and mixed-integer formulations~\cite{CKSM22}. Exact second-price budget pacing is $\PPAD$-complete~\cite{CKK24,FHHH24}. Weak approximation is also $\PPAD$-complete, with constant inapproximability in~\cite{CL25,Liu26}. ROI weak-approximation hardness is studied in~\cite{LT24}, while optimizing welfare or revenue over ROI equilibria faces further inapproximability barriers~\cite{AGPS26}. For second-price throttling, \cite{CKK21} gives weak-approximation algorithms with two bidders per good, and \cite{Liu26} proves $\PPAD$-hardness for every constant approximation parameter $\delta\in(0,1)$. Our results classify exact computation and approximation in distance.

\paragraph{GSP and structural restrictions.}
GSP originated in search position auctions \cite{EOS07,Varian07}. For value-maximizing bidders, its welfare guarantee can vanish and depends on the slot CTR discounts~\cite{DMMMZZ23}. D{\"u}tting, Li, Paes Leme, Spendlove, and Teng extended it to sponsored shopping and proved ROI equilibrium existence and welfare guarantees \cite{DLPST26}. Hansen, Li, and Nie proved $\FIXP$ membership, irrational equilibria, and weak-approximation $\PPAD$-completeness \cite{HLN26}. We use their identical-catalog comparison, aggregate-payment arguments, and quantile coupling \cite{HLN26}, adding exact sender conservation, independently forced ranges, and completion at all admissible lower prices to compose exact arithmetic (Section~\ref{sec:overview}). Exact budget and canonical-cap ROI GSP admit polynomial-time algorithms for fixed numbers of advertisers or slots \cite{HLYW26}.

\paragraph{Circuit Methods and Broader Connections.}
Our source problem is the exact generalized-circuit problem of~\cite{FGHLP23}. Pure-Circuit is used to prove $\PPAD$-hardness with constant error~\cite{DFHM24}. Complement copying, square identities, and polarization are standard circuit tools used here. For GSP, we also use these tools from~\cite{HLN26}. Our auction constructions add the preservation, range, and extension guarantees needed for exact composition. Studies of learning and dynamics address complementary questions about regret, welfare, and convergence~\cite{Survey24,BG19,GLLLS23,LPSZ24,PLPSZ24}. Dynamic budget throttling is studied for a single advertiser in repeated second-price auctions~\cite{CWW24}. Auction-design work studies reserve prices and efficiency guarantees for value-maximizing autobidders~\cite{BDMMZ21,DMMZ21,Mehta22}. Constrained no-regret bidding~\cite{AFZ24,DLTZ25} and diffuse Bayesian valuations~\cite{CMY26} address other objectives and information structures.

\section{Preliminaries}\label{sec:preliminaries}
\subsection{Auction Models and Equilibrium Conditions}\label{sec:auction-definitions}
Each market contains bidders $A_1,\ldots,A_n$ and goods $G_j$ indexed by a finite set $\mathcal J$. Pacing bidder $A_i$ submits $\alpha_i v_{ij}$ on each eligible good $G_j$. Throttling bidder $A_i$ instead has fixed bids $b_{ij}$ and participates with probability $\theta_i$. In every model, feasibility is accompanied by \emph{no unnecessary pacing}: a control below its upper bound requires a tight aggregate constraint.

\paragraph{Ordinary Second-Price ROI Pacing.}
Let bidder $A_i$ have nonnegative value $v_{ij}$ for good $G_j$ and submit bid $\alpha_i v_{ij}$, where $\alpha_i\in[1,A]$ and $A\ge1$. We sell each good separately, and only highest bidders can receive it. If several bidders tie for the highest bid, they may share the good, with bidder $A_i$ receiving a share $x_{ij}\ge0$ and paying its corresponding share of the price. These shares sum to one for every good with a positive bid.

The second-price rule determines the price $p_j$: sort all bids on $G_j$ in nonincreasing order, retaining repeated values, and take the second entry as the price. If there is only one bidder, $p_j=0$. When at least two bidders tie for the highest bid, $p_j$ equals that highest bid. Bidder $A_i$ pays $x_{ij}p_j$, so its acquired value and total payment are $V_i=\sum_jx_{ij}v_{ij}$ and $P_i=\sum_jx_{ij}p_j$. We call $V_i-P_i$ the bidder's \emph{ROI surplus}.

\begin{definition}[Second-price ROI pacing equilibrium]\label{def:spa-roi-equilibrium}
Following \cite{LT24,FHHH24}, we say $(\alpha,x)$, with $\alpha=(\alpha_i)\in[1,A]^n$ and a valid allocation $x=(x_{ij})\in[0,1]^{n\times|\mathcal J|}$, is a second-price ROI pacing equilibrium with ROI target one if
\begin{equation}\label{eq:roi-equilibrium}\left\{\begin{aligned}&1\le\alpha_i\le A,\\&P_i\le V_i,\\&P_i<V_i\ \Longrightarrow\ \alpha_i=A.\end{aligned}\right.\qquad(i=1,\ldots,n).\end{equation}
\end{definition}

\paragraph{GSP Sponsored Shopping.}
Unlike ordinary second-price auctions, GSP sponsored shopping assigns each good a designated owner \cite{DLPST26,HLN26}. Bidder $A_i$ owns a catalog indexed by $\mathcal J_i\subseteq\mathcal J$, has value $v_{ij}>0$ for each owned good, and submits bid $\alpha_i v_{ij}$. We order all goods by nonincreasing bid; such an order is valid. Under the \emph{no-self-pricing} rule, a good's price per click is the bid of the first later good owned by a different bidder, or zero if no such good exists.

For example, suppose four goods belong respectively to $(A_1,A_1,A_2,A_3)$ and have ordered bids $(12,10,8,5)$. Their prices are $(8,8,5,0)$: the first good skips the second because they have the same owner. Adding a good can therefore change the prices of other goods in the ranking.

Let $\pi$ be a distribution over valid orders, $r_j(\omega)$ the position of good $G_j$ in order $\omega$, and $p_{ij}(\omega)$ its price per click. Write $c_k$ for the click-through rate (CTR) of position $k$, with $c_k=0$ at undisplayed positions. The expected acquired value and payment of $A_i$ are
\[V_i=\mathbb E_{\omega\sim\pi}\!\left[\sum_{j\in\mathcal J_i}c_{r_j(\omega)}v_{ij}\right],\qquad P_i=\mathbb E_{\omega\sim\pi}\!\left[\sum_{j\in\mathcal J_i}c_{r_j(\omega)}p_{ij}(\omega)\right].\]
Our constructions use one auction with every good displayed at unit CTR. In this case, $V_i=\sum_{j\in\mathcal J_i}v_{ij}$ and $P_i=\mathbb E_{\omega\sim\pi}[\sum_{j\in\mathcal J_i}p_{ij}(\omega)]$. Under a budget constraint, each bidder has a positive budget $B_i$; an ROI-constrained bidder has no budget parameter. Both models satisfy their corresponding no-unnecessary-pacing conditions.

\begin{definition}[Budget-constrained GSP equilibrium]\label{def:budget-gsp-equilibrium}
We say $(\alpha,\pi)$, with $\alpha=(\alpha_i)\in[0,1]^n$ and a probability distribution $\pi$ over orders valid at $\alpha$, is a budget-constrained GSP equilibrium if
\begin{equation}\label{eq:budget-equilibrium}\left\{\begin{aligned}&0\le\alpha_i\le1,\\&P_i\le B_i,\\&P_i<B_i\ \Longrightarrow\ \alpha_i=1.\end{aligned}\right.\qquad(i=1,\ldots,n).\end{equation}
\end{definition}

\begin{definition}[ROI-constrained GSP equilibrium]\label{def:roi-gsp-equilibrium}
We say $(\alpha,\pi)$, with $\alpha=(\alpha_i)\in[1,A]^n$ and a probability distribution $\pi$ over orders valid at $\alpha$, is an ROI-constrained GSP equilibrium with ROI target one and cap $A\ge1$ if
\begin{equation}\label{eq:roi-gsp-equilibrium}\left\{\begin{aligned}&1\le\alpha_i\le A,\\&P_i\le V_i,\\&P_i<V_i\ \Longrightarrow\ \alpha_i=A.\end{aligned}\right.\qquad(i=1,\ldots,n).\end{equation}
\end{definition}

We consider two conventions for the ROI-GSP cap. The first is the canonical instance-defined cap of \cite{HLN26}:
\[A_0=1+\frac{\sum_i\sum_{j\in\mathcal J_i}v_{ij}}{\min\{v_{ij}:1\le i\le n,\ j\in\mathcal J_i\}}.\]
The second is any prescribed rational cap $A>1$ fixed independently of the auction instance.

\paragraph{Second-Price Throttling.}
Unlike ordinary second-price pacing, throttling uses a participation probability $\theta_i\in[0,1]$ instead of a multiplier. Bidder $A_i$ has fixed nonnegative bids $b_{ij}$ and a positive budget $B_i$. Participation is independent across bidders. Each good is allocated separately by a second-price auction among its participating bidders, and all auctions use the same fixed bidder priority to resolve ties.

Let $P_i(\theta_{-i})$ denote the expected total payment of $A_i$ conditional on its participation, where $\theta_{-i}$ lists the other bidders' probabilities.

\begin{definition}[Second-price throttling equilibrium]\label{def:throttling-equilibrium}
We say $\theta=(\theta_i)\in[0,1]^n$ is a second-price throttling equilibrium if, for every bidder,
\begin{equation}\label{eq:throttle-conditions}\left\{\begin{aligned}&0\le\theta_i\le1,\\&\theta_iP_i(\theta_{-i})\le B_i,\\&\theta_iP_i(\theta_{-i})<B_i\ \Longrightarrow\ \theta_i=1.\end{aligned}\right.\end{equation}
\end{definition}
The last condition is analogous to no-unnecessary-pacing: a bidder whose budget is not exhausted cannot reduce its participation probability. For convenience, we use the following equivalent expression for the participation probability \cite{CKK21}:
\begin{equation}\label{eq:throttle-response}\theta_i=\frac{B_i}{\max\{B_i,P_i(\theta_{-i})\}}.\end{equation}
Since $B_i>0$, every bidder has positive participation probability at equilibrium.

\subsection{FIXP Completeness and Strong Approximation}\label{sec:complexity}
We now introduce the complexity framework and the exact and approximate computational tasks used in our results. In the framework of Etessami and Yannakakis \cite{EY10}, $\FIXP$ membership is established by reducing a problem to finding a fixed point of a continuous map $F:[0,1]^d\to[0,1]^d$ represented by an algebraic circuit with rational constants and operations $+,-,\times,\div,\min,\max$. Brouwer's fixed-point theorem guarantees that such a map has a fixed point. A problem that is both in $\FIXP$ and $\FIXP$-hard is $\FIXP$-complete.

The membership results for these models follow from prior work, as discussed in Section~\ref{sec:related-work}, together with the prescribed-cap ROI-GSP result in Appendix~\ref{sec:prescribed-membership}. We focus on proving hardness. Our reductions are \emph{separable linear} ($\SL$): they construct a target instance in polynomial time and decode each source variable from one coordinate of a target solution $y$ as $L_i(y)=a_i y_{\sigma(i)}+b_i$. Here $\sigma(i)$ selects one coordinate, and $a_i,b_i$ are rational numbers computable in polynomial time. Note that every target solution must decode to a source solution. To prove hardness, we reduce the exact generalized-circuit problem in Lemma~\ref{lem:source} to each auction equilibrium problem, decoding each source variable from the multiplier or participation probability of one designated bidder.

We also establish $\FIXPa$-completeness for each model. Unlike exact computation in $\FIXP$, $\FIXPa$ concerns \emph{strong approximation}, also called approximation in distance \cite{EY10}. This task asks for a rational vector within a supplied rational $\varepsilon>0$ of some exact solution in the maximum norm $\|z\|_\infty=\max_i|z_i|$. Thus $\FIXPa$ captures the complexity of approximating exact fixed points to a specified accuracy. Note that our approximation concerns only the multiplier or participation vector.

\subsection{Source Circuits and Outdegree Control}\label{sec:source-equations}
We now specify the source problem used in the hardness reductions. A \emph{generalized circuit} has variables $x_1,\ldots,x_m\in[0,1]$, each designated as the output of exactly one gate. A gate imposes an equation $x_i=f(x_j,x_k)$ or its unary version. The network may contain directed cycles, and a gate may use its own output as an input. A solution assigns values to all variables so that every gate equation holds simultaneously. Equivalently, it is a fixed point of the continuous map from the cube to itself specified by the gates. We call such an assignment a \emph{source solution}.

Our reductions implement the following complete gate set.

\begin{lemma}[Exact Generalized Circuits, {\cite{FGHLP23}}]\label{lem:source}
Finding an exact solution of a generalized circuit with gate functions
\[C(x)=1-x,\qquad D(x)=\min\{2x,1\},\qquad \phi(x,y)=\frac{(1+x)(1+y)}4\]
is $\FIXP$-complete under $\SL$ reductions.
\end{lemma}

Among these gate functions, the complement gate $C$ and capped-doubling gate $D$ are piecewise linear, whereas $\phi$ uses multiplication. Each reduction must implement these three functions and preserve their equations when the corresponding auction gadgets are connected.

Consider the situation in which a gate's output is used as input to many other gates. The bidder representing this output must transmit its control to the corresponding gadgets, potentially requiring many goods and preventing the desired bound on goods per bidder. We use the following method to address this problem. The \emph{indegree} of a gate is its number of input positions, and its \emph{outdegree} is the number of input positions of gates that use its output; its degree is the sum of its indegree and outdegree. Two successive applications of the complement gate $C$ from Lemma~\ref{lem:source} produce an exact copy because $C(C(x))=x$. The next lemma uses these copies to bound the outdegree of every gate by two.

\begin{lemma}[Bounding Gate Outdegree]\label{lem:bounded-use}
A bounded-indegree gate network with complement gates admits a linear-size expansion with outdegree at most two and no self-loops. Restriction to the original variables gives a bijection between the expanded and original solution sets.
\end{lemma}

\begin{proof}
Let $x$ be the output of a gate with outdegree $d$, and set $c_0=x$. For each $j=1,\ldots,d$, add two complement gates with outputs $a_j=C(c_{j-1})$ and $c_j=C(a_j)$, and replace the $j$th input occurrence of $x$ by $c_j$. Since $C(C(t))=t$, every solution has $c_j=x$ and $a_j=1-x$. Each $c_j$ is used by its assigned gate and, except for the last, by the next copy pair. Each $a_j$ has outdegree one, and $x$ is used only by the first copy pair. Thus every gate has outdegree at most two.

If a gate uses its own output, that input is replaced by a fresh copy $c_j$, so the self-loop is removed. Since the indegrees are bounded, the sum of the outdegrees is linear in the number of gates. The number of copies is therefore also linear in the size of the network. Restricting a solution of the expanded network to the original variables recovers the original equations. Conversely, every original solution uniquely extends by setting $c_j=x$ and $a_j=1-x$.
\end{proof}

For later formulas, we write $[t]_+=\max\{t,0\}$ and $\operatorname{clip}_{[a,b]}(t)=\min\{b,\max\{a,t\}\}$. 

\section{Technical Overview}
\label{sec:overview}
We now outline the main difficulties and methods in our hardness reductions. The reductions represent exact generalized circuits through bidders' controls and auction payments, establishing a correspondence between source solutions and auction equilibria.

\paragraph{Our approach and technical challenges.}
We use the exact generalized circuits of \cite{FGHLP23}, complement copying (Lemma~\ref{lem:bounded-use}), and square and polarization identities as circuit tools. For GSP, we also use the identical-catalog comparisons, aggregate-payment bounds, and quantile coupling of \cite{HLN26}. But these tools do not directly provide the exact gates needed here: threshold constraints from approximate reductions do not specify exact outputs throughout an interval, and an isolated irrational equilibrium does not supply a gate with a freely varying input. Our reductions require exact input--output identities, completion for every permitted input, and preservation of these identities when gadgets are connected.

We implement the source gates through exact payment identities, deriving the required control ranges and price orders from equilibrium conditions before using these identities. Every permitted source assignment admits a completion by suitable auxiliary controls and auction outcomes. To represent the whole source circuit, we must connect the auction gadgets so that the output of one can serve as an input to others. The difficulty is that a bidder uses the same control in its upstream and downstream gadgets. A new signal makes this control affect another bidder's payment, but can also change the sender's own payment or acquired value. Its upstream equation may then cease to hold, even if both gadgets implement the intended relations in isolation. This issue is especially important on a cyclic circuit, where there is no order in which we can fix each gadget's output and then treat it as an unaffected input. Section~\ref{sec:roi-overview} explains the different ways in which our four models address this effect.

Our argument has four parts: we first represent circuit variables by bidders, then explain how payment equations produce affine and nonlinear relations and implement the source gates, show how signals preserve the input bidders' equilibrium conditions when gadgets are connected, and finally prove the correspondence with complete auction equilibria.

\subsection{Representing Variables and Connecting Gadgets}
To represent circuit values through bidders, we choose an encoding for each model. In the three pacing models, a source value $x\in[0,1]$ is represented by a multiplier $\alpha_{\mathrm{out}}=a+bx$, with fixed rational $a$ and $b>0$. For throttling, the internal wire encoding uses the ratio of absence probability to participation probability; a separate decoding bidder will provide an affine decoding rule.

Having described the representation of variables, we now introduce the bidder roles in a generic arithmetic gadget. Such a gadget must enforce a relation between its input values and its output value through the equilibrium conditions of its bidders, enabling transmission and connection in the circuit. For a gadget with $k$ inputs, we use the following notation:
\begin{itemize}
\item $A_{\mathrm{in},j}$: input bidder carrying input $j$ through its multiplier $\alpha_{\mathrm{in},j}$ in a pacing model or its participation probability $\theta_{\mathrm{in},j}$ in throttling, for $j=1,\ldots,k$.
\item $A_{\mathrm{out}}$: output bidder whose multiplier $\alpha_{\mathrm{out}}$ or participation probability $\theta_{\mathrm{out}}$ represents the output determined by the gadget.
\item $A_{\mathrm{aux}}$: auxiliary bidder introduced to help enforce the required relation between the inputs and output. If a gadget uses $t$ auxiliary bidders, we label them $A_{\mathrm{aux},\ell}$ for $\ell=1,\ldots,t$.
\end{itemize}
We collectively call the input and output bidders \emph{wire bidders}. Note that a wire bidder may act as $A_{\mathrm{out}}$ in one gadget and as $A_{\mathrm{in},j}$ in another. We call the former its \emph{upstream gadget} and the latter its \emph{downstream gadget}. A wire bidder's control is determined in its upstream gadget and then used as an input in its downstream gadgets, connecting the gadgets into an arithmetic circuit. Our task is to transmit the input bidders' multipliers or participation probabilities to the output bidder through intermediate connections.

How do we connect the inputs to the output? An auxiliary bidder helps establish this connection: it interacts with the input and output bidders so that the input can affect the output. We call this transmission a \emph{signal}, and the way it is transmitted depends on the model. In pacing, the input bidder's scaled bid can determine a price paid by the other bidder. In throttling, the input bidder's presence or absence changes an event in which the other bidder pays.

With the variables and bidder roles specified, we next explain how auction gadgets implement the operations needed by the source circuit. We begin with the three source gates and the two types of arithmetic they require. After describing the local mechanisms, we explain how to connect them while preserving the input bidders' equilibrium conditions.

\subsection{Variable Transmission by Gadgets}
\label{subsec:overview-arithmetic-relations}
\label{subsec:overview-source-gates}
Recall the three source gates from Lemma~\ref{lem:source} \cite{FGHLP23}: $C(x)=1-x$, $D(x)=\min\{2x,1\}$, and $\phi(x,y)=(1+x)(1+y)/4$. We use two types of arithmetic to implement these gates: affine operations and nonlinear operations. Affine operations handle complement $C(x)$ and capped doubling $D(x)$ by scaling inputs with fixed constants, adding or subtracting them, and applying an output cap. Nonlinear operations supply the dependence between variables needed for the shifted product $\phi(x,y)$. They do not directly compute multiplication. Instead, we combine their responses with the other operations to construct a square, then obtain multiplication through polarization. We explain this step at the end of this subsection.

These two types of arithmetic guide our gadget design. To construct either type, we first specify how its output should depend on its inputs. In the three pacing models, this dependence is carried by multipliers; in throttling, it is carried by participation probabilities. We then choose item values, bids, and financial parameters so that the equilibrium conditions enforce the required dependence. As introduced above, our constructions use $A_{\mathrm{aux}}$ as a bridge, giving the transmission chain $A_{\mathrm{in},j}\to A_{\mathrm{aux}}\to A_{\mathrm{out}}$. In the pacing constructions, we call the goods carrying the first connection \emph{signal goods} and those linking the latter two bidders \emph{main goods}. We next explain how this chain produces affine and nonlinear operations in the four models.

\paragraph{Ordinary second-price ROI pacing.}
\emph{From $A_{\mathrm{in},j}$ to $A_{\mathrm{aux}}$:} we let them bid for a signal good and make $A_{\mathrm{aux}}$ always win. It therefore always pays the bid of $A_{\mathrm{in},j}$, bringing the input multiplier into its payment. Raising this multiplier increases that payment, leaving less ROI surplus for the next good.

\emph{From $A_{\mathrm{aux}}$ to $A_{\mathrm{out}}$:} we let $A_{\mathrm{out}}$ and $A_{\mathrm{aux}}$ compete for a main good at tied bids. In the affine gadget, they value this good equally, so their common multiplier determines its price, and an allocation share divides its value and payment between them. Their ROI conditions require their losses on this good to offset their respective outside surpluses, and the outside surplus of $A_{\mathrm{aux}}$ depends on the input multiplier. The input multiplier therefore determines the price of this good and hence the output multiplier.
The nonlinear gadget keeps the same transmission chain but gives $A_{\mathrm{out}}$ and $A_{\mathrm{aux}}$ different values for the shared good. The input still changes the outside surplus of $A_{\mathrm{aux}}$, but a given shared price now creates different losses for the two bidders. The ROI condition of $A_{\mathrm{out}}$ determines how much of the good it needs, while the condition of $A_{\mathrm{aux}}$ links the remaining share to its input-dependent surplus. These two requirements make the shared price, and therefore the output multiplier, depend nonlinearly on the input.

\paragraph{Budget-constrained GSP.}
Unlike ordinary second-price auctions, each good in GSP has an owner. To put the input multiplier into the signal bid, we give the signal good to $A_{\mathrm{in},j}$, while $A_{\mathrm{aux}}$ owns receiving goods priced by this signal. The input multiplier therefore determines part of the spending of $A_{\mathrm{aux}}$, leaving the rest of its budget for its main good. To pass this dependence to $A_{\mathrm{out}}$, we use tied main goods owned by $A_{\mathrm{aux}}$ and $A_{\mathrm{out}}$. Their common multiplier sets the upper main price, while their order determines which bidder pays it. In the affine gadget, the combined main payment is determined by the common multiplier, independently of the order. As the spending of $A_{\mathrm{aux}}$ that depends on the input multiplier changes, the two budget conditions determine the remaining main payment and hence the output multiplier.

For nonlinearity, we replace the fixed lower main price with the bid of a middle good owned by $A_{\mathrm{aux}}$. This bid varies with the multiplier shared by the main goods. The same tied-order probability now divides payments between two prices that both depend on the output. Requiring both budgets to be spent gives a nonlinear relation between the input-dependent payment and the output multiplier. Similar tied-order payment equations appear in the irrational GSP examples of \cite{HLN26}.

\paragraph{ROI-constrained GSP.}
We use the same path from the bid of $A_{\mathrm{in},j}$ to the payment of receiving goods owned by $A_{\mathrm{aux}}$, then from the main-order payments to the multiplier of $A_{\mathrm{out}}$. The difference is that ROI conditions compare payment with acquired value rather than a fixed budget. We use private signal pairs to supply chosen ROI-surplus adjustments, setting the payment that the remaining goods must make. Input payments reduce the amount available for the main payment of $A_{\mathrm{aux}}$, and the two core ROI equalities transmit this change to $A_{\mathrm{out}}$. The affine gadget offsets the variation in the lower main price with compensation goods. The nonlinear gadget instead uses a middle good to make that price depend on the output multiplier. These choices give the two types of response along the same transmission chain.

\paragraph{Second-price throttling.}
\emph{From $A_{\mathrm{in},j}$ to $A_{\mathrm{aux}}$:} we use the participation or absence of $A_{\mathrm{in},j}$ to control the payment of $A_{\mathrm{aux}}$. A presence signal makes $A_{\mathrm{aux}}$ pay a fixed amount when $A_{\mathrm{in},j}$ participates, while an absence signal makes it pay a fixed amount when $A_{\mathrm{in},j}$ is absent. Conditional on $A_{\mathrm{aux}}$ participating, these signals contribute a constant times $\theta_{\mathrm{in},j}$ or $1-\theta_{\mathrm{in},j}$, respectively. By combining the signal goods for its inputs and adding a constant-payment good, we realize a prescribed conditional-payment expression of the form
\[P_{\mathrm{aux}}=c+\sum_j\bigl(a_j\theta_{\mathrm{in},j}+d_j(1-\theta_{\mathrm{in},j})\bigr),\qquad a_j,d_j\ge0.\]
The coefficients specify how strongly each input affects the payment. Presence signals provide increasing contributions, and absence signals provide decreasing contributions. The budget condition then determines the participation probability of $A_{\mathrm{aux}}$: when full participation is unaffordable, it equals its budget divided by this conditional payment \cite{CKK21}.

\emph{From $A_{\mathrm{aux}}$ to $A_{\mathrm{out}}$:} we use the participation probability of $A_{\mathrm{aux}}$ in the same way to determine the conditional payment of $A_{\mathrm{out}}$, whose budget response determines the output. By choosing the signal coefficients, constant payments, and budgets, we implement complement, capped doubling, halving, a reciprocal, and a shifted average. Thus the signal goods build the required payment expressions, and the budget responses turn those expressions into the operations needed for both types of arithmetic.

\paragraph{Implementing the exact source gates.}
We now combine the local operations to implement the three source gates. The affine constructions provide complement $C(x)$ and capped doubling $D(x)$. To obtain the shifted product $\phi(x,y)$, we combine affine operations with the nonlinear responses supplied by the gadgets. The following theorem states the arithmetic operations sufficient for this step.

\begin{theorem}[Exact Arithmetic from a Nonlinear Response]
\label{thm:criterion}
Either of the following gate sets implements the source functions of \Cref{lem:source} by fixed-size rational subcircuits:
\begin{enumerate}[label=(\roman*),itemsep=2pt]
\item Signed clipped affine expressions and a continuous strictly
monotone response $h$ on a fixed compact rational interval $I_0$,
whose inverse on a nondegenerate closed rational interval
$J\subset\operatorname{int}h(I_0)$ is a nonaffine
quadratic-over-linear function with rational coefficients
and no pole on $J$.
\item Complement, capped doubling, halving, capped addition,
and $I(x)=1/(1+x)$ on $[0,1]$.
\end{enumerate}
Every source input assignment extends through the subcircuits, and every solution of their equations gives the required source outputs.
\end{theorem}

The proof is given in Appendix~\ref{sec:reciprocal}. Ordinary ROI pacing and budget-constrained GSP use alternative~(i). ROI-constrained GSP follows the same inverse-feedback and square-extraction approach, with additional range and endpoint checks. Affine feedback compares the nonlinear output with a desired target and adjusts its input. At a solution, this forces the output to equal the target, so the input gives the inverse value. A quadratic inverse yields a square after removing its constant and linear terms. Otherwise, removing the affine part yields a reciprocal operation, and differences and further reciprocals produce a quadratic expression from which a square can be extracted. Polarization then gives multiplication by subtracting the individual squares from the square of a sum \cite{FGHLP23}. Applying multiplication to $(1+x)/2$ and $(1+y)/2$ gives $\phi(x,y)$. Section~\ref{subsec:roi-arithmetic-rewritten} gives these checks and the normalization map into the source-encoding interval for ROI GSP.

Throttling uses alternative~(ii): its reciprocal response supplies the nonlinear operation needed to construct squares and multiplication. Its internal encoding is nonlinear in participation probabilities, so an additional budget response converts the encoded value into a participation probability affine in the source value. This gives the required single-coordinate affine decoder, with the payment from this additional use included in the outgoing-charge accounting.

These constructions give fixed-size rational implementations of the source gates with both solution directions. To assemble them into a circuit, we must also ensure that transmission from $A_{\mathrm{in},j}$ to $A_{\mathrm{aux}}$ does not change the equilibrium condition of $A_{\mathrm{in},j}$. The next subsection explains how we handle this effect.

\subsection{Preserving the Equilibrium Condition}
\label{sec:roi-overview}
We now explain how each pricing rule lets us preserve the input bidder's equilibrium condition while supplying the payment dependence needed for the local arithmetic.

\paragraph{Ordinary second-price ROI pacing.}
For each input bidder $A_{\mathrm{in},j}$, we make it always lose the signal good when bidding against $A_{\mathrm{aux}}$. The second-price rule then makes $A_{\mathrm{aux}}$ pay the bid of $A_{\mathrm{in},j}$. Also, $A_{\mathrm{in},j}$ receives no value and makes no payment on the signal good, so its ROI condition is preserved.

\paragraph{Budget-constrained GSP.}
Unlike ordinary second-price pacing, GSP cannot use the same construction. This is because, as we just explained, the signal good must belong to $A_{\mathrm{in},j}$. Displaying this good therefore adds a payment for $A_{\mathrm{in},j}$, changing its budget condition.

To control this added payment, we introduce a guard bidder for each input bidder. We choose the value of its guard good and its budget and arrange the bids so that the guard good is the first good after the signal good owned by another bidder. The no-self-pricing rule then makes the signal good pay the guard's fixed bid. For each such fixed payment, we increase the budget of $A_{\mathrm{in},j}$ by the same amount. The added payment and budget increase cancel, preserving its budget condition. The use of guard bids and budget adjustments is similar to the budget-GSP construction of \cite{HLN26}.

\paragraph{ROI-constrained GSP.}
Preserving the equilibrium condition is more difficult for ROI GSP. This is because adding a signal good to $A_{\mathrm{in},j}$ changes not only its payment but also its acquired value, and we cannot even adjust a budget to balance the added payment. We therefore need to supply additional payment that offsets the added value and preserves its ROI equilibrium condition.

Our solution is to give $A_{\mathrm{in},j}$ a compensation good alongside the signal good, making their total payment equal their total value. We introduce two supplier groups, each with $n$ bidders, where $n$ depends only on the fixed cap $A$. Within each group, bidders have identical valuation catalogs. The first group sets the signal price, and the second sets the compensation price. Some goods of the first group are also priced by the second, linking its total payment to both groups' multipliers.

The first group's acquired value is fixed because all its goods are displayed. We force the multipliers of this group to lie strictly below the cap, so its total payment must equal that fixed value. Part of this payment comes from equal-bid goods within the group, and the rest comes from prices set by the second group. If the first part increases, the second must decrease by the same amount. We choose the catalog values and multiplicities so that these two parts are proportional, by the same factor, to the signal price and the compensation price. The fixed total payment then forces the sum of these two prices to equal the pair's total value.

The suppliers therefore link two prices that would otherwise affect $A_{\mathrm{in},j}$ separately. A rise in the signal good's price is matched by an equal fall in the compensation good's price. Since their sum equals the combined value of the signal good and the compensation good, adding these two goods increases the payment and acquired value of $A_{\mathrm{in},j}$ by the same amount. Its ROI surplus is unchanged. At the same time, its multiplier still scales the signal bid, which sets the receiving bidder's price. This lets the multiplier affect another bidder without changing the ROI condition of $A_{\mathrm{in},j}$.

\paragraph{Second-price throttling.}
The presence and absence signals from Section~\ref{subsec:overview-arithmetic-relations} have different effects on $A_{\mathrm{in},j}$. A presence signal adds no payment: it loses whenever $A_{\mathrm{aux}}$ participates and otherwise can win only at price zero. An absence signal instead gives $A_{\mathrm{in},j}$ the highest bid, with an always-participating dummy providing a fixed lower bid. Whenever $A_{\mathrm{in},j}$ participates, it wins and pays that bid, adding a fixed conditional charge.

We handle this charge by letting it supply part of a payment that the upstream gadget already requires. The prescribed conditional-payment equation of $A_{\mathrm{in},j}$ contains a constant term. Without outgoing absence signals, we would realize the whole constant using a separate constant-payment good. With these signals connected, their fixed charges already supply part of that constant, so we reduce the payment on the separate good by exactly their total.

We choose the coefficients and limit the number of connections so that this reduction never exceeds the required constant. The outgoing signals and the separate good then supply exactly the same total constant payment as before, preserving the full conditional-payment equation and budget response of $A_{\mathrm{in},j}$.

With the input bidders' equilibrium conditions preserved, we can now assemble the gadgets into a full auction.

\subsection{Completing the Reduction}
\label{sec:roi-range-overview}
We connect the gadgets according to the source circuit and verify both directions of the reduction. Every auction equilibrium satisfies the local gate equations once the required control ranges and price orders are established. Shared wire bidders make these equations hold simultaneously, so their controls decode to a source solution, including on directed cycles.

Conversely, every source solution determines wire controls that extend to local equilibria through suitable auxiliary controls and allocations. Input preservation keeps these conditions valid when the gadgets are connected. For GSP, we combine local ordering choices into a distribution of legal global orders with polynomial support and unchanged expected payments, using quantile coupling as in \cite{HLN26}. Thus every source solution extends to a full auction equilibrium.

For a source circuit $\mathcal C$, let $\mathcal R_{\mathcal C}$ be the constructed auction and $L_{\mathcal C}$ its coordinatewise affine decoder. The two directions give
\begin{equation}\label{eq:global-correspondence}L_{\mathcal C}\bigl(\operatorname{Eq}(\mathcal R_{\mathcal C})\bigr)=\operatorname{Fix}(\mathcal C).\end{equation}
Here $\operatorname{Eq}$ and $\operatorname{Fix}$ denote the auction equilibrium set and source solution set, respectively. Fixed-size gate implementations and copying yield polynomial-size auctions with the stated local bounds. The affine decoder gives a separable linear reduction and transfers distance guarantees. Together with membership, this proves $\FIXP$-completeness and $\FIXPa$-completeness.

\section{Second-Price ROI Pacing}
\label{sec:ordinary-roi}
\label{subsec:spa-arbitrary-cap}
We first consider ordinary second-price ROI pacing with a fixed rational cap $A>1$.

\begin{theorem}[Every Nontrivial Fixed Cap Is Hard]\label{thm:spa-arbitrary-cap}
For every fixed rational cap $A>1$, exact second-price ROI pacing equilibrium computation is $\FIXP$-complete under separable linear reductions. Approximating the multiplier vector in distance to an exact equilibrium is $\FIXPa$-complete under the same restrictions. Hardness holds with at most four positive valuations per bidder, two per good, and positive valuations in a fixed finite rational set depending only on $A$.
\end{theorem}

At $A=1$, all multipliers equal one. Allocate each good to a highest-value bidder; the second price is at most its value. ROI is feasible and all bidders are at the cap, giving a rational equilibrium computable in polynomial time.

\subsection{A Common Auction Structure}\label{subsec:spa-common}
Fix a rational $A>1$. We reduce the generalized circuits of Lemma~\ref{lem:source}, with variables in $[0,1]$ and gate functions
\[C(x)=1-x,\qquad D(x)=\min\{2x,1\},\qquad \phi(x,y)=\frac{(1+x)(1+y)}4.\]

An affine gadget supplies $C$, $D$, and rescaling; a nonlinear gadget supplies the additional operation needed for $\phi$. We specify the wire encoding after proving their output ranges.

Let $k\in\{0,1,2\}$ be the number of gadget inputs. Its bidder roles are:
\begin{itemize}
\item $A_{\mathrm{in},i}$: input bidder with multiplier $\alpha_{\mathrm{in},i}$, for $i=1,\ldots,k$.
\item $A_{\mathrm{out}}$: output bidder with multiplier $\alpha_{\mathrm{out}}$.
\item $A_{\mathrm{aux}}$: auxiliary bidder with multiplier $\alpha_{\mathrm{aux}}$.
\end{itemize}
Wire roles follow Section~\ref{sec:overview}. Each $A_{\mathrm{aux}}$ is fresh and distinct from all wire bidders.

We next define the goods:
\begin{itemize}
\item $G_{\mathrm{out}}$: shared output good.
\item $G_{\mathrm{in},j}$: input good, for $j=1,\ldots,k$.
\item $G_{\mathrm{pout}}$: private good of $A_{\mathrm{out}}$.
\item $G_{\mathrm{paux}}$: private good of $A_{\mathrm{aux}}$.
\end{itemize}

The table shows $k=2$. For $k=1$, omit $G_{\mathrm{in},2}$ and $A_{\mathrm{in},2}$; for $k=0$, omit both input rows and columns.

\begin{table}[H]
\centering
\begingroup
\setlength{\tabcolsep}{5pt} \renewcommand{\arraystretch}{1.2}
\sbox{\TableBody}{%
\begin{tabular}{|c|c|c|c|c|}
\hline
Goods $\backslash$ Bidders & $A_{\mathrm{in},1}$ & $A_{\mathrm{in},2}$ & $A_{\mathrm{out}}$ & $A_{\mathrm{aux}}$\\ \hline
$G_{\mathrm{out}}$ & $0$ & $0$ & $1$ & $v$\\ \hline
$G_{\mathrm{in},1}$ & $d_1$ & $0$ & $0$ & $r_1$\\ \hline
$G_{\mathrm{in},2}$ & $0$ & $d_2$ & $0$ & $r_2$\\ \hline
$G_{\mathrm{pout}}$ & $0$ & $0$ & $t$ & $0$\\ \hline
$G_{\mathrm{paux}}$ & $0$ & $0$ & $0$ & $x$\\ \hline
\end{tabular}%
}
\usebox{\TableBody}
\caption{Valuations in the common second-price ROI gadget with two inputs.}
\label{tab:spa-common-valuations}
\endgroup
\end{table}

In the constructions below, the affine gadget uses $v=1$, while the nonlinear gadget uses $v>1$. 
If $r_i>Ad_i>0$, $A_{\mathrm{aux}}$ wins $G_{\mathrm{in},i}$ and pays $d_i\alpha_{\mathrm{in},i}$ for every feasible multiplier vector. The input bidder receives no value and pays nothing, preserving its ROI condition.

The private goods $G_{\mathrm{pout}}$ and $G_{\mathrm{paux}}$ have single positive bidders $A_{\mathrm{out}}$ and $A_{\mathrm{aux}}$, respectively. At price zero they supply surpluses $t$ and $x$, which, together with input-good surplus, offset losses on $G_{\mathrm{out}}$.
On $G_{\mathrm{out}}$, equal bids let the ROI conditions of $A_{\mathrm{out}}$ and $A_{\mathrm{aux}}$ link $\alpha_{\mathrm{out}}$ to $\alpha_{\mathrm{in},i}$. The local completions assign positive shares to both bidders.

\subsubsection{Affine gadget}
\begin{figure}[H]
\centering
\begin{tikzpicture}[
bidder/.style={ellipse,draw,line width=0.55pt,fill=blue!4,minimum width=24mm,minimum height=11mm,inner sep=3pt},
good/.style={rectangle,draw,rounded corners=1pt,line width=0.4pt,fill=black!2,minimum width=12mm,minimum height=6mm,inner sep=2pt},
edge/.style={->,>=stealth,line width=0.4pt},
val/.style={midway,fill=white,inner sep=1.2pt}
]
\node[bidder] (i1) at (-6.4,1.7) {$A_{\mathrm{in},1}$};
\node[bidder] (i2) at (-6.4,-1.7) {$A_{\mathrm{in},2}$};
\node[bidder] (a) at (0,0) {$A_{\mathrm{aux}}$};
\node[bidder] (o) at (6.4,0) {$A_{\mathrm{out}}$};

\node[good] (g1) at (-3.2,1.7) {$G_{\mathrm{in},1}$};
\node[good] (g2) at (-3.2,-1.7) {$G_{\mathrm{in},2}$};
\node[good] (g) at (3.2,0) {$G_{\mathrm{out}}$};
\node[good] (pa) at (0,1.5) {$G_{\mathrm{paux}}$};
\node[good] (po) at (6.4,1.5) {$G_{\mathrm{pout}}$};

\draw[edge] (g1.west)--node[val] {$d_1$}(i1.east);
\draw[edge] (g1.south east)--node[val] {$r_1$}(a.north west);
\draw[edge] (g2.west)--node[val] {$d_2$}(i2.east);
\draw[edge] (g2.north east)--node[val] {$r_2$}(a.south west);
\draw[edge] (g.west)--node[val] {$1$}(a.east);
\draw[edge] (g.east)--node[val] {$1$}(o.west);
\draw[edge] (pa.south)--node[val] {$x$}(a.north);
\draw[edge] (po.south)--node[val] {$t$}(o.north);
\end{tikzpicture}
\caption{Affine gadget with two inputs. Ellipses represent bidders and rectangles represent goods. Here $v=1$, so $A_{\mathrm{aux}}$ and $A_{\mathrm{out}}$ both value $G_{\mathrm{out}}$ at $1$.}
\label{fig:spa-affine}
\end{figure}
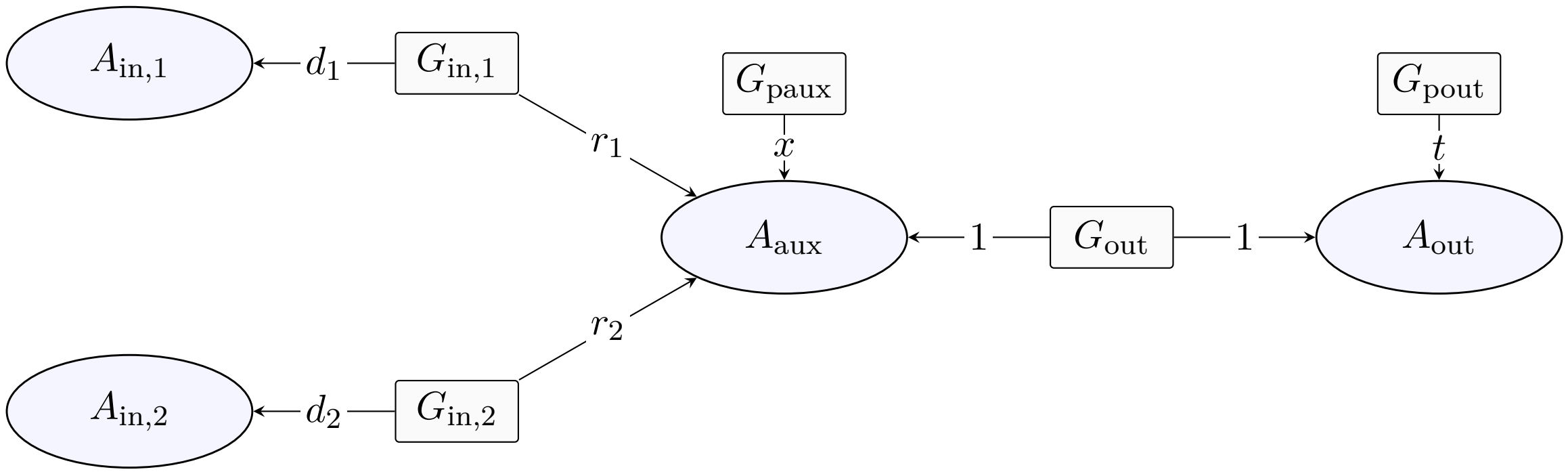

For the affine gadget, we set $v=1$, so that $A_{\mathrm{out}}$ and $A_{\mathrm{aux}}$ have the same valuation $1$ for $G_{\mathrm{out}}$. Figure~\ref{fig:spa-affine} shows the construction.

Raising $\alpha_{\mathrm{in},i}$ reduces the outside surplus of $A_{\mathrm{aux}}$. Equal bids on $G_{\mathrm{out}}$ turn this change into the output relation.
With $v=1$, both bidders obtain value $1$ per unit of $G_{\mathrm{out}}$, so adding their ROI conditions eliminates the allocation shares.

\paragraph{Choosing the valuations.} Set

\[m=A-1>0,\qquad L=1+\frac m4.\]
We represent a source value $z\in[0,1]$ by the output multiplier

\[\alpha_{\mathrm{out}}=L+(A-L)z.\]
We encode values in $[L,A]$ within the feasible domain $[1,A]$. The construction below forces this range, of width $A-L=3m/4>0$. Since $L>1$, losses on $G_{\mathrm{out}}$ can offset the positive private surplus, and the allocation denominator $\alpha_{\mathrm{out}}-1$ is positive.

On $G_{\mathrm{out}}$, keep valuation $1$ for $A_{\mathrm{out}}$ and $A_{\mathrm{aux}}$. On each input good $G_{\mathrm{in},i}$, give $A_{\mathrm{in},i}$ valuation $d_i>0$, where $i=1,\ldots,k$ and $k\in\{0,1,2\}$.  Choose a rational intercept $C$ so that
\[\alpha_{\mathrm{out}}=\min\left\{A,C-\sum_{i=1}^{k}d_i\alpha_{\mathrm{in},i}\right\},\qquad C-A\sum_{i=1}^{k}d_i\ge L.\]
The inequality makes the untruncated expression at least $L$ for every input in $[1,A]$, leaving a margin of at least $L-1=m/4$ above the common valuation $1$.

Still on $G_{\mathrm{in},i}$, give $A_{\mathrm{aux}}$ valuation
\[r_i=Ad_i+\frac{m}{16k}\qquad(i=1,\ldots,k;\ k>0).\]
Since $r_i>Ad_i$, $A_{\mathrm{aux}}$ strictly wins and pays $d_i\alpha_{\mathrm{in},i}$, while $A_{\mathrm{in},i}$ receives no value and pays nothing. The added margins total $m/16<m/4$, allowing positive private values.

For $G_{\mathrm{pout}}$, give $A_{\mathrm{out}}$ private valuation
\[t=\frac m8=\frac{L-1}{2}.\]
This assigns $A_{\mathrm{out}}$ half the margin $L-1$. We choose the private value of $G_{\mathrm{paux}}$ to leave $A_{\mathrm{aux}}$ outside surplus at least $t$ as well.

For $G_{\mathrm{paux}}$, give $A_{\mathrm{aux}}$ private valuation
\[x=C-1-t-\sum_{i=1}^{k}r_i.\]
 This completes the constant part: $t+x+\sum_i r_i=C-1$, and deducting the input payments gives combined outside surplus $C-1-\sum_i d_i\alpha_{\mathrm{in},i}$. For $k>0$,
\[x=C-1-A\sum_{i=1}^{k}d_i-\frac m8-\frac m{16}\ge\frac m{16}>0.\]
For $k=0$, we have $x=C-1-t\ge m/8>0$. All empty sums are zero.

\paragraph{Deriving the equilibrium relation.} Let $E_{\mathrm{out}}$ and $E_{\mathrm{aux}}$ denote the outside ROI surpluses of $A_{\mathrm{out}}$ and $A_{\mathrm{aux}}$. The chosen valuations give
\[\begin{aligned}E_{\mathrm{out}}&=t,\\E_{\mathrm{aux}}&=x+\sum_i r_i-\sum_i d_i\alpha_{\mathrm{in},i}\\&=C-1-t-\sum_i d_i\alpha_{\mathrm{in},i}\ge L-1-t=t>0.\end{aligned}\]
These positive outside surpluses prevent either $A_{\mathrm{out}}$ or $A_{\mathrm{aux}}$ from strictly losing $G_{\mathrm{out}}$. Lemma~\ref{lem:spa-affine-equal-bids} uses these positive outside surpluses to show that the two main bids must be equal.

\begin{lemma}[Equal bids in the affine second-price ROI gadget]\label{lem:spa-affine-equal-bids}
In every equilibrium of the constructed auction, $A_{\mathrm{out}}$ and $A_{\mathrm{aux}}$ submit equal bids on $G_{\mathrm{out}}$. Thus $\alpha_{\mathrm{out}}=\alpha_{\mathrm{aux}}$, and the second price is $\alpha_{\mathrm{out}}$.
\end{lemma}

\begin{proof}
If $A_{\mathrm{out}}$ strictly lost $G_{\mathrm{out}}$, its positive outside surplus would force $\alpha_{\mathrm{out}}=A$. Since both valuations on this good are $1$, $A_{\mathrm{aux}}$ could not strictly outbid it. If $A_{\mathrm{aux}}$ strictly lost, its positive outside surplus would similarly force $\alpha_{\mathrm{aux}}=A$, which $A_{\mathrm{out}}$ could not strictly outbid. Hence the bids are equal, giving the stated multiplier equality and second price.
\end{proof}

Let $q$ be the share of $G_{\mathrm{out}}$ allocated to $A_{\mathrm{out}}$, with $1-q$ allocated to $A_{\mathrm{aux}}$. Their payments on this good are $q\alpha_{\mathrm{out}}$ and $(1-q)\alpha_{\mathrm{out}}$, and their allocated values are $q$ and $1-q$. The ROI inequalities are
\[\left\{\begin{aligned}&q(\alpha_{\mathrm{out}}-1)\le t,\\&(1-q)(\alpha_{\mathrm{out}}-1)\le E_{\mathrm{aux}}.\end{aligned}\right.\]
Below the cap, Lemma~\ref{lem:spa-affine-equal-bids} puts both multipliers below $A$, so both inequalities bind, and adding gives $\alpha_{\mathrm{out}}-1=t+E_{\mathrm{aux}}$. At the cap, adding gives $A\le C-\sum_i d_i\alpha_{\mathrm{in},i}$. Thus
\begin{equation}\label{eq:anycap-affine} \alpha_{\mathrm{out}}=\min\left\{A,C-\sum_{i=1}^{k}d_i\alpha_{\mathrm{in},i}\right\}\in[L,A]. \end{equation}

\paragraph{Completing the gadget.} The following lemma also shows that every permitted input can realize this equation.

\begin{lemma}[Exact realization of the affine second-price ROI gadget]\label{lem:spa-affine-realization}
Every equilibrium satisfies \eqref{eq:anycap-affine}. Conversely, every input assignment in $[1,A]^k$ admits a local completion realizing this output and satisfying both core ROI conditions. The input bidders' upstream ROI conditions remain unchanged.
\end{lemma}

\begin{proof}

For any input multipliers in $[1,A]$, choose $\alpha_{\mathrm{out}}$ by \eqref{eq:anycap-affine} and set $\alpha_{\mathrm{aux}}=\alpha_{\mathrm{out}}$ and $q=t/(\alpha_{\mathrm{out}}-1)$. Since $\alpha_{\mathrm{out}}\ge L$, we have $0<q\le1/2$. Allocate $G_{\mathrm{pout}}$ to $A_{\mathrm{out}}$, allocate $G_{\mathrm{paux}}$ and all $G_{\mathrm{in},i}$ to $A_{\mathrm{aux}}$, and split $G_{\mathrm{out}}$ between $A_{\mathrm{out}}$ and $A_{\mathrm{aux}}$ in shares $q$ and $1-q$. Every allocation goes to a highest bidder.
The total surplus of $A_{\mathrm{out}}$ is $t-q(\alpha_{\mathrm{out}}-1)=0$. The total surplus of $A_{\mathrm{aux}}$ is
\[\begin{aligned}E_{\mathrm{aux}}-(1-q)(\alpha_{\mathrm{out}}-1)&=E_{\mathrm{aux}}+t-(\alpha_{\mathrm{out}}-1)\\&=C-\sum_i d_i\alpha_{\mathrm{in},i}-\alpha_{\mathrm{out}}.\end{aligned}\]
When $\alpha_{\mathrm{out}}<A$, \eqref{eq:anycap-affine} makes this zero. When $\alpha_{\mathrm{out}}=A$, it is nonnegative, as allowed because $\alpha_{\mathrm{aux}}=A$. Thus $A_{\mathrm{out}}$ and $A_{\mathrm{aux}}$ satisfy their ROI conditions. The strict input-good loss preserves each input bidder's ROI condition. 
\end{proof}

\subsubsection{Nonlinear gadget}
\begin{figure}[H]
\centering
\begin{tikzpicture}[
bidder/.style={ellipse,draw,line width=0.55pt,fill=blue!4,minimum width=24mm,minimum height=11mm,inner sep=3pt},
good/.style={rectangle,draw,rounded corners=1pt,line width=0.4pt,fill=black!2,minimum width=12mm,minimum height=6mm,inner sep=2pt},
edge/.style={->,>=stealth,line width=0.4pt},
val/.style={midway,fill=white,inner sep=1.2pt}
]
\node[bidder] (i1) at (-6.4,0) {$A_{\mathrm{in},1}$};
\node[bidder] (a) at (0,0) {$A_{\mathrm{aux}}$};
\node[bidder] (o) at (6.4,0) {$A_{\mathrm{out}}$};

\node[good] (g1) at (-3.2,0) {$G_{\mathrm{in},1}$};
\node[good] (g) at (3.2,0) {$G_{\mathrm{out}}$};
\node[good] (pa) at (0,1.5) {$G_{\mathrm{paux}}$};
\node[good] (po) at (6.4,1.5) {$G_{\mathrm{pout}}$};

\draw[edge] (g1.west)--node[val] {$d_1$}(i1.east);
\draw[edge] (g1.east)--node[val] {$r_1$}(a.west);
\draw[edge] (g.west)--node[val] {$L$}(a.east);
\draw[edge] (g.east)--node[val] {$1$}(o.west);
\draw[edge] (pa.south)--node[val] {$x$}(a.north);
\draw[edge] (po.south)--node[val] {$t$}(o.north);
\end{tikzpicture}
\caption{Nonlinear gadget with one input. Here $v=L$, so $A_{\mathrm{aux}}$ and $A_{\mathrm{out}}$ value $G_{\mathrm{out}}$ at $L$ and $1$, respectively.}
\label{fig:spa-nonlinear}
\end{figure}
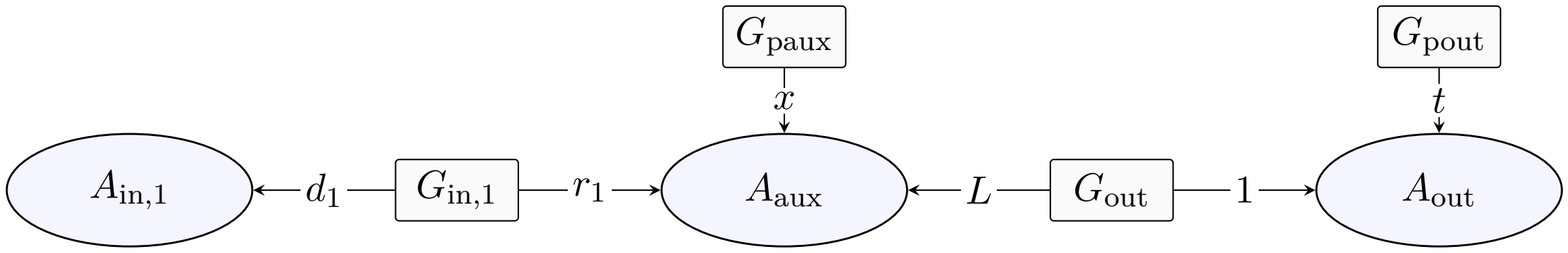

The nonlinear gadget has one input ($k=1$) and uses $v=L$, giving $A_{\mathrm{out}}$ and $A_{\mathrm{aux}}$ values $1$ and $L$ for $G_{\mathrm{out}}$ (Figure~\ref{fig:spa-nonlinear}).

\paragraph{Choosing the valuations.} Keep $m=A-1$ and $L=1+m/4$. We choose positive outside surpluses and exclude the cap so that equal bids make both ROI constraints tight.

On $G_{\mathrm{out}}$, keep valuations $1$ for $A_{\mathrm{out}}$ and $L$ for $A_{\mathrm{aux}}$. For $G_{\mathrm{in},1}$, choose
\[d_1=\frac1{16},\qquad r_1=Ad_1+\frac m{64},\]
as the values for $A_{\mathrm{in},1}$ and $A_{\mathrm{aux}}$. Since $r_1>Ad_1$, $A_{\mathrm{aux}}$ always wins and pays $d_1\alpha_{\mathrm{in},1}$, while $A_{\mathrm{in},1}$ receives no value and pays nothing. The coefficient $d_1$ makes the input payment vary by at most $m/16$ over $[1,A]$.

For $G_{\mathrm{pout}}$, give $A_{\mathrm{out}}$ private valuation
\[t=L-1=\frac m4.\]
 Matching the valuation difference $L-1$ makes the share of $A_{\mathrm{aux}}$ proportional to $\alpha_{\mathrm{out}}-L$, producing the square in the ROI calculation below.

For $G_{\mathrm{paux}}$, give $A_{\mathrm{aux}}$ private valuation
\[x=\frac{5m}{64}.\]
 Together with the valuations $d_1$ and $r_1$ on $G_{\mathrm{in},1}$, this choice keeps the outside surplus of $A_{\mathrm{aux}}$ positive but small enough to rule out a strict win on $G_{\mathrm{out}}$ by $A_{\mathrm{aux}}$ against $\alpha_{\mathrm{out}}=A$ and to rule out a common bid of $A$. 

\paragraph{Deriving the equilibrium relation.} Using the same outside-surplus notation gives
\[\begin{aligned}E_{\mathrm{out}}&=t=L-1,\\E_{\mathrm{aux}}&=r_1+x-d_1\alpha_{\mathrm{in},1}\\&=\frac{A-\alpha_{\mathrm{in},1}}{16}+\frac{3m}{32}.\end{aligned}\]
Thus $3m/32\le E_{\mathrm{aux}}\le5m/32$ for every input multiplier in $[1,A]$. The next lemma forces equal bids despite the unequal valuations.

\begin{lemma}[Equal bids in the nonlinear second-price ROI gadget]\label{lem:spa-nonlinear-equal-bids}
In every equilibrium of the constructed auction, $A_{\mathrm{out}}$ and $A_{\mathrm{aux}}$ submit equal bids on $G_{\mathrm{out}}$. Thus $\alpha_{\mathrm{out}}=L\alpha_{\mathrm{aux}}$, and the second price is $\alpha_{\mathrm{out}}$.
\end{lemma}

\begin{proof}
If $A_{\mathrm{out}}$ strictly lost, its positive outside surplus would force $\alpha_{\mathrm{out}}=A$. The winning $A_{\mathrm{aux}}$ would then pay $A$, but $L+E_{\mathrm{aux}}\le1+13m/32<A$, contradicting its ROI feasibility. If $A_{\mathrm{aux}}$ strictly lost, its positive outside surplus would force $\alpha_{\mathrm{aux}}=A$, giving bid $LA>A$, above every possible bid of $A_{\mathrm{out}}$. Both strict orders are impossible. Equal bids therefore give $\alpha_{\mathrm{out}}=L\alpha_{\mathrm{aux}}$ and the stated second price.
\end{proof}

By Lemma~\ref{lem:spa-nonlinear-equal-bids}, the common price is $\alpha_{\mathrm{out}}$; we next exclude $\alpha_{\mathrm{out}}=A$. Otherwise the total payment for $G_{\mathrm{out}}$ is $A$, the total allocated value from $G_{\mathrm{out}}$ is at most $L$, and adding the ROI inequalities of $A_{\mathrm{out}}$ and $A_{\mathrm{aux}}$ gives
\[A\le L+t+E_{\mathrm{aux}}\le1+\frac{21m}{32}<A.\]
Thus $\alpha_{\mathrm{out}}<A$ and $\alpha_{\mathrm{aux}}=\alpha_{\mathrm{out}}/L<A$, so both ROI constraints bind.

Let $q$ be the share of $G_{\mathrm{out}}$ allocated to $A_{\mathrm{out}}$, with $1-q$ allocated to $A_{\mathrm{aux}}$. Their payments are $q\alpha_{\mathrm{out}}$ and $(1-q)\alpha_{\mathrm{out}}$, and their allocated values are $q$ and $(1-q)L$. Hence
\[\left\{\begin{aligned}&q(\alpha_{\mathrm{out}}-1)=L-1,\\&(1-q)(\alpha_{\mathrm{out}}-L)=E_{\mathrm{aux}}.\end{aligned}\right.\]
The positive right sides give $0<q<1$ and $\alpha_{\mathrm{out}}>L$. The first equality gives $q=(L-1)/(\alpha_{\mathrm{out}}-1)$ and therefore $1-q=(\alpha_{\mathrm{out}}-L)/(\alpha_{\mathrm{out}}-1)$. Substitution into the second equality yields
\begin{equation}\label{eq:anycap-root-inverse} E_{\mathrm{aux}}=\frac{(\alpha_{\mathrm{out}}-L)^2}{\alpha_{\mathrm{out}}-1}. \end{equation}
Substituting $E_{\mathrm{aux}}=r_1+x-d_1\alpha_{\mathrm{in},1}$ gives
\[\frac{A-\alpha_{\mathrm{in},1}}{16}+\frac{3m}{32}=\frac{(\alpha_{\mathrm{out}}-L)^2}{\alpha_{\mathrm{out}}-1}.\]

\paragraph{Completing the gadget.} We next prove its unique output and show that this output always has a local completion.

\begin{lemma}[Exact realization of the nonlinear second-price ROI gadget]\label{lem:spa-nonlinear-realization}
Every equilibrium satisfies \eqref{eq:anycap-root-inverse}. For each input in $[1,A]$, the unique output lies in $(L,A)$ and is continuous and strictly decreasing in the input. Conversely, every such input admits a local completion realizing this output and satisfying both core ROI conditions. The input bidder's upstream ROI condition remains unchanged.
\end{lemma}

\begin{proof}

For $u\in[L,A]$, define $f(u)=(u-L)^2/(u-1)$. We have
\[f(L)=0,\qquad f(A)=\frac{9m}{16},\qquad f'(u)=1-\frac{(L-1)^2}{(u-1)^2}>0\quad(L<u\le A).\]
Since every allowed $E_{\mathrm{aux}}$ lies strictly between $f(L)$ and $f(A)$, \eqref{eq:anycap-root-inverse} determines a unique $\alpha_{\mathrm{out}}\in(L,A)$ for each input multiplier $\alpha_{\mathrm{in},1}$. The outside surplus $E_{\mathrm{aux}}$ of $A_{\mathrm{aux}}$ is continuous and strictly decreasing in $\alpha_{\mathrm{in},1}$, so the input--output function $\alpha_{\mathrm{out}}=h(\alpha_{\mathrm{in},1})$ is continuous and strictly decreasing on $[1,A]$.

For any input, choose this output and set $\alpha_{\mathrm{aux}}=\alpha_{\mathrm{out}}/L$ and $q=(L-1)/(\alpha_{\mathrm{out}}-1)$. Then $1<\alpha_{\mathrm{aux}}<A$ and $0<q<1$. Allocate $G_{\mathrm{pout}}$ to $A_{\mathrm{out}}$, allocate $G_{\mathrm{paux}}$ and $G_{\mathrm{in},1}$ to $A_{\mathrm{aux}}$, and split $G_{\mathrm{out}}$ in shares $q$ and $1-q$ between $A_{\mathrm{out}}$ and $A_{\mathrm{aux}}$. All allocations go to highest bidders.
The total surplus of $A_{\mathrm{out}}$ is $L-1-q(\alpha_{\mathrm{out}}-1)=0$. The total surplus of $A_{\mathrm{aux}}$ is
\[E_{\mathrm{aux}}-(1-q)(\alpha_{\mathrm{out}}-L)=E_{\mathrm{aux}}-\frac{(\alpha_{\mathrm{out}}-L)^2}{\alpha_{\mathrm{out}}-1}=0.\]
Thus $A_{\mathrm{out}}$ and $A_{\mathrm{aux}}$ satisfy the required ROI equalities below the cap. The strict input-good loss preserves the input bidder's ROI condition. 
\end{proof}

\subsection{Encoding and Exact Gate Implementation}\label{subsec:spa-encoding}
Lemmas~\ref{lem:spa-affine-realization} and~\ref{lem:spa-nonlinear-realization} apply to inputs throughout $[1,A]$ and force both outputs into $[L,A]$. We now use this proved range to encode source variables. Its width is

\[W=A-L=\frac{3m}{4}>0.\]

Represent a source value $z\in[0,1]$ by $\alpha_{\mathrm{out}}=L+Wz$ and decode it by

\[z=\frac{\alpha_{\mathrm{out}}-L}{W}.\]
Only wire multipliers are decoded; the auxiliary multipliers $\alpha_{\mathrm{aux}}$ do not represent source values.

\paragraph{Affine source operations.} Suppose the inputs encode source values $y,w$, so that $\alpha_{\mathrm{in},1}=L+Wy$ and $\alpha_{\mathrm{in},2}=L+Ww$. To realize

\[z=\min\{1,c-ay-bw\},\qquad a,b\ge0,\qquad c\ge a+b,\]
choose $d_1=a$, $d_2=b$, and $C=L+Wc+L(a+b)$. These choices meet the construction's lower-bound condition because $C-A(a+b)=L+W(c-a-b)\ge L$. Also,

\[C-d_1\alpha_{\mathrm{in},1}-d_2\alpha_{\mathrm{in},2}=L+W(c-ay-bw).\]

Using $A=L+W$, the capped response \eqref{eq:anycap-affine} is exactly $\alpha_{\mathrm{out}}=L+Wz$. A zero coefficient is implemented by omitting that input good.
Taking $a=c=1$ gives source complement $z=1-y$. To obtain capped doubling, first complement the source input and then apply $z=\min\{1,2-2y\}$ to that complement. The result is $\min\{1,2y\}$. Lemma~\ref{lem:clipped-affine} extends these constructions to every fixed signed clipped affine expression with at most two inputs per basic gadget. 

\paragraph{Using the nonlinear response for multiplication.} Theorem~\ref{thm:criterion}(i) constructs the source product from the inverse of the nonlinear response $\alpha_{\mathrm{out}}=h(\alpha_{\mathrm{in},1})$. Rearranging \eqref{eq:anycap-root-inverse}, with $\mu=(1+A)/2$, gives

\[\alpha_{\mathrm{in},1}=p(\alpha_{\mathrm{out}}):=\mu+2m-16\frac{(\alpha_{\mathrm{out}}-L)^2}{\alpha_{\mathrm{out}}-1}.\]
The inverse has rational coefficients. Polynomial division gives

\[\frac{(u-L)^2}{u-1}=u-2L+1+\frac{(L-1)^2}{u-1}.\]
Since $L>1$, the reciprocal term is nonzero, so $p$ is nonaffine. Its only possible pole is $1$, outside $(L,A)$.
We must also find an interval on which inverse values remain valid encoded multipliers. At $\mu=1+m/2$, direct substitution gives

\[p(\mu)=\mu+2m-16\frac{(m/4)^2}{m/2}=\mu.\]
Both $\mu$ and $p(\mu)$ are interior to $[L,A]$. To specify a rational interval on which the inverse can be used, take

\[J=\left[\mu-\frac{m}{128},\mu+\frac{m}{128}\right].\]
Since $L=1+m/4$ and $\mu=1+m/2$, we have $J\subset(L,A)$. On this interval, $p'(u)=-16\bigl(1-(L-1)^2/(u-1)^2\bigr)$ lies strictly between $-16$ and $0$. Using $p(\mu)=\mu$, the mean value theorem therefore gives

\[p(J)\subset\left(\mu-\frac m8,\mu+\frac m8\right)\subset(L,A).\]
Thus both the prescribed output in $J$ and its inverse value are valid encoded multipliers. Equivalently, $J\subset\operatorname{int}h([L,A])$.

The signed affine operations above and this rational inverse satisfy Theorem~\ref{thm:criterion}(i). Its inverse-feedback construction therefore implements $p$ on $J$ and then the source product $\phi$, with both solution directions. The constants depend only on the fixed cap $A$ and the source gate set.

\subsection{Completing the Proof}
We now combine the local realizations and the source arithmetic to prove Theorem~\ref{thm:spa-arbitrary-cap}.

\begin{proof}[Proof of Theorem~\ref{thm:spa-arbitrary-cap}]
For $A>1$, membership follows from \cite{FHHH24}, and Section~\ref{sec:complexity} gives the upper bound for approximation in distance.

Lemmas~\ref{lem:spa-affine-realization} and~\ref{lem:spa-nonlinear-realization} give the local equations, output ranges, and completions, while preserving the input bidders' ROI conditions. Section~\ref{subsec:spa-encoding} combines these relations into implementations of the source gates in Lemma~\ref{lem:source}. Expand these implementations, then apply Lemma~\ref{lem:bounded-use} so each wire has at most two input uses. Create one bidder per wire, acting as $A_{\mathrm{out}}$ in its defining gadget and $A_{\mathrm{in},i}$ downstream, with fresh auxiliaries and goods. Input preservation makes all local equations hold simultaneously, including on directed cycles, so every equilibrium decodes to a source solution. Conversely, extend a source solution through the arithmetic and copies, then use the local completions from the two realization lemmas. They combine into an equilibrium because the goods are auctioned separately.

Each wire bidder values $G_{\mathrm{out}}$ and $G_{\mathrm{pout}}$ in its defining gadget and at most two outgoing goods $G_{\mathrm{in},i}$, hence at most four goods. An affine auxiliary $A_{\mathrm{aux}}$ values at most four goods; a nonlinear auxiliary $A_{\mathrm{aux}}$ values three. Each good has at most two positive valuations. For fixed $A$, the fixed-size arithmetic implementations and complement copies use a fixed finite rational valuation set depending only on $A$. Auction size is linear in source size, with polynomial encoding length. Each source variable is decoded from one designated multiplier by $z=(\alpha_{\mathrm{out}}-L)/W$, giving an SL reduction and hence $\FIXP$-hardness. Together with membership, this proves $\FIXP$-completeness under all stated restrictions.

A rational multiplier vector within $W\varepsilon$ in $\ell_\infty$ of an exact equilibrium decodes to a vector within $\varepsilon$ of a source solution. Since $W$ is a fixed positive rational, this tolerance conversion preserves polynomial encoding length. The reduction therefore also gives distance hardness; the matching upper bound proves $\FIXPa$-completeness for the multiplier vector.
\end{proof}

\section{Budget-Constrained GSP}\label{sec:budget}
We next consider budget-constrained no-self-pricing GSP with multipliers in $[0,1]$. Goods have designated owners and compete in one ranking. Full display and unit click-through rates fix acquired values, while payments must exhaust the budget whenever the multiplier is below $1$.

Unlike Section~\ref{sec:ordinary-roi}, input bidder $A_{\mathrm{in},j}$ owns and pays for $G_{\mathrm{in},j}$. Its price may depend on other multipliers, so adding this signal without compensation would change the bidder's upstream budget equation.
We place a guard good $G_{\mathrm{guard},j}$ with a fixed bid after $G_{\mathrm{in},j}$ and add that bid to the budget of $A_{\mathrm{in},j}$. Once the guard is proved to set the signal price, the extra payment and budget increase cancel exactly.

\begin{theorem}[Exact Budget-Constrained GSP Equilibria]\label{thm:budget-gsp}
Exact budget-constrained no-self-pricing GSP equilibrium computation is $\FIXP$-complete under separable linear reductions. Approximating the multiplier vector in distance to an exact equilibrium is $\FIXPa$-complete under the same restrictions. Hardness holds in one auction with unit CTRs, full display, positive rational values and budgets, a universal constant bound on goods per bidder, and polynomial-support equilibrium distributions over legal tied orders.
\end{theorem}

We use affine and nonlinear gadgets for Lemma~\ref{lem:source}. Their equations use prices and compensation proved independently of the gate equations in Section~\ref{subsec:budget-completing}, so the construction also applies to cycles.

\subsection{A Common Auction Structure}\label{subsec:budget-common}
We use the same input, output, and auxiliary bidder roles from Section~\ref{sec:ordinary-roi}.  In particular, $A_{\mathrm{out}}$ and $A_{\mathrm{aux}}$ now own distinct main goods $G_{\mathrm{out}}$ and $G_{\mathrm{aux}}$, rather than competing for a shared good $G_{\mathrm{out}}$.

Let $k$ denote the number of inputs. The affine gadget allows $k\in\{0,1,2\}$, and the nonlinear gadget uses $k=1$. The bidder roles are:
\begin{itemize}
\item $A_{\mathrm{in},j}$: input bidder with multiplier $\alpha_{\mathrm{in},j}$ and budget $B_{\mathrm{in},j}$, for $j=1,\ldots,k$.
\item $A_{\mathrm{out}}$: output bidder with multiplier $\alpha_{\mathrm{out}}$ and budget $B_{\mathrm{out}}$.
\item $A_{\mathrm{aux}}$: auxiliary bidder with multiplier $\alpha_{\mathrm{aux}}$ and budget $B_{\mathrm{aux}}$.
\item $A_{\mathrm{guard},j}$: guard bidder with multiplier $\alpha_{\mathrm{guard},j}$ and budget $B_{\mathrm{guard},j}$, for $j=1,\ldots,k$, and also $j=0$ in the affine gadget.
\end{itemize}

Wire bidders use the upstream and downstream roles of Section~\ref{sec:overview}, sharing one multiplier and actual budget across all roles. Each auxiliary $A_{\mathrm{aux}}$ and guard $A_{\mathrm{guard},j}$ is fresh and distinct from all wire bidders.

For a wire bidder, let $B$ be its actual budget, denoted $B_{\mathrm{out}}$ or $B_{\mathrm{in},j}$ according to its role. Choose a base budget $B^0$ for its upstream main good $G_{\mathrm{out}}$, then add the intended guard price for each outgoing signal $G_{\mathrm{in},j}$. Once these prices are verified, cancelling the outgoing payments leaves main payment at most $B^0$, with equality when $\alpha_{\mathrm{out}}<1$. Without downstream uses, $B=B^0$.
The auxiliary $A_{\mathrm{aux}}$ owns no outgoing signal goods, so its actual budget is simply $B_{\mathrm{aux}}=B_{\mathrm{aux}}^0$.

The goods and their owners are:
\begin{itemize}
\item $G_{\mathrm{out}}$: main good owned by $A_{\mathrm{out}}$.
\item $G_{\mathrm{aux}}$: main good owned by $A_{\mathrm{aux}}$.
\item $G_{\mathrm{in},j}$: signal good owned by $A_{\mathrm{in},j}$, for $j=1,\ldots,k$.
\item $G_{\mathrm{guard},j}$: guard good owned by $A_{\mathrm{guard},j}$, for $j=1,\ldots,k$, and also $j=0$ in the affine gadget.
\item $G_{\mathrm{recv},j,\ell}$: receiving good owned by $A_{\mathrm{aux}}$ in the affine gadget, for $j=1,\ldots,k$ and $\ell=1,\ldots,k_j$, where the positive integers $k_j$ are specified below.
\item $G_{\mathrm{mid}}$: middle good owned by $A_{\mathrm{aux}}$ in the nonlinear gadget.
\end{itemize}

At a main tie, we choose an order distribution meeting both payment targets, as allowed by Section~\ref{sec:auction-definitions}. This differs from the fixed tie priority in throttling.
Equal values on $G_{\mathrm{out}}$ and $G_{\mathrm{aux}}$ will force $\alpha_{\mathrm{out}}=\alpha_{\mathrm{aux}}$. We choose probabilities for their two tied orders to meet both base budgets. In the affine gadget, the combined main payment is the common bid plus the fixed bid of $G_{\mathrm{guard},0}$, regardless of order. Adding the tight budget conditions will therefore give an affine relation after including the input-dependent payments of $A_{\mathrm{aux}}$.

The affine receiving goods $G_{\mathrm{recv},j,\ell}$ belong to $A_{\mathrm{aux}}$ and precede $G_{\mathrm{in},j}$ with no foreign good intervening. If the signal value is $v_j>0$, each pays $v_j\alpha_{\mathrm{in},j}$. Their multiplicity $k_j$ sets the input coefficient through total payment $k_jv_j\alpha_{\mathrm{in},j}$.

The nonlinear middle good $G_{\mathrm{mid}}$ belongs to $A_{\mathrm{aux}}$ and has half its main-good value. It follows both main goods and precedes $G_{\mathrm{in},1}$. Besides paying the signal bid, it prices $G_{\mathrm{out}}$ when that good follows $G_{\mathrm{aux}}$, making the lower main price depend on the output multiplier.

Each guard $A_{\mathrm{guard},j}$ owns only $G_{\mathrm{guard},j}$ of value $w>0$ and has budget $2w$, forcing multiplier $1$ and fixed bid $w$. We will prove that it is the first good after $G_{\mathrm{in},j}$ owned by a bidder other than $A_{\mathrm{in},j}$, fixing the signal payment at $w$. The affine guard $G_{\mathrm{guard},0}$ instead prices the second main good.

\subsubsection{Affine gadget}
The affine gadget has $k\in\{0,1,2\}$ inputs, with rational coefficients $a_j>0$ for $j=1,\ldots,k$ and rational constant term $C$. Its target relation is

\begin{equation}\label{eq:budget-affine} \alpha_{\mathrm{out}}=\min\left\{1,C-\sum_{j=1}^{k}a_j\alpha_{\mathrm{in},j}\right\},\qquad a_j>0,\qquad C-\sum_{j=1}^{k}a_j\ge\frac34. \end{equation}

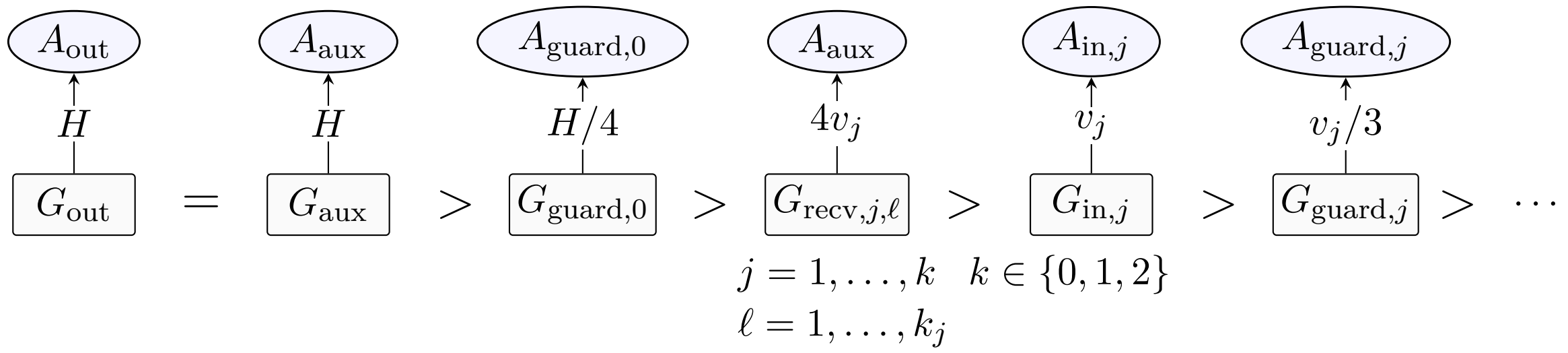
\begin{figure}[H]
\centering
\begin{tikzpicture}[
bidder/.style={ellipse,draw,line width=0.55pt,fill=blue!4,inner xsep=2.5pt,inner ysep=1.5pt},
good/.style={rectangle,draw,rounded corners=1pt,line width=0.4pt,fill=black!2,minimum width=12mm,minimum height=6mm,inner sep=2pt},
edge/.style={->,>=stealth,line width=0.4pt},
val/.style={midway,fill=white,inner sep=1.2pt},
relation/.style={font=\Large,inner sep=0pt}
]
\node[good] (go) at (0,0) {$G_{\mathrm{out}}$};
\node[good] (ga) at (2.5,0) {$G_{\mathrm{aux}}$};
\node[good] (gg0) at (5.0,0) {$G_{\mathrm{guard},0}$};
\node[good] (gr) at (7.5,0) {$G_{\mathrm{recv},j,\ell}$};
\node[good] (gi) at (10.0,0) {$G_{\mathrm{in},j}$};
\node[good] (gg) at (12.5,0) {$G_{\mathrm{guard},j}$};

\node[relation] at (1.25,0) {$=$};
\node[relation] at (3.75,0) {$>$};
\node[relation] at (6.25,0) {$>$};
\node[relation] at (8.75,0) {$>$};
\node[relation] at (11.25,0) {$>$};
\node[relation] at (13.6,0) {$>$};
\node at (14.4,0) {$\cdots$};

\node[bidder] (ao) at (0,1.6) {$A_{\mathrm{out}}$};
\node[bidder] (aa) at (2.5,1.6) {$A_{\mathrm{aux}}$};
\node[bidder] (ag0) at (5.0,1.6) {$A_{\mathrm{guard},0}$};
\node[bidder] (ar) at (7.5,1.6) {$A_{\mathrm{aux}}$};
\node[bidder] (ai) at (10.0,1.6) {$A_{\mathrm{in},j}$};
\node[bidder] (ag) at (12.5,1.6) {$A_{\mathrm{guard},j}$};

\draw[edge] (go.north)--node[val] {$H$}(ao.south);
\draw[edge] (ga.north)--node[val] {$H$}(aa.south);
\draw[edge] (gg0.north)--node[val] {$H/4$}(ag0.south);
\draw[edge] (gr.north)--node[val] {$4v_j$}(ar.south);
\draw[edge] (gi.north)--node[val] {$v_j$}(ai.south);
\draw[edge] (gg.north)--node[val] {$v_j/3$}(ag.south);

\node[anchor=north,inner sep=0pt,yshift=-2mm] (jrange)
  at (gr.south) {$j=1,\ldots,k$};
\node[anchor=north west,inner sep=0pt,yshift=-1.5mm]
  at (jrange.south west) {$\ell=1,\ldots,k_j$};
\node[anchor=base west,inner sep=0pt,xshift=3mm]
  at (jrange.base east) {$k\in\{0,1,2\}$};
\end{tikzpicture}
\caption{Affine gadget in decreasing bid order. The displayed input branch is repeated for $j=1,\ldots,k$, with $k\in\{0,1,2\}$, and is omitted when $k=0$. The receiving node represents $k_j$ equal-bid goods.}
\label{fig:budget-affine}
\end{figure}

A zero coefficient requires no input branch, so $k$ counts only positive coefficients. Since $\alpha_{\mathrm{in},j}\le1$, the condition on $C$ places the output in $[3/4,1]$. Each branch contains $G_{\mathrm{recv},j,\ell}$ for $\ell=1,\ldots,k_j$, $G_{\mathrm{in},j}$, and $G_{\mathrm{guard},j}$. Figure~\ref{fig:budget-affine} shows their intended order, proved in Section~\ref{subsec:budget-completing}.

\paragraph{Choosing the valuations and budgets.} We represent a source value $z\in[0,1]$ by the output multiplier

\[ \alpha_{\mathrm{out}}=\frac34+\frac z4. \]
Both gadgets encode outputs in $[3/4,1]$ within the feasible domain $[0,1]$. The affine condition $C-\sum_j a_j\ge3/4$ forces this range, placing signals above their guards. To justify prices in arbitrary equilibria, we first prove $\alpha_{\mathrm{in},j}>1/3$ from budgets alone.

Choose a rational scale $H>0$ and let $S=\sum_{j=1}^{k}a_j$, with $S=0$ when $k=0$. The receiving payments will total at most $SH$, which determines the auxiliary budget below.

Give both main goods value $H$, so equal bids imply $\alpha_{\mathrm{out}}=\alpha_{\mathrm{aux}}$. Give $G_{\mathrm{guard},0}$ value $H/4$ and its owner budget $H/2$, forcing multiplier $1$ and bid $H/4$. At a main tie immediately before this guard, the first main good pays the common bid and the second pays $H/4$. Their total payment is $H\alpha_{\mathrm{out}}+H/4$ in either order.

For the receiving goods $G_{\mathrm{recv},j,\ell}$, give their owner $A_{\mathrm{aux}}$ value $4v_j$ for each $\ell=1,\ldots,k_j$, where $v_j>0$ and $k_j$ is a positive integer. Give the signal good $G_{\mathrm{in},j}$, owned by $A_{\mathrm{in},j}$, value $v_j$. Our construction makes each $G_{\mathrm{recv},j,\ell}$ pay $v_j\alpha_{\mathrm{in},j}$. To get the coefficient $a_j$ after dividing payments by the main value $H$, choose

\[ k_jv_j=a_jH\qquad(j=1,\ldots,k). \]

The factor $4$ in their values will ensure that their bids exceed the signal bid once we prove $\alpha_{\mathrm{aux}}>1/4$.

Give $G_{\mathrm{guard},j}$ value $v_j/3$ and $A_{\mathrm{guard},j}$ budget $B_{\mathrm{guard},j}=2v_j/3$, fixing its bid at $v_j/3$. The later order proof will show that this is the price of $G_{\mathrm{in},j}$. Add $v_j/3$ to the actual budget of $A_{\mathrm{in},j}$.

To obtain the required order between input branches, number the positive coefficients consecutively. When $k\ge1$, choose

\[ k_1=\lceil32a_1\rceil,\qquad v_1=\frac{a_1H}{k_1}. \]
When $k=2$, also choose

\[ k_2=\left\lceil\frac{24a_2H}{v_1}\right\rceil,\qquad v_2=\frac{a_2H}{k_2}. \]
These choices preserve $k_jv_j=a_jH$ and give $4v_1\le H/8<H/4$ and, when the second branch is present, $4v_2\le v_1/6<v_1/3$. Thus each branch lies below the preceding guard; the within-branch order will follow from the multiplier bounds. For fixed coefficients $a_j$, these formulas make $k_j$ independent of $H$. Since only a fixed finite collection of affine gate types is used, the number of receiving goods is bounded by a constant.

For $A_{\mathrm{out}}$ and $A_{\mathrm{aux}}$, choose the base budgets together as

\[ B_{\mathrm{out}}^0=(C-S-1/4)H,\qquad B_{\mathrm{aux}}^0=(S+1/2)H. \]
Their sum is

\[ B_{\mathrm{out}}^0+B_{\mathrm{aux}}^0=(C+1/4)H. \]
Subtracting the guard payment $H/4$ from this budget sum leaves $CH$, giving the desired relation after including receiving payments.

\paragraph{Deriving the equilibrium relation.} Using Lemma~\ref{lem:budget-multiplier-bounds} and the independently proved prices and compensation of Section~\ref{subsec:budget-completing}, we first establish equal main bids.

\begin{lemma}[Equal main bids in the affine budget-GSP gadget]\label{lem:budget-affine-equal-bids}
For the constructed affine gadget with $C-S\ge3/4$, every equilibrium has equal main bids and hence $\alpha_{\mathrm{out}}=\alpha_{\mathrm{aux}}$.
\end{lemma}

\begin{proof}
Suppose $G_{\mathrm{out}}$ bids strictly more than $G_{\mathrm{aux}}$. The established prices make $A_{\mathrm{aux}}$ pay $H/4$ for its main good and at most $SH$ for its receiving goods. Its total payment is at most $(S+1/4)H<(S+1/2)H=B_{\mathrm{aux}}^0$, forcing $\alpha_{\mathrm{aux}}=1$. Equal main values $H$ make the assumed strict order impossible. Conversely, if $G_{\mathrm{aux}}$ bids strictly more, $A_{\mathrm{out}}$ pays $H/4<B_{\mathrm{out}}^0$ for its main good. After outgoing compensation is cancelled, unspent budget forces $\alpha_{\mathrm{out}}=1$, again contradicting the strict order. Hence the main bids and multipliers are equal.
\end{proof}

By Lemma~\ref{lem:budget-affine-equal-bids}, write $t=H\alpha_{\mathrm{out}}$ for the common main bid.

Whichever main good appears first, the combined payment for $G_{\mathrm{out}}$ and $G_{\mathrm{aux}}$ is $t+H/4$. The receiving goods $G_{\mathrm{recv},j,\ell}$ contribute total payment $H\sum_{j=1}^{k}a_j\alpha_{\mathrm{in},j}$ to $A_{\mathrm{aux}}$. If $\alpha_{\mathrm{out}}<1$, then $\alpha_{\mathrm{aux}}<1$ as well, so both base budgets must be spent in full. Adding their payment equations gives

\[ H\alpha_{\mathrm{out}}+\frac H4+H\sum_{j=1}^{k}a_j\alpha_{\mathrm{in},j}=B_{\mathrm{out}}^0+B_{\mathrm{aux}}^0=(C+1/4)H. \]
Subtracting $H/4$ from both sides and dividing both sides by $H$ gives $\alpha_{\mathrm{out}}=C-\sum_{j=1}^{k}a_j\alpha_{\mathrm{in},j}$. If instead $\alpha_{\mathrm{out}}=\alpha_{\mathrm{aux}}=1$, adding the two budget inequalities gives $C-\sum_{j=1}^{k}a_j\alpha_{\mathrm{in},j}\ge1$. Thus every equilibrium satisfies exactly \eqref{eq:budget-affine}.

\paragraph{Completing the gadget.} The next lemma adds the local completion for every encoded input.

\begin{lemma}[Exact realization of the affine budget-GSP gadget]\label{lem:budget-affine-realization}
Under the independently established prices and compensation of Section~\ref{subsec:budget-completing}, every equilibrium satisfies \eqref{eq:budget-affine}. Conversely, every input assignment in $[3/4,1]^k$ admits a local completion realizing this output and satisfying both core budget conditions. Outgoing signal payments are offset by the prescribed budget increases.
\end{lemma}

\begin{proof}

For any encoded inputs $\alpha_{\mathrm{in},j}\in[3/4,1]$, choose $\alpha_{\mathrm{out}}$ by \eqref{eq:budget-affine}, set $\alpha_{\mathrm{aux}}=\alpha_{\mathrm{out}}$, and set every guard multiplier to $1$. These choices give the required local order: both main bids are at least $3H/4$, receiving bids are at least $3v_j$, and signal bids are at least $3v_j/4$. Write $t=H\alpha_{\mathrm{out}}$ and choose the probability that $G_{\mathrm{out}}$ appears before $G_{\mathrm{aux}}$ as

\[ q=\frac{\min\{B_{\mathrm{out}}^0,t\}-H/4}{t-H/4}. \]
Since $B_{\mathrm{out}}^0\ge H/2$ and $t\ge3H/4$, we have $0<q\le1$. If $\alpha_{\mathrm{out}}<1$, then $t-B_{\mathrm{out}}^0=H\bigl(\sum_j a_j(1-\alpha_{\mathrm{in},j})+1/4\bigr)>0$. Thus the expected payment of $A_{\mathrm{out}}$ for $G_{\mathrm{out}}$ equals $B_{\mathrm{out}}^0$, and the combined-payment identity makes the total payment of $A_{\mathrm{aux}}$ equal $B_{\mathrm{aux}}^0$. If $\alpha_{\mathrm{out}}=1$, the expected payment of $A_{\mathrm{out}}$ for $G_{\mathrm{out}}$ is $\min\{B_{\mathrm{out}}^0,H\}$. When $B_{\mathrm{out}}^0\le H$, feasibility for $A_{\mathrm{aux}}$ follows from $C-\sum_j a_j\alpha_{\mathrm{in},j}\ge1$. When $B_{\mathrm{out}}^0>H$, we have $q=1$, and $A_{\mathrm{aux}}$ pays at most $H/4+SH<B_{\mathrm{aux}}^0$. Both multipliers equal $1$ in this case, so unspent budget is allowed.

Adding back the outgoing signal payments and their equal budget increases preserves these budget conditions. Every guard bidder $A_{\mathrm{guard},j}$ has multiplier $1$ and payment below its budget. Hence every encoded input has a local completion.
\end{proof}

\subsubsection{Nonlinear gadget}\label{subsubsec:budget-nonlinear}
The nonlinear gadget has $k=1$, so it uses one input bidder $A_{\mathrm{in},1}$.  We aim to enforce

\[ \alpha_{\mathrm{out}}=\frac12+\sqrt{\frac14-\frac8{45}\alpha_{\mathrm{in},1}}. \]

The nonlinear output must share the encoding interval $[3/4,1]$ and lie strictly below $1$, so both tied bidders exhaust their budgets. These two equalities will determine the nonlinear relation.

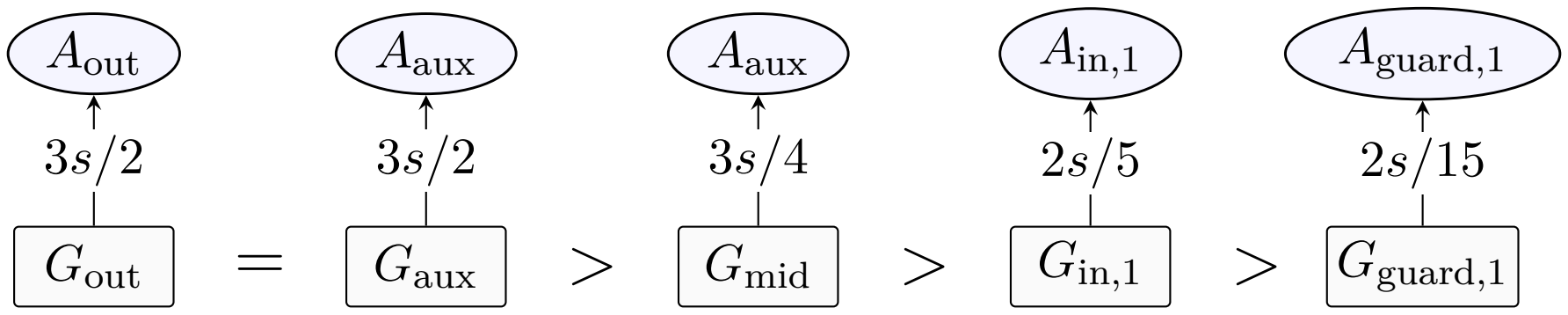
\begin{figure}[H]
\centering
\begin{tikzpicture}[
bidder/.style={ellipse,draw,line width=0.55pt,fill=blue!4,inner xsep=2.5pt,inner ysep=1.5pt},
good/.style={rectangle,draw,rounded corners=1pt,line width=0.4pt,fill=black!2,minimum width=12mm,minimum height=6mm,inner sep=2pt},
edge/.style={->,>=stealth,line width=0.4pt},
val/.style={midway,fill=white,inner sep=1.2pt},
relation/.style={font=\Large,inner sep=0pt}
]
\node[good] (go) at (0,0) {$G_{\mathrm{out}}$};
\node[good] (ga) at (2.5,0) {$G_{\mathrm{aux}}$};
\node[good] (gm) at (5.0,0) {$G_{\mathrm{mid}}$};
\node[good] (gi) at (7.5,0) {$G_{\mathrm{in},1}$};
\node[good] (gg) at (10.0,0) {$G_{\mathrm{guard},1}$};

\node[relation] at (1.25,0) {$=$};
\node[relation] at (3.75,0) {$>$};
\node[relation] at (6.25,0) {$>$};
\node[relation] at (8.75,0) {$>$};

\node[bidder] (ao) at (0,1.6) {$A_{\mathrm{out}}$};
\node[bidder] (aa) at (2.5,1.6) {$A_{\mathrm{aux}}$};
\node[bidder] (am) at (5.0,1.6) {$A_{\mathrm{aux}}$};
\node[bidder] (ai) at (7.5,1.6) {$A_{\mathrm{in},1}$};
\node[bidder] (ag) at (10.0,1.6) {$A_{\mathrm{guard},1}$};

\draw[edge] (go.north)--node[val] {$3s/2$}(ao.south);
\draw[edge] (ga.north)--node[val] {$3s/2$}(aa.south);
\draw[edge] (gm.north)--node[val] {$3s/4$}(am.south);
\draw[edge] (gi.north)--node[val] {$2s/5$}(ai.south);
\draw[edge] (gg.north)--node[val] {$2s/15$}(ag.south);

\end{tikzpicture}
\caption{Nonlinear gadget with one input, in decreasing bid order.}
\label{fig:budget-nonlinear}
\end{figure}

For each $\alpha_{\mathrm{in},1}\in(1/3,1]$, the expression lies in $(3/4,1)$. The bound $\alpha_{\mathrm{in},1}>1/3$, proved in Section~\ref{subsec:budget-completing}, also places the signal above its guard. We replace $G_{\mathrm{guard},0}$ and $G_{\mathrm{recv},j,\ell}$ by a middle good $G_{\mathrm{mid}}$ of $A_{\mathrm{aux}}$, whose bid sets the lower main price and whose payment depends on the input. Figure~\ref{fig:budget-nonlinear} shows the intended order.

\paragraph{Choosing the valuations and budgets.}

Choose a rational scale $s>0$.

For the main goods $G_{\mathrm{out}}$ and $G_{\mathrm{aux}}$, give their respective owners $A_{\mathrm{out}}$ and $A_{\mathrm{aux}}$ the same value $3s/2$. Equal main bids will then imply $\alpha_{\mathrm{out}}=\alpha_{\mathrm{aux}}$. Write their common bid as $t=(3s/2)\alpha_{\mathrm{out}}$.

Give $G_{\mathrm{mid}}$, owned by $A_{\mathrm{aux}}$, value $3s/4$. Its bid is half the auxiliary main bid, hence $t/2$ at a main tie. It will follow both main goods, price $G_{\mathrm{out}}$ in the second main position, and pay the input bid.

Give $G_{\mathrm{in},1}$ of $A_{\mathrm{in},1}$ value $2s/5$, hence bid $(2s/5)\alpha_{\mathrm{in},1}$. When $G_{\mathrm{out}}$ precedes $G_{\mathrm{aux}}$, the no-self-pricing rule skips $G_{\mathrm{mid}}$ and makes both goods of $A_{\mathrm{aux}}$ pay this signal bid. Their total $(4s/5)\alpha_{\mathrm{in},1}\le4s/5<s$ will exclude unequal main bids against the auxiliary budget $s$.

Give $G_{\mathrm{guard},1}$ value $2s/15$ and $A_{\mathrm{guard},1}$ budget $B_{\mathrm{guard},1}=4s/15$, fixing its bid at $2s/15$. The order proof will make this the signal price. Add $2s/15$ to the actual budget of $A_{\mathrm{in},1}$ to compensate that payment.

For $A_{\mathrm{out}}$ and $A_{\mathrm{aux}}$, choose the base budgets together as

\[ B_{\mathrm{out}}^0=B_{\mathrm{aux}}^0=s. \]

We eliminate the main-order probability as in \cite{HLN26}, now with a variable input bid to obtain the nonlinear response.

\paragraph{Deriving the equilibrium relation.} We use Lemma~\ref{lem:budget-multiplier-bounds} and the prices and compensation proved independently in Section~\ref{subsec:budget-completing}, including that $G_{\mathrm{mid}}$ precedes $G_{\mathrm{in},1}$. We first establish equal main bids.

\begin{lemma}[Equal main bids in the nonlinear budget-GSP gadget]\label{lem:budget-nonlinear-equal-bids}
In every equilibrium of the constructed auction, the nonlinear gadget has equal bids on $G_{\mathrm{out}}$ and $G_{\mathrm{aux}}$. Thus, $\alpha_{\mathrm{out}}=\alpha_{\mathrm{aux}}$.
\end{lemma}

\begin{proof}
Lemma~\ref{lem:budget-multiplier-bounds} gives $2s/15<(2s/5)\alpha_{\mathrm{in},1}\le2s/5$. If $G_{\mathrm{out}}$ bids strictly more than $G_{\mathrm{aux}}$, both goods of $A_{\mathrm{aux}}$ pay the input bid, for total payment $(4s/5)\alpha_{\mathrm{in},1}\le4s/5<s$. Unspent budget forces $\alpha_{\mathrm{aux}}=1$, contradicting the strict order because both main values are $3s/2$. If $G_{\mathrm{aux}}$ bids strictly more, the price of $G_{\mathrm{out}}$ is at most $\max\{3s/4,(2s/5)\alpha_{\mathrm{in},1}\}=3s/4<s$. Cancelling outgoing compensation then forces $\alpha_{\mathrm{out}}=1$, again contradicting the strict order. Hence the main bids and multipliers are equal.
\end{proof}

By Lemma~\ref{lem:budget-nonlinear-equal-bids}, the common main bid is $t=(3s/2)\alpha_{\mathrm{out}}$. The middle bid is $t/2$, so $G_{\mathrm{mid}}$ follows both main goods.

Let $q$ be the probability that $G_{\mathrm{out}}$ appears before $G_{\mathrm{aux}}$ at their tie. In this order, $G_{\mathrm{out}}$ pays $t$, while $G_{\mathrm{aux}}$ and $G_{\mathrm{mid}}$ each pay $(2s/5)\alpha_{\mathrm{in},1}$. In the opposite order, $G_{\mathrm{aux}}$ pays $t$, $G_{\mathrm{out}}$ pays $t/2$, and $G_{\mathrm{mid}}$ pays $(2s/5)\alpha_{\mathrm{in},1}$. Therefore the expected payment of $A_{\mathrm{out}}$ for $G_{\mathrm{out}}$ is $t(1+q)/2$, and the total expected payment of $A_{\mathrm{aux}}$ is $(2s/5)\alpha_{\mathrm{in},1}(1+q)+t(1-q)$.

The common multiplier cannot equal $1$. Otherwise $t=3s/2$, and the budget inequality of $A_{\mathrm{out}}$ would require $q\le1/3$. Since the total expected payment of $A_{\mathrm{aux}}$ decreases with $q$, it would then be at least $s+(8s/15)\alpha_{\mathrm{in},1}>s$, violating its budget. Thus both multipliers are below $1$, and both base budgets must be spent in full:

\begin{equation}\label{eq:budget-nonlinear-payments}
\left\{\begin{aligned}&\frac t2(1+q)=s,\\&\frac{2s}{5}\alpha_{\mathrm{in},1}(1+q)+t(1-q)=s.\end{aligned}\right.
\end{equation}
The first equation gives $q=2s/t-1$. Substituting this expression into the second equation eliminates $q$ and gives $(2s/5)\alpha_{\mathrm{in},1}=3t/2-t^2/s$. Substituting $t=(3s/2)\alpha_{\mathrm{out}}$ yields

\[ \frac{2s}{5}\alpha_{\mathrm{in},1}=\frac{9s}{4}\alpha_{\mathrm{out}}(1-\alpha_{\mathrm{out}}). \]

The payment for $G_{\mathrm{out}}$ is at most $t$, so its base-budget equality gives $t\ge s$ and hence $\alpha_{\mathrm{out}}\ge2/3$. This bound selects the admissible quadratic root in the following lemma.

\paragraph{Completing the gadget.} The following lemma gives that output and shows that it has a local completion.

\begin{lemma}[Exact realization of the nonlinear budget-GSP gadget]\label{lem:budget-nonlinear-realization}
Under the prices and compensation of Lemma~\ref{lem:budget-signal-prices}, every equilibrium satisfies
\[
\alpha_{\mathrm{aux}}=\alpha_{\mathrm{out}}
=\frac12+\sqrt{\frac14-\frac{8}{45}\alpha_{\mathrm{in},1}}
\in(3/4,1).
\]
For inputs in $(1/3,1]$, this is the unique quadratic root at least $2/3$ and is continuous and strictly decreasing. Conversely, every input in $[3/4,1]$ admits a local completion realizing this output and satisfying both core budget conditions. Outgoing signal payments are offset by the prescribed budget increases.
\end{lemma}

\begin{proof}

The quadratic relation above has a unique root at least $2/3$, namely the stated output in $(3/4,1)$. Its formula gives continuity and strict monotonicity.

For any encoded input $\alpha_{\mathrm{in},1}\in[3/4,1]$, choose this output, set $\alpha_{\mathrm{aux}}=\alpha_{\mathrm{out}}$, and set the guard multiplier to $1$. With $t=(3s/2)\alpha_{\mathrm{out}}$, choose $q=2s/t-1$ as the probability that $G_{\mathrm{out}}$ appears before $G_{\mathrm{aux}}$. Since $s<t<3s/2$, we have $0<q<1$. The bid of $G_{\mathrm{mid}}$ satisfies $t/2>9s/16>2s/5\ge(2s/5)\alpha_{\mathrm{in},1}$, and the bid of $G_{\mathrm{in},1}$ is at least $3s/10>2s/15$. Thus the required local order holds, and both orders of the tied main goods are valid.

Substitution into~\eqref{eq:budget-nonlinear-payments} verifies both base-budget equalities. Outgoing signal payments are exactly compensated, and the guard pays less than its budget at multiplier $1$. Thus the local order distribution gives the required extension.
\end{proof}

\subsection{Encoding and Exact Gate Implementation}
\label{subsec:budget-encoding}
Lemmas~\ref{lem:budget-affine-realization} and~\ref{lem:budget-nonlinear-realization} put every wire output in $[3/4,1]$. Represent a source value $z\in[0,1]$ by

\[\alpha_{\mathrm{out}}=\frac34+\frac z4,\qquad z=4\alpha_{\mathrm{out}}-3.\]

\paragraph{Affine source operations.} Suppose the inputs encode source values $y,w$, so that $\alpha_{\mathrm{in},1}=3/4+y/4$ and $\alpha_{\mathrm{in},2}=3/4+w/4$. To realize

\[z=\min\{1,c-ay-bw\},\qquad a,b\ge0,\qquad c\ge a+b,\]
choose $a_1=a$, $a_2=b$, and $C=3/4+c/4+3(a+b)/4$, omitting any input whose coefficient is zero. These choices satisfy the lower-bound condition of the affine gadget because

\[C-a-b=\frac34+\frac{c-a-b}{4}\ge\frac34.\]

Also, substituting the encoded inputs gives

\[C-a\alpha_{\mathrm{in},1}-b\alpha_{\mathrm{in},2}=\frac34+\frac{c-ay-bw}{4}.\]
Since $1=3/4+1/4$, the local relation \eqref{eq:budget-affine} therefore gives $\alpha_{\mathrm{out}}=3/4+z/4$, as required.
Taking $a=c=1$ and $b=0$ gives source complement $z=1-y$. To obtain capped doubling, first complement the input and then apply the operation $z=\min\{1,2-2y\}$ to that complement. The result is $\min\{1,2y\}$. As in Section~\ref{subsec:spa-encoding}, Lemma~\ref{lem:clipped-affine} extends these operations to every fixed signed clipped affine expression, with at most two inputs per basic gadget.

\paragraph{Using the nonlinear relation for multiplication.} To express the nonlinear gadget's operation in source coordinates, write $\alpha_{\mathrm{in},1}=3/4+x/4$ and $\alpha_{\mathrm{out}}=3/4+y/4$. Substituting these expressions into the square-root relation of Section~\ref{subsubsec:budget-nonlinear} gives

\[y=h(x),\qquad h(x)=4\sqrt{\frac7{60}-\frac{2x}{45}}-1.\]
The function $h$ is continuous and strictly decreasing on $[0,1]$, and its values lie in $(0,1)$. Solving this equation for $x$ gives

\[h^{-1}(y)=\frac{21}{8}-\frac{45}{32}(y+1)^2.\]
This inverse is a nonaffine quadratic polynomial with rational coefficients. To use it in the feedback construction of Section~\ref{subsec:spa-encoding}, we need an interval whose inverse values remain valid source inputs. Choose

\[J=\left[\frac18,\frac14\right].\]
The inverse is strictly decreasing on $J$, since $(h^{-1})'(y)=-45(y+1)/16<0$. Evaluating the endpoints gives

\[h^{-1}(J)=\left[\frac{219}{512},\frac{1731}{2048}\right]\subset(0,1).\]
Thus $J\subset\operatorname{int}h([0,1])$: each target in $J$ has an inverse strictly inside the source interval, as required to exclude endpoint solutions of the feedback equations below.

To implement the inverse for a prescribed $\tau\in J$, impose

\[\left\{\begin{aligned}&y=h(x),\\&x=\mathrm{clip}_{[0,1]}(x+y-\tau).\end{aligned}\right.\]
Here $\operatorname{clip}_{[0,1]}(r)=\min\{1,[r]_+\}$, which is implemented by the affine operations above. If $x=0$, then $h(0)>\tau$, so the feedback right side is positive. If $x=1$, then $h(1)<\tau$, so the feedback right side is less than $1$. Therefore $x\in(0,1)$, and the feedback equation forces $y=\tau$. The nonlinear equation then gives $x=h^{-1}(\tau)$. Conversely, these values satisfy both equations.
These affine operations and the verified inverse satisfy Theorem~\ref{thm:criterion}(i), giving a fixed-size rational implementation of $\phi(y,w)=(1+y)(1+w)/4$ with both solution directions. All constants depend only on the fixed source gate set.

\subsection{Completing the Proof}
\label{subsec:budget-completing}
We first assemble the local arithmetic implementations into one auction. Replace each source gate by its arithmetic implementation, then apply Lemma~\ref{lem:bounded-use} to the expanded circuit so that every wire is used as an input at most twice. Create one wire bidder per wire, acting as $A_{\mathrm{out}}$ in its upstream gadget and as $A_{\mathrm{in},j}$ in downstream gadgets. Each gadget has fresh goods, a fresh auxiliary $A_{\mathrm{aux}}$, and fresh guards $A_{\mathrm{guard},j}$. Set each wire bidder's actual budget to its upstream base budget plus one third of the value of each outgoing signal good it owns.

\paragraph{Multiplier bounds.} We first obtain the preliminary bounds needed to prove prices without using any gate equation.

\begin{lemma}[Order-independent multiplier bounds]\label{lem:budget-multiplier-bounds}
In every equilibrium of the assembled budget-GSP auction, each wire bidder has multiplier strictly greater than $1/3$, each affine auxiliary has $\alpha_{\mathrm{aux}}\ge(S+1/2)/(1+4S)>1/4$, each nonlinear auxiliary has $\alpha_{\mathrm{aux}}\ge4/9$, and every guard has multiplier $1$.
\end{lemma}

\begin{proof}
Let $D$ be its main-good value, $V$ its total owned value, and $P$ its total expected payment. The construction gives $B^0>D/3$: in an affine gadget, $B^0\ge H/2>H/3$, while in a nonlinear gadget, $B^0=s>(3s/2)/3$. The prescribed compensation therefore gives

\[B=B^0+\frac{V-D}{3}>\frac V3.\]
Every good pays at most its own bid under any valid order, so $P\le\alpha V$. If $\alpha<1$, budget exhaustion gives $B=P\le\alpha V$, hence $\alpha>1/3$. The same bound is immediate when $\alpha=1$. Since each $A_{\mathrm{in},j}$ is a wire bidder used in a downstream role, this proves $\alpha_{\mathrm{in},j}>1/3$ without assuming any signal price or gate equation.

We also bound the auxiliary multipliers and fix the guard bids. For an affine auxiliary $A_{\mathrm{aux}}$, its total owned value is $(1+4S)H$ and its budget is $(S+1/2)H$, where $S=\sum_{j=1}^k a_j$. The same payment bound gives $\alpha_{\mathrm{aux}}\ge(S+1/2)/(1+4S)>1/4$. For a nonlinear auxiliary $A_{\mathrm{aux}}$, the total owned value is $9s/4$ and the budget is $s$, giving $\alpha_{\mathrm{aux}}\ge4/9$. Both bounds also hold at multiplier $1$. Each guard $A_{\mathrm{guard},j}$ with value $w$ and budget $2w$ pays at most $w$, forcing $\alpha_{\mathrm{guard},j}=1$ and fixing its bid at $w$.
\end{proof}

The next lemma establishes the signal prices and local bid orders before any gate equation is used.
\begin{lemma}[Signal prices]\label{lem:budget-signal-prices}
In every equilibrium, each signal good of value $v$ pays $v/3$, matching its owner's budget increment. Affine receiving goods pay their input signal bids, and the last affine main good pays $H/4$. Each nonlinear middle good precedes its input signal.
\end{lemma}
\begin{proof}
 The finite collection of gadget types and Lemma~\ref{lem:budget-multiplier-bounds} give fixed positive lower and upper bounds on bids relative to each gadget's scale. Choose successive scales using a fixed rational geometric factor large enough to make their possible bid intervals disjoint. Within an affine gadget, the main goods $G_{\mathrm{out}}$ and $G_{\mathrm{aux}}$ bid more than $H/4$, and each receiving bid satisfies $4v_j\alpha_{\mathrm{aux}}>v_j\ge v_j\alpha_{\mathrm{in},j}>v_j/3$. Together with $4v_1\le H/8$ and $4v_2\le v_1/6$, these inequalities give the branch order in Figure~\ref{fig:budget-affine}. Thus each $G_{\mathrm{recv},j,\ell}$ pays $v_j\alpha_{\mathrm{in},j}$, and each $G_{\mathrm{in},j}$ pays $v_j/3$.

We next prove that $G_{\mathrm{mid}}$ precedes $G_{\mathrm{in},1}$. Both main goods already precede the signal: their bids exceed $s/2$ and are at least $2s/3$, respectively, while the signal bids at most $2s/5$. Suppose $(3s/4)\alpha_{\mathrm{aux}}\le(2s/5)\alpha_{\mathrm{in},1}$, so $\alpha_{\mathrm{aux}}\le8/15<1$. If $G_{\mathrm{out}}$ outbid $G_{\mathrm{aux}}$, bidder $A_{\mathrm{aux}}$ would pay at most $4s/5<s$, contradicting full budget spending. Thus $\alpha_{\mathrm{out}}\le\alpha_{\mathrm{aux}}\le8/15$. All outgoing affine signals of this $A_{\mathrm{out}}$ already pay their guard prices. In any downstream nonlinear gadget of scale $s'$, its signal bids at most $16s'/75<s'/3$, below both main bids and the middle bid, which are at least $s'/3$. Outgoing nonlinear signals also pay their guards. Cancelling these payments and matching budget increases leaves budget $s$ for upstream $G_{\mathrm{out}}$, whose payment is at most $(3s/2)\alpha_{\mathrm{out}}\le4s/5<s$. This contradicts full spending because $\alpha_{\mathrm{out}}<1$. Hence $G_{\mathrm{mid}}$ precedes $G_{\mathrm{in},1}$, which pays its guard price $2s/15$. Every signal price equals its prescribed budget increment.
\end{proof}

\paragraph{Fixed-price compensation and global correspondence.} Every outgoing signal good of value $v$ pays exactly $v/3$, matching the increase in its owner's actual budget. Subtracting these amounts restores the upstream base-budget condition, including equality below multiplier $1$. Lemmas~\ref{lem:budget-affine-realization} and~\ref{lem:budget-nonlinear-realization} therefore apply simultaneously, placing every wire multiplier in $[3/4,1]$ and enforcing all arithmetic and copy equations, including on directed cycles. Every auction equilibrium thus decodes to a solution of the expanded circuit and, after removing intermediate arithmetic and copy wires, to a solution of the original source circuit.

To extend a source solution, extend through the arithmetic and copies, then encode all wires. Apply Lemmas~\ref{lem:budget-affine-realization} and~\ref{lem:budget-nonlinear-realization} to choose each $\alpha_{\mathrm{aux}}$ and tied-main order probability, setting all guard multipliers to $1$. Bid separation and exact compensation preserve all local budgets simultaneously. Lemma~\ref{lem:local-coupling} couples the local probabilities using one shared uniform number in $[0,1]$. For $N$ gadgets, the thresholds give at most $N+1$ intervals, each defining a valid global order. This distribution preserves every local expected payment, giving an equilibrium with polynomial support.

\paragraph{Sparsity and complexity.} We check size and ownership. Each wire bidder owns upstream $G_{\mathrm{out}}$ and at most two outgoing $G_{\mathrm{in},j}$, hence at most three goods. An affine auxiliary $A_{\mathrm{aux}}$ owns $G_{\mathrm{aux}}$ and $\sum_{j=1}^k k_j$ receiving goods; a nonlinear auxiliary $A_{\mathrm{aux}}$ owns two goods; each guard $A_{\mathrm{guard},j}$ owns one. The integers $k_j$ depend only on the fixed finite gate types, so all these bounds are universal constants. Gate replacement and copying increase size by a constant factor. Geometric scales and compensation sums have polynomial bit length. Thus auction size is linear in source size, with positive rational values and budgets of polynomial encoding length.

We now combine the established correspondence and instance bounds to prove Theorem~\ref{thm:budget-gsp}.

\begin{proof}[Proof of Theorem~\ref{thm:budget-gsp}]
Membership follows from \cite{HLN26}; Section~\ref{sec:complexity} gives the corresponding distance upper bound.

Lemma~\ref{lem:budget-multiplier-bounds} and the pricing argument above justify Lemmas~\ref{lem:budget-affine-realization} and~\ref{lem:budget-nonlinear-realization}. With Section~\ref{subsec:budget-encoding}, these give the source equations and both global solution directions proved above. Lemma~\ref{lem:local-coupling} gives polynomial support; the ownership and scale counts give the stated constant inventory and polynomial encoding length.

For each source variable, the decoder reads one designated multiplier and applies $z=4\alpha_{\mathrm{out}}-3$. Thus the correspondence is an SL reduction from Lemma~\ref{lem:source}, proving $\FIXP$-hardness under the stated restrictions and hence $\FIXP$-completeness. A rational multiplier vector within $\varepsilon/4$ in $\ell_\infty$ of an exact equilibrium decodes to a vector within $\varepsilon$ of a source solution. This fixed rational tolerance conversion gives distance hardness; the matching upper bound proves $\FIXPa$-completeness for the multiplier vector.
\end{proof}

\section{ROI-Constrained GSP}
\label{sec:roi-gsp-rewritten}\label{sec:roi}
We now replace the budget constraint of Section~\ref{sec:budget} by ROI. Full display and unit click-through rates again fix acquired values. A bidder uses a multiplier in $[1,\widehat A]$ and must pay its total value whenever its multiplier is below $\widehat A$.
Displaying $G_{\mathrm{in},j}$ now adds both payment and value to $A_{\mathrm{in},j}$, with no independent budget to increase. We add a compensation good so that the pair's payment equals its value, preserving the upstream ROI condition.

\begin{figure}[H]
\centering
\begin{tikzpicture}[
box/.style={draw,rounded corners=2pt,align=center,text width=14.6cm,inner xsep=5pt,inner ysep=4pt,font=\normalsize},
side/.style={box,text width=6.7cm},
arr/.style={-{Stealth[length=1.7mm]},line width=0.45pt}
]
\node[box] (pre) at (0,0) {Lemma~\ref{lem:roi-explicit-parameters}; Appendix~\ref{sec:identical-suppliers}: Parameters and supplier bounds};
\node[box,below=2.5mm of pre] (sep) {ROI feasibility: Core bound $\alpha_i<B$ and prescribed scale gaps (Section~\ref{subsec:roi-suppliers-rewritten})};
\node[box,below=2.5mm of sep] (same) {Lemma~\ref{lem:roi-common-supplier-multipliers}: Common supplier multipliers and block separation};
\node[box,below=2.5mm of same] (pair) {Lemma~\ref{lem:roi-signal-pair-payments}: Pair-payment identities and one-sided bounds};
\node[box,below=2.5mm of pair] (range) {Lemmas~\ref{lem:roi-core-upper-bound} and~\ref{lem:roi-wire-lower-bound}: Core upper bounds; wire and nonlinear auxiliary lower bounds};
\node[side,below=3mm of range,xshift=-3.75cm] (aff) {Lemmas~\ref{lem:roi-affine-aux-lower}--\ref{lem:roi-affine-realization}: Affine realization\\Requires \eqref{eq:roi-affine-promise}};
\node[side,below=3mm of range,xshift=3.75cm] (non) {Lemmas~\ref{lem:roi-nonlinear-equal-bids}--\ref{lem:roi-nonlinear-realization}: Nonlinear realization};
\node[box,below=3mm of aff,xshift=3.75cm] (enc) {Lemma~\ref{lem:roi-restricted-affine}: Restricted affine construction\\Check input ranges and endpoints before using each equation};
\node[box,below=2.5mm of enc] (close) {Lemma~\ref{lem:roi-encoding-closure}: Source arithmetic and simultaneous encoding closure};
\node[box,below=2.5mm of close] (rev) {Proposition~\ref{prop:roi-block-interface}: Complete suppliers at the actual lower price\\Then choose and couple main orders};
\foreach \a/\b in {pre/sep,sep/same,same/pair,pair/range,range/aff,range/non,aff/enc,non/enc,enc/close,close/rev}{\draw[arr] (\a)--(\b);}
\end{tikzpicture}
\caption{Logical order of the ROI-GSP proof.}
\label{fig:roi-proof-dependencies}
\end{figure}
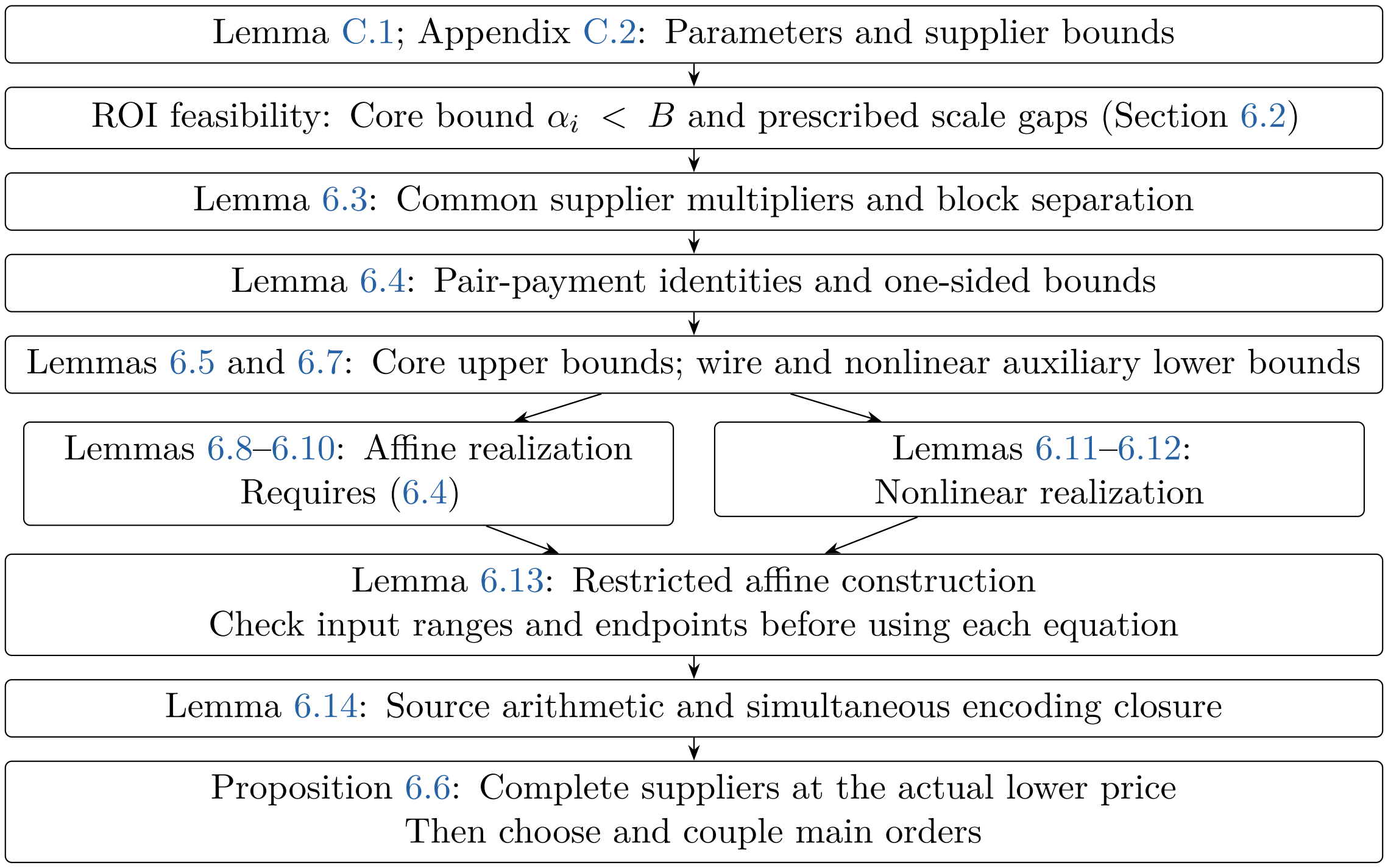

\begin{theorem}[Exact ROI-Constrained GSP Equilibria]
\label{thm:roi-gsp-rewritten}
For the canonical instance-defined cap or any prescribed fixed rational cap $A>1$, exact ROI-constrained no-self-pricing GSP equilibrium computation is $\FIXP$-complete under separable linear reductions. Approximating the multiplier vector in distance to an exact equilibrium is $\FIXPa$-complete under the same restrictions. Hardness holds in one auction with unit CTRs, full display, positive rational values, polynomial-support equilibrium distributions over legal tied orders, and a constant bound on goods per bidder, universal for the canonical cap and depending only on $A$ otherwise.
\end{theorem}

\paragraph{Proof dependencies.} Figure~\ref{fig:roi-proof-dependencies} summarizes the proof dependencies. Lemma~\ref{lem:roi-explicit-parameters} and Lemma~\ref{lem:identical-suppliers} first give the supplier parameters and preliminary multiplier bounds. Lemma~\ref{lem:roi-common-supplier-multipliers} establishes common supplier multipliers and block separation, and Lemma~\ref{lem:roi-signal-pair-payments} gives the pair-payment bounds used in Lemmas~\ref{lem:roi-core-upper-bound} and~\ref{lem:roi-wire-lower-bound}. These working bounds support the affine and nonlinear realizations in Lemmas~\ref{lem:roi-affine-aux-lower}--\ref{lem:roi-nonlinear-realization}. Lemma~\ref{lem:roi-restricted-affine} checks the affine calls, and Lemma~\ref{lem:roi-encoding-closure} closes the source encoding. Proposition~\ref{prop:roi-block-interface} supplies the supplier completions used in the reverse extension.

\subsection{A Common Auction Structure}
\label{subsec:roi-common-rewritten}
Use the wire roles of Section~\ref{sec:overview}, with main good $G_{\mathrm{out}}$ owned by $A_{\mathrm{out}}$ and main good $G_{\mathrm{aux}}$ owned by a fresh auxiliary $A_{\mathrm{aux}}$ with multiplier $\alpha_{\mathrm{aux}}$. Wire and auxiliary bidders are \emph{core bidders}; $A_i$ denotes a generic core bidder.

For each input bidder $A_{\mathrm{in},j}$, we add a corresponding compensation good $G_{\mathrm{comp},j}$. This good balances the added payment and value of $A_{\mathrm{in},j}$, preserving its ROI condition in the upstream gadget. We call $G_{\mathrm{in},j}$ and $G_{\mathrm{comp},j}$ together an ordinary signal pair.
For each core bidder $A_i$, introduce a \emph{private signal pair} $G_{\mathrm{psig},i},G_{\mathrm{pcomp},i}$. Its value minus payment is a prescribed adjustment $\Delta_i$ on the intended interval, setting the payment target for the other goods of $A_i$. Its payment bounds also exclude large multipliers. Private pairs do not transmit inputs downstream.

For each ordinary or private pair, add two fresh supplier groups to set its signal and compensation prices. Each group has $n\ge2$ bidders, chosen below:
\begin{itemize}
\item $A_{\mathrm{sup},b}$: first-group supplier, $b=1,\ldots,n$, owning three goods $G_{\mathrm{sup},b,\ell}$, $\ell=1,2,3$. These goods set the signal price level. Summing this group's ROI equalities fixes the sum of signal and compensation prices.
\item $A_{\mathrm{cmp},b}$: second-group supplier, $b=1,\ldots,n$, owning six goods $G_{\mathrm{cmp},b,\ell}$, $\ell=1,\ldots,6$. One gives an order-independent core multiplier bound; the others set compensation price levels. Solving both groups' ROI equalities gives supplier multipliers consistent with the signal, compensation, and other constructed bids.
\end{itemize}
Suppliers are distinct from core bidders and do not represent circuit variables.

\paragraph{Choosing the multiplier interval.} We need ordinary pairs to preserve sender surplus and private pairs to supply fixed adjustments. Choose a working interval for inputs and a smaller interval for constant private payments, both below the actual cap so the core ROI conditions are equalities.

Fix a rational design parameter $A>1$, and set

\[h=\frac1{16}\min\{A-1,1\},\qquad L=1+4h,\qquad U=1+8h.\]
Here $h$ places $[L,U]$ strictly inside $(1,A)$ and keeps it narrow through $h\le1/16$. We will prove that all core equilibrium multipliers lie in $[L,U)$, and check arithmetic on $[L,U]$. The same construction works for every actual cap $\widehat A\ge U$; the prescribed-cap result uses $\widehat A=A$. The feasible domain remains $[1,\widehat A]$.

To specify where the private signal pairs must supply their fixed adjustments, let $s$ denote the relative width of the working interval and choose an inner upper endpoint $F$ by

\[s=\frac{U-L}{L},\qquad F=U-\frac{Ls^2}{2}.\]
These choices give $0<s\le1/5$ and $L+3h<F<U$. Ordinary pairs must conserve surplus on $[L,U]$, and private pairs must contribute $\Delta_i$ on $[L,F]$. Arithmetic outputs will lie in $[L,F)$, where both properties apply; source encoding uses a still smaller interval. We establish the supplier bounds and payments before using gate equations.

\subsection{Supplier Construction}
\label{subsec:roi-suppliers-rewritten}

\paragraph{Choosing the supplier values.} Supplier ROI equations will fix the combined pair payment; bid separation will justify the prices in those equations.
A \emph{block} consists of one signal pair, its suppliers, and the additional core goods specified by its gadget. Its scale is $g$. Values divided by $g$ are called normalized values; in particular, a compensation good has normalized value $e+\Delta/g$. A supplier \emph{level} consists of one equal-value good per supplier in a group. Common supplier multipliers give equal bids within a level.

A second-group supplier good of value $Gg$, with $G=401/100$, will give the preliminary core bound $B=401/50$ before the signal prices are known. Its role is proved below using only ROI feasibility and the supplier bound from Lemma~\ref{lem:identical-suppliers}.

We can therefore fix the upper supplier values independently of the supplier count. Choose

\[T=\frac{47}{2},\quad N=48,\quad M=96,\quad G=\frac{401}{100},\quad Q=\frac83,\quad B=2G=\frac{401}{50}.\]

\[\sigma=4h^2,\quad \rho=\frac{h^4}{16},\quad n=\left\lceil\frac1{4h^2}\right\rceil+2,\quad D=145,\quad a=(n-1)D+N,\quad K=nD,\quad S=K/a.\]

For the first group, $D=M+N+1$ is each supplier's normalized value and $K=nD$ is the group total. The coefficient $a$ comes from its total payment, and $Sg$ will be the combined pair payment. The inequalities $M>N>2T$ separate the upper supplier levels under the preliminary multiplier bounds.

We next choose how the compensation bid sits between the lower supplier levels. Set

\[q=\frac{1+1/L}{2}=1-\frac s2,\qquad
\vartheta=\frac12\left((q+\rho)F+\frac{(q-\rho)U}{1+\sigma}\right).\]

Use a parameter $\theta=2$ for ordinary signal pairs and $\theta=\vartheta$ for private signal pairs. The constant $q$ will set the compensation value relative to the lower pricing level, while $\theta$ will set the upper level that its bid may cross. Since $qL>1$ and $qF<\vartheta<qU$, the private choice leaves room to keep compensation between the two levels throughout $[L,F]$ and move it above the upper level before multiplier $U$. The ordinary choice will keep it between the levels throughout $[L,U]$.

The margins $\sigma$ and $\rho$ allow variation in supplier multipliers and private valuations. Lemma~\ref{lem:roi-explicit-parameters} proves that the interval defining $\vartheta$ is nonempty and gives

\begin{equation}\label{eq:small-cap-supplier-margins}\left\{\begin{aligned}&(q-\rho)L>1+\sigma,\\&(q+\rho)F<\vartheta,\\&(q-\rho)U>\vartheta(1+\sigma).\end{aligned}\right.\end{equation}
All these parameters depend only on the fixed design cap.

For either $\theta\in\{\vartheta,2\}$, define the remaining supplier levels and the signal and compensation coefficients by

\begin{equation}\label{eq:small-cap-supplier-data}\begin{aligned}&c(\theta)=\frac{T}{a-\theta},\qquad u(\theta)=\theta c(\theta),\qquad e(\theta)=qc(\theta),\\&v(\theta)=S-e(\theta),\qquad \delta(\theta)=\frac{49c(\theta)}{100}.\end{aligned}\end{equation}

Within a block, abbreviate these quantities by $c,u,e,v,\delta$. The six goods of each $A_{\mathrm{cmp},b}$ will have values $(T,G,u,c,2\delta,\delta)g$. The levels $u$ and $c$ determine the intended position and price of the compensation good, while $2\delta$ and $\delta$ provide the lower supplier levels used to complete the block. The identity $u=\theta c$ explains the choice $c=T/(a-\theta)$: it gives $T+u=ac$. Also, choosing $v=S-e$ gives $v+e=S$, so the ordinary signal pair has total value $Sg$ and the private signal pair has total value $Sg+\Delta$.

We will prove below that the first and second supplier groups have common multipliers $\alpha_{\mathrm{sup}}$ and $\alpha_{\mathrm{cmp}}$, respectively. Under the intended order, the first group's aggregate ROI equality will give

\[a\alpha_{\mathrm{sup}}+(T+u)\alpha_{\mathrm{cmp}}=K,\qquad \alpha_{\mathrm{sup}}+c\alpha_{\mathrm{cmp}}=S.\]

The intended signal and compensation prices are $\alpha_{\mathrm{sup}}g$ and $c\alpha_{\mathrm{cmp}}g$, whose sum is $Sg$. This gives ordinary surplus zero and private surplus $\Delta$. The payment proof below also treats private senders outside the constant-payment interval.

Additional core goods in an ordinary block have normalized values at most $Q=8/3$; affine receiving goods have value $rg$ with $r=8/5$. The fixed levels satisfy $G>UQ$ and $T>BQ+2$. These comparisons place the new supplier good above inserted core goods in the working range, and keep all inserted core bids below the $T$ level even under the preliminary bound $B$. Define the private excess-payment margin by

\[\gamma=\frac{c(\vartheta)(\vartheta-1)}2.\]

Lemma~\ref{lem:roi-explicit-parameters} in Appendix~\ref{subsec:roi-parameter-existence} verifies all parameter inequalities, including~\eqref{eq:small-cap-supplier-margins}. For the design cap $A=2$, each supplier group has $n=66$ bidders.

All counts and normalized value coefficients depend only on the fixed design cap $A$ and can be fixed before the source circuit is given. 

\paragraph{Values and multiplier bounds.} We now assign the chosen values to the supplier goods and choose the private block scales. For each signal block and each $b=1,\ldots,n$, the supplier ownership and values are as follows.

\begin{table}[H]
\centering
\begingroup
\setlength{\tabcolsep}{6pt}
\renewcommand{\arraystretch}{1.15}
\sbox{\TableBody}{%
\begin{tabular}{ccc}
\hline
Bidder & Owned goods & Values \\
\hline
$A_{\mathrm{sup},b}$ & $G_{\mathrm{sup},b,1},G_{\mathrm{sup},b,2},G_{\mathrm{sup},b,3}$ & $Mg,Ng,g$ \\
$A_{\mathrm{cmp},b}$ & $G_{\mathrm{cmp},b,1},\ldots,G_{\mathrm{cmp},b,6}$ & $Tg,Gg,ug,cg,2\delta g,\delta g$ \\
\hline
\end{tabular}%
}
\usebox{\TableBody}
\caption{Goods and values owned by each supplier in a signal block. Within each row, goods and values are listed in corresponding order.}
\label{tab:roi-supplier-ownership}
\endgroup
\end{table}

An ordinary block may also contain the receiving or other core goods prescribed by its gadget, with values in $(\delta g,Qg]$. In the affine gadget, $G_{\mathrm{out}}$ and $G_{\mathrm{aux}}$ occur before all goods in the ordinary block used to determine their lower price. Their placement will be specified in the affine construction.

We choose the private scale of each core bidder $A_i$ to protect the valuation adjustment and make the private payment sufficient to exclude large multipliers. Let $W_i$ be the total value of all goods owned by $A_i$ outside $G_{\mathrm{psig},i}$ and $G_{\mathrm{pcomp},i}$. After these other values and the required adjustment $\Delta_i$ have been fixed, choose $g_i$ so that

\begin{equation}\label{eq:roi-private-scale-choice}\left\{\begin{aligned}&\frac{|\Delta_i|}{g_i}\le c(\vartheta)\rho,\\&\gamma g_i>W_i+|\Delta_i|.\end{aligned}\right.\end{equation}

The first inequality keeps the compensation value within the reserved margin. Since a private block uses $e=qc(\vartheta)$, it gives $e+\Delta_i/g_i\ge c(\vartheta)(q-\rho)>0$. Thus $G_{\mathrm{pcomp},i}$ has positive value, and its normalized value remains within the bounds used to determine its position among the supplier goods. The second inequality ensures that the total value of all goods owned by $A_i$ satisfies

\[Sg_i+\Delta_i+W_i\le Sg_i+|\Delta_i|+W_i<(S+\gamma)g_i<2g_i.\]
Both scale conditions can be met by increasing $g_i$, since $W_i$ and $\Delta_i$ are fixed and $c(\vartheta)\rho,\gamma>0$. All scales are chosen rationally from the constructed data. The preliminary bounds and price induction below justify the resulting block order.

Apply Lemma~\ref{lem:bounded-use} to the source circuit and call the resulting circuit $\mathcal C'$. Replace every gate of $\mathcal C'$, including its complement-copy gates, by the fixed implementation in Section~\ref{subsec:roi-arithmetic-rewritten}, with a fresh normalization subcircuit at each input occurrence as specified there. Each circuit gate has outdegree at most two; normalization and arithmetic expansion turn each input occurrence into only constantly many internal connections. Create one bidder per resulting wire and fresh auxiliaries and suppliers. Choose ordinary patterns at geometrically separated scales, including the main goods before each affine floor block. After fixing all these ordinary values and private adjustments, enumerate the core bidders as $A_i$ and choose their private scales by

\[g_i=D_0\left(g_{\mathrm{base}}+1+W_i+|\Delta_i|+\sum_{j<i}g_j\right),\]
where $g_{\mathrm{base}}>0$ exceeds all ordinary pattern scales and $D_0$ is a sufficiently large fixed rational depending only on $A$. This enforces \eqref{eq:roi-private-scale-choice} and separates successive private blocks. The bounds below justify choosing all separating factors independently of the actual cap.

\paragraph{Bounds independent of bid order.} We adapt the ratio-two comparison and total-payment arguments of \cite{HLN26}. Identical catalogs have the same multiset of positive values, but distinct owners and goods. The following lemma bounds supplier multipliers using only ROI feasibility and these catalogs.

\begin{lemma}[Identical-Supplier Bounds]
\label{lem:identical-suppliers}
In a unit-CTR, full-display no-self-pricing GSP auction, consider $n\ge2$ suppliers with identical catalogs whose positive values include a pair in ratio two. Suppose their multipliers are at least one and they satisfy ROI feasibility. Unequal multipliers force the minimum multiplier in the group to be one. Every supplier multiplier is at most two. If all $n$ suppliers share a common multiplier $\alpha_{\mathrm{grp}}$, then $\alpha_{\mathrm{grp}}\le n/(n-1)$.
\end{lemma}

The proof is in Appendix~\ref{sec:identical-suppliers}.

We first apply Lemma~\ref{lem:identical-suppliers} to both supplier groups. Each group has identical catalogs containing a ratio-two pair, $Mg,Ng$ in the first group and $2\delta g,\delta g$ in the second. Every supplier multiplier is therefore at most two, before any separation or gate equation is used.

We next bound the core multipliers. If $\alpha_i\ge B=2G$, its private signal of value $vg_i>g_i$ bids more than $Bg_i$. A foreign second-group good of value $Gg_i$ has bid in $[Gg_i,Bg_i]$ and occurs later. The private signal therefore pays at least $Gg_i$, exceeding the core bidder's total value, which is below $2g_i$ by \eqref{eq:roi-private-scale-choice}. This violates ROI feasibility, so $\alpha_i<B$. Write the construction as complete arithmetic patterns, each consisting of a nonlinear block or an affine floor block with its input branches. Separate different patterns and private blocks by fixed geometric factors large enough for their preliminary bid bounds; affine main goods precede their floor block as proved below. Within an affine pattern, choose the branch scales by \eqref{eq:roi-explicit-receiving-count}, so that $98g_{\mathrm{lower}}<\delta_{\mathrm{upper}}g_{\mathrm{upper}}$. All normalized values within a block are at least $\delta$; inserted core bids are below $Tg$. The next lemma derives common supplier multipliers and complete block separation together by induction from lower to higher scales. Thus the local gap uses the sharper bound $98$, rather than the preliminary supplier bound $2M$.

\paragraph{Supplier multipliers and bid order.} The following price induction proves common supplier multipliers and block separation without any core gate equation.

\begin{lemma}[Common supplier multipliers]\label{lem:roi-common-supplier-multipliers}
In every equilibrium, the two supplier groups in each block have common multipliers satisfying
\[
1<\alpha_{\mathrm{sup}},\alpha_{\mathrm{cmp}}
\le\frac{n}{n-1}<1+\sigma<U\le\widehat A,
\]
and all supplier ROI constraints are tight. Block bids are below $98g$, blocks follow the prescribed order, and the actual highest lower bid satisfies $p<\delta g$.
\end{lemma}

\begin{proof}
Proceed from lower to higher price patterns, and within each affine pattern from its lowest branch to its floor block. Different complete patterns and private blocks satisfy the coarse separation chosen above. For adjacent blocks within an affine pattern, the scale gap is $98g_{\mathrm{lower}}<\delta_{\mathrm{upper}}g_{\mathrm{upper}}$. By induction, every lower block has bids below $98g_{\mathrm{lower}}$, so the highest lower bid is $p<\delta g$ for the current block. Before proving the current bound, every good of a higher block bids at least its value and hence more than $98g>Ng$. Affine main goods precede their entire floor block under the preliminary bounds. These facts suffice for the following minimum-one arguments; full separation of the current block is their conclusion, not a premise.

If a supplier group's multipliers differ, Lemma~\ref{lem:identical-suppliers} makes their minimum one. In the second group, the last value-$\delta g$ good among minimum-one suppliers pays less than $\delta g$: all other current-block goods have bids above $\delta g$, by $c(q-\rho)>\delta$, and only the lower blocks can follow this last good at smaller bids. In the first group, the last value-$Ng$ good among minimum-one suppliers also pays less than $Ng$. Lower first-group levels bid at most $2g$; second-group goods bid at most $2Tg<Ng$; inserted core bids are below $Tg$; lower blocks bid below $\delta g$; and higher-block goods exceed $Ng$ and therefore cannot price it. All other goods of a minimum-one supplier pay at most their values. This gives strict aggregate ROI surplus among the minimum-one suppliers, impossible below the cap. The same argument excludes a common multiplier of one.

Both groups therefore have common multipliers strictly above one. Lemmas~\ref{lem:identical-suppliers} and~\ref{lem:roi-explicit-parameters} give

\[1<\alpha_{\mathrm{sup}},\alpha_{\mathrm{cmp}}\le\frac n{n-1}<1+\sigma<U\le\widehat A.\]

Every supplier bid in the current block is now at most $M(1+\sigma)g<98g$; every inserted core bid is below $Tg<98g$. Thus the current block is completely separated from the higher blocks, and its upper bound supplies the induction hypothesis for the next block. This proves the stated order and lower-price bound for all blocks. This numerical induction uses no source equation.
\end{proof}

By Lemma~\ref{lem:roi-common-supplier-multipliers}, these common supplier bounds determine the signal positions before using any core gate equation. The first group's equation is $K=a\alpha_{\mathrm{sup}}+T\alpha_{\mathrm{cmp}}+z_*$ with $z_*\ge0$, so

\[v-\alpha_{\mathrm{sup}}\ge c(1-q-\theta/a)>0.\]
Since $\alpha_{\mathrm{send}}\ge1$, every signal precedes the first group's value-one goods. Also, the compensation bid is at most $Bc(q+\rho)<1/4$, while $u\alpha_{\mathrm{cmp}}<1/4<\alpha_{\mathrm{sup}}$. The signal lies below the $T$ level by $Bv\le BQ<T$. These preliminary comparisons, including compensation positions outside the constant-payment interval, justify the first-group equations used next.

\paragraph{Deriving the pair payments.} Using the preliminary bound $B=401/50$, block separation, and common supplier multipliers, we calculate pair payments. The first group's equation gives conservation; the private-payment estimate then yields the sharper bound $U$.

\begin{lemma}[Signal-pair payment identities and bounds]\label{lem:roi-signal-pair-payments}
Under the preliminary bounds, block order, and supplier multipliers established above, let $P_{\Sigma}$ denote a signal pair's payment at scale $g$, including ties. An ordinary pair has value $Sg$ and, if its additional upper core goods precede $G_{\mathrm{sig}}$, satisfies
\[
P_{\Sigma}\le Sg\quad(1\le\alpha_{\mathrm{send}}<L),
\qquad
P_{\Sigma}=Sg\quad(L\le\alpha_{\mathrm{send}}\le U).
\]
A private pair has value $Sg+\Delta$ and satisfies
\[\left\{\begin{aligned}&P_{\Sigma}\le Sg &&(1\le\alpha_{\mathrm{send}}<L),\\&P_{\Sigma}=Sg &&(L\le\alpha_{\mathrm{send}}\le F),\\&P_{\Sigma}\ge Sg &&(F<\alpha_{\mathrm{send}}<U),\\&P_{\Sigma}>(S+\gamma)g &&(U\le\alpha_{\mathrm{send}}\le B).\end{aligned}\right.\]
\end{lemma}

\begin{proof}
First consider constant-payment order, where compensation bids lie strictly between $c\alpha_{\mathrm{cmp}}g$ and $u\alpha_{\mathrm{cmp}}g$. The full table also assumes inserted core goods lie below level $G$, as in permitted completions with core multipliers at most $U$. The first-group rows proving conservation need only that these goods precede the signal and lie below level $T$; the preliminary bound gives the latter. These rows thus apply before the strict core upper bound. Let $p$ be the highest bid after the block, with $0\le p<\delta g$. Divide values, bids, and prices by $g$, and write

\[E=e+\frac{\Delta}{g},\qquad y=\alpha_{\mathrm{send}}E,\qquad z=\frac pg.\]
Thus $E$ is the normalized value of $G_{\mathrm{comp}}$, $y$ is its normalized bid, and $z$ is the normalized highest bid after the block. Let $J_{\max}$ be the highest normalized core bid between the supplier goods of values $Tg$ and $g$, including $G_{\mathrm{sig}}$. Each row of the table contains $n$ equal-bid goods, one for each $b=1,\ldots,n$. The last two columns distinguish the first $n-1$ goods in that row's bid order from its last good; they do not refer to fixed bidder indices.

The table follows from the no-self-pricing rule. Each of the first $n-1$ goods in a row has a later equal-bid good with a different owner and therefore pays that common bid. The last good must use a later pricing good. If that good belongs to the same owner, another supplier owns a good with the same value and bid at that position, so skipping the owner's own good does not change the listed price. The resulting group totals therefore do not depend on the order of supplier identities.

\begin{table}[H]
\centering
\begingroup
\setlength{\tabcolsep}{4pt}
\renewcommand{\arraystretch}{1.15}
\sbox{\TableBody}{%
\begin{tabular}{|c|c|c|c|}
\hline
Good & Value/$g$ & First $n-1$: price/$g$ & Last: price/$g$ \\
\hline
\multicolumn{4}{|c|}{First supplier group: owner $A_{\mathrm{sup},b}$} \\ \hline
$G_{\mathrm{sup},b,1}$ & $M$ & $M\alpha_{\mathrm{sup}}$ & $N\alpha_{\mathrm{sup}}$ \\ \hline
$G_{\mathrm{sup},b,2}$ & $N$ & $N\alpha_{\mathrm{sup}}$ & $T\alpha_{\mathrm{cmp}}$ \\ \hline
$G_{\mathrm{sup},b,3}$ & $1$ & $\alpha_{\mathrm{sup}}$ & $u\alpha_{\mathrm{cmp}}$ \\ \hline
\multicolumn{4}{|c|}{Second supplier group: owner $A_{\mathrm{cmp},b}$} \\ \hline
$G_{\mathrm{cmp},b,1}$ & $T$ & $T\alpha_{\mathrm{cmp}}$ & $G\alpha_{\mathrm{cmp}}$ \\ \hline
$G_{\mathrm{cmp},b,2}$ & $G$ & $G\alpha_{\mathrm{cmp}}$ & $J_{\max}$ \\ \hline
$G_{\mathrm{cmp},b,3}$ & $u$ & $u\alpha_{\mathrm{cmp}}$ & $y$ \\ \hline
$G_{\mathrm{cmp},b,4}$ & $c$ & $c\alpha_{\mathrm{cmp}}$ & $2\delta\alpha_{\mathrm{cmp}}$ \\ \hline
$G_{\mathrm{cmp},b,5}$ & $2\delta$ & $2\delta\alpha_{\mathrm{cmp}}$ & $\delta\alpha_{\mathrm{cmp}}$ \\ \hline
$G_{\mathrm{cmp},b,6}$ & $\delta$ & $\delta\alpha_{\mathrm{cmp}}$ & $z$ \\ \hline
\end{tabular}%
}
\usebox{\TableBody}
\caption{Supplier values and prices in the constant-payment order. Values and prices are divided by $g$; each row contains one good per supplier.}
\label{tab:roi-supplier-payments}
\endgroup
\end{table}

Summing the first supplier group's three rows gives the combined payment identity. Its total value is $Kg$, and its total payment divided by $g$ is $a\alpha_{\mathrm{sup}}+(T+u)\alpha_{\mathrm{cmp}}$. The suppliers' ROI equalities and the choices $T+u=ac$ and $S=K/a$ give

\[a\alpha_{\mathrm{sup}}+(T+u)\alpha_{\mathrm{cmp}}=K,\qquad \alpha_{\mathrm{sup}}+c\alpha_{\mathrm{cmp}}=S.\]

The first later foreign bid for $G_{\mathrm{sig}}$ is $\alpha_{\mathrm{sup}}g$, and that for $G_{\mathrm{comp}}$ is $c\alpha_{\mathrm{cmp}}g$. Their combined payment is thus $Sg$. An ordinary signal pair has total value $Sg$, so its ROI surplus is zero. A private signal pair has total value $Sg+\Delta$, so its ROI surplus is $\Delta$. The compensation-value conditions place $G_{\mathrm{comp}}$ between its two pricing levels throughout $[L,U]$ for an ordinary sender and throughout $[L,F]$ for a private sender. The required position of any additional core goods will be verified in the gadget and global range proofs.

We must also bound the private payment outside the constant-payment order. Write $\phi g$ for the compensation good's expected price and $z_*g$ for the last first-group value-one price. The unchanged upper first-group prices give $a\alpha_{\mathrm{sup}}+T\alpha_{\mathrm{cmp}}+z_*=K$. The private signal pays at least $\alpha_{\mathrm{sup}}g$, and pays exactly that below multiplier $U$, because then its bid is below $G\alpha_{\mathrm{cmp}}g$. Thus,

\[\frac{P_{\Sigma}}g-S\ge\phi-c\alpha_{\mathrm{cmp}}-\frac{z_*-u\alpha_{\mathrm{cmp}}}{a},\qquad z_*\ge u\alpha_{\mathrm{cmp}}.\]
Equality holds whenever the sender multiplier is below $U$.

If $y<u\alpha_{\mathrm{cmp}}$, then $z_*=u\alpha_{\mathrm{cmp}}$ and $\phi\le c\alpha_{\mathrm{cmp}}$, with equality when $y>c\alpha_{\mathrm{cmp}}$. At $y=u\alpha_{\mathrm{cmp}}$, the last first-group price is unchanged and $\phi\in[c\alpha_{\mathrm{cmp}},u\alpha_{\mathrm{cmp}}]$. If $y>u\alpha_{\mathrm{cmp}}$, then $z_*=y$ and $\phi=u\alpha_{\mathrm{cmp}}$. Since $\alpha_{\mathrm{send}}\le B$ and $E\le c(q+\rho)$, this gives

\[\begin{aligned}\frac{P_{\Sigma}}g-S&\ge c(\vartheta-1)-\frac{Bc(q+\rho)}a\\&>\frac{c(\vartheta-1)}2=\gamma.\end{aligned}\]
The strict inequality uses $q+\rho<1$ and $a(\vartheta-1)>2B$.
The compensation-value margins now place the private pair in the cases stated in the lemma. Below $L$ its compensation bid is below the upper pricing level; on $[L,F]$ it lies between the two levels; above $F$ it stays above the lower level; and at or above $U$ it strictly exceeds the upper level.
For an ordinary signal pair, the compensation bid stays below $u\alpha_{\mathrm{cmp}}g$ up to multiplier $U$. Provided the additional upper core goods occur before $G_{\mathrm{sig}}$, its combined payment is at most $Sg$ below $L$ and exactly $Sg$ on $[L,U]$.
\end{proof}

The payment bounds now let us sharpen the preliminary core bound to the working upper endpoint.

\begin{lemma}[Upper bound on core multipliers]
\label{lem:roi-core-upper-bound}
With the private scales chosen according to \eqref{eq:roi-private-scale-choice}, every equilibrium of the constructed auction satisfies $\alpha_i<U$ for every core bidder $A_i$.
\end{lemma}

\begin{proof}
Suppose that a core bidder $A_i$ has $\alpha_i\ge U$. Lemma~\ref{lem:roi-signal-pair-payments} gives a combined payment for $G_{\mathrm{psig},i}$ and $G_{\mathrm{pcomp},i}$ strictly greater than $(S+\gamma)g_i$. By \eqref{eq:roi-private-scale-choice}, the total value of all goods owned by $A_i$ is strictly less than $(S+\gamma)g_i$. Thus the payment for its private signal pair alone exceeds its total value. Payments on its other goods are nonnegative, so its total payment also exceeds its total value, contradicting ROI feasibility. Hence $\alpha_i<U$.
\end{proof}

\paragraph{Completing the supplier block.} For fixed permitted core bids, the following proposition supplies all supplier multipliers and tied supplier orders. Main and receiving payments are checked in the arithmetic gadgets.

\begin{proposition}[Signal-block composition interface]\label{prop:roi-block-interface}
Fix the parameters of Lemma~\ref{lem:roi-explicit-parameters}, a scale $g>0$, a foreign lower bid $0\le p<\delta g$, and the numerical core bids. An ordinary sender has multiplier in $[L,U]$ and zero adjustment; a private sender has multiplier in $[L,F]$ and $|\Delta|/g\le c(\vartheta)\rho$. Additional upper core goods must precede the signal, lie below the $G$ level, and have normalized bids at most $UQ$, as in Table~\ref{tab:roi-supplier-payments}. The only other core goods allowed are prescribed sender-owned compensation copies in ordinary floor blocks, with the same value as the compensation good; private blocks have no additional core goods. Then there exist supplier multipliers in $(1,n/(n-1))$ and a distribution over valid supplier-identity orders under which every supplier pays its value and core prices are unchanged. Ordinary and private pairs contribute surplus zero and $\Delta$, respectively.

The completions compose across separated blocks from lower to higher bids, with any valid main-order distributions preserving the numerical bids and block order. If each core bidder's remaining goods have payment minus value equal to its private adjustment, all core and supplier ROI equalities hold. Lemma~\ref{lem:local-coupling} couples the local marginals with polynomial support.
\end{proposition}

\begin{proof}
Fix the prescribed core bids and the actual lower bid $p$. We first solve for the two group multipliers using the total payments in Table~\ref{tab:roi-supplier-payments}. Let

\[D'=T+G+u+c+3\delta,\qquad a'=(n-1)D'+G+3\delta,\qquad J=J_{\max}+y,\qquad z=p/g.\]

Each $A_{\mathrm{cmp},b}$ owns value $D'g$. Summing the six rows of the second group gives total payment $g(a'\alpha_{\mathrm{cmp}}+J+z)$. Equating this payment to the group value $nD'g$, and then using the first group's equation, gives

\[\alpha_{\mathrm{cmp}}=\frac{nD'-J-z}{a'},\qquad \alpha_{\mathrm{sup}}=\frac{K-(T+u)\alpha_{\mathrm{cmp}}}{a}.\]
We check that these choices preserve the required order. The normalized core bounds give $0<J+z<U(Q+1)+1<T$, hence $1<\alpha_{\mathrm{cmp}}<n/(n-1)$. Also $K-a=M+1>(T+u)\alpha_{\mathrm{cmp}}$, so $1<\alpha_{\mathrm{sup}}<n/(n-1)$. Lemma~\ref{lem:roi-explicit-parameters} places both multipliers below $1+\sigma$, and the remaining parameter margins give the required supplier and core comparisons. The total payments of the two groups are exactly $Kg$ and $nD'g$, equal to their respective total values.

Fix an order of the $n$ supplier identities and apply one uniform cyclic shift at every level in both groups. Each supplier occupies each position with probability $1/n$. Since the table's prices are independent of identities, its expected payment is one $n$th of the group total: $Dg$ for $A_{\mathrm{sup},b}$ and $D'g$ for $A_{\mathrm{cmp},b}$, equal to its value. The shift preserves core prices. Supplier equations use numerical core bids, so main-order probabilities remain free to meet core targets.
Repeated receiving goods or compensation copies of one owner do not multiply $J$, because suppliers use the first relevant core bid. A nonlinear main tie has the same numerical bid and $J_{\max}$ in either order. Thus we can complete suppliers at the actual lower bid $p$ before choosing main-order probabilities, then combine both ordering choices.

Lemma~\ref{lem:roi-signal-pair-payments} gives the pair surpluses. Complete separated blocks from lower to higher bids. Adding a higher block leaves every lower good's first later foreign bid unchanged, preserving lower prices pointwise. Supplier-identity shifts preserve core prices pointwise and each supplier's expected payment at its value. For a core bidder, ordinary pairs contribute zero surplus, its private pair contributes $\Delta$, and the stipulated remaining goods contribute $-\Delta$. Its total surplus is therefore zero, and coupling preserves all expected payments.
\end{proof}

\subsection{Affine and Nonlinear Gadgets}
\label{subsec:roi-gadgets-rewritten}
Before deriving gate equations, we use the supplier payment bounds to establish the common lower bound for wire multipliers.

The argument uses the following local data, which are specified in detail in the two constructions below. Both gadgets use main value $m>0$ and give every output wire the private adjustment $(L-1)m$, so its main-payment target is $Lm$ when its private pair contributes its base adjustment. Affine receiving goods have value $rg$ with $r=8/5$. The nonlinear auxiliary owns a main good of value $m$ and a middle good of value $m/2$; its input signal has value $\kappa m$, where

\[\kappa=\frac{1-s-s^2}{2}\in[19/50,1/2).\]

Its private adjustment is $(L-3/2)m$, giving the same remaining target $Lm$. Thus $r>Uv$ and $\kappa U<L/2$. These are specified valuations and adjustments, not consequences of a gate equation.

\paragraph{Wire ranges and conserved signals.} The following proof uses the minimum multiplier across the auction, without an evaluation order for the circuit.

\begin{lemma}[Wire and nonlinear auxiliary lower bounds]\label{lem:roi-wire-lower-bound}
In every equilibrium, wire bidders and nonlinear auxiliaries have multipliers at least $L$. Every outgoing ordinary signal pair of a wire bidder has zero ROI surplus, preserving its upstream ROI condition.
\end{lemma}

\begin{proof}
Use Lemma~\ref{lem:roi-core-upper-bound}'s upper bound and Lemma~\ref{lem:roi-signal-pair-payments}'s payment bounds. Let $\alpha_*$ be the minimum multiplier over wire bidders and nonlinear auxiliaries, and suppose $\alpha_*<L$. If a wire bidder attains it, downstream affine receiving goods precede its signal since $r>Uv$. Downstream nonlinear main bids are at least $m\alpha_*$, and the middle bid at least $m\alpha_*/2>\kappa m\alpha_*$. Each outgoing ordinary pair thus pays at most its value. Its private pair contributes at least the required adjustment, but its upstream main pays at most $m\alpha_*<Lm$, giving strict ROI slack below the cap, a contradiction.

Otherwise, a nonlinear auxiliary attains the minimum and its output wire has a larger multiplier. Its main good follows the output main good, skips its own middle good, and pays the input bid. Its middle good pays at most the same. Together they pay at most $2\kappa mU<Lm$: the input bid is below the auxiliary main bid by $\kappa U<L/2<\alpha_*$, using $L\le5/4$ and $\alpha_*\ge1$. Its private payment is at most the base amount, contradicting target $Lm$. Thus all wire bidders and nonlinear auxiliaries have multiplier at least $L$. Receiving and middle-good orders follow, and each outgoing ordinary pair pays exactly its added value.
\end{proof}

\subsubsection{Affine gadget}
\label{subsubsec:roi-affine-rewritten}
Let $C$ be a rational constant term and let $d_j>0$ be rational input weights. We aim to enforce $\alpha_{\mathrm{out}}=H$, where

\[H=C-\sum_{j=1}^{k}d_j\alpha_{\mathrm{in},j}.\]
Here $k\in\{0,1,2\}$ is the number of input branches; the binary construction below reduces every three-input call to this form. An input with coefficient zero is omitted. Each use of the gadget must establish these conditions before invoking its affine equation:

\begin{equation}\label{eq:roi-affine-promise}L\le H<F,\qquad H=L\ \Longrightarrow\ \alpha_{\mathrm{out}}=L.\end{equation}

The first condition keeps the intended output inside the private constant-payment interval. The second requires the already known constraints to force $\alpha_{\mathrm{out}}=L$ whenever the proposed right-hand side reaches $L$. It is used below to rule out a low auxiliary multiplier, and must be verified independently of the equation $\alpha_{\mathrm{out}}=H$. If $H>L$ throughout the relevant range, it holds automatically. Section~\ref{subsec:roi-arithmetic-rewritten} verifies it for feedback constructions that can reach the endpoint.

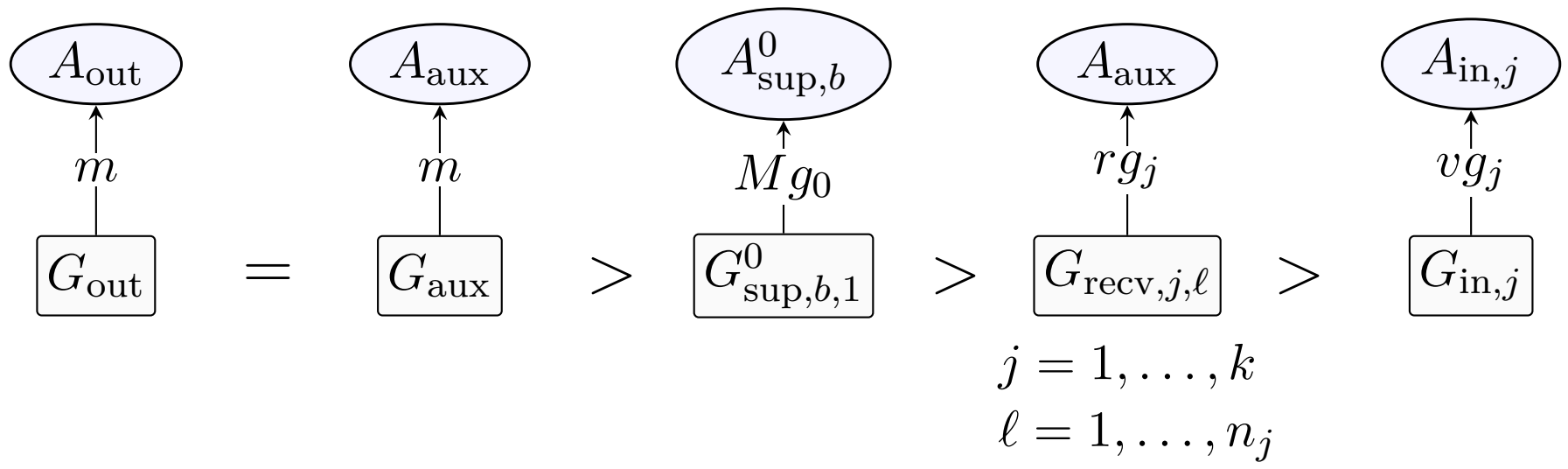
\begin{figure}[H]
\centering
\begin{tikzpicture}[
bidder/.style={ellipse,draw,line width=0.55pt,fill=blue!4,inner xsep=2.5pt,inner ysep=1.5pt},
good/.style={rectangle,draw,rounded corners=1pt,line width=0.4pt,fill=black!2,minimum height=6mm,inner sep=2pt},
edge/.style={->,>=stealth,line width=0.4pt},
val/.style={midway,fill=white,inner sep=1.2pt},
relation/.style={font=\Large,inner sep=0pt}
]
\node[good] (go) at (0,0) {$G_{\mathrm{out}}$};
\node[good] (ga) at (2.6,0) {$G_{\mathrm{aux}}$};
\node[good] (gf) at (5.2,0) {$G_{\mathrm{sup},b,1}^{0}$};
\node[good] (gr) at (7.8,0) {$G_{\mathrm{recv},j,\ell}$};
\node[good] (gi) at (10.4,0) {$G_{\mathrm{in},j}$};
\node[relation] at (1.3,0) {$=$};
\node[relation] at (3.9,0) {$>$};
\node[relation] at (6.5,0) {$>$};
\node[relation] at (9.1,0) {$>$};
\node[bidder] (ao) at (0,1.6) {$A_{\mathrm{out}}$};
\node[bidder] (aa) at (2.6,1.6) {$A_{\mathrm{aux}}$};
\node[bidder] (af) at (5.2,1.6) {$A_{\mathrm{sup},b}^{0}$};
\node[bidder] (ar) at (7.8,1.6) {$A_{\mathrm{aux}}$};
\node[bidder] (ai) at (10.4,1.6) {$A_{\mathrm{in},j}$};
\draw[edge] (go.north)--node[val] {$m$}(ao.south);
\draw[edge] (ga.north)--node[val] {$m$}(aa.south);
\draw[edge] (gf.north)--node[val] {$Mg_0$}(af.south);
\draw[edge] (gr.north)--node[val] {$rg_j$}(ar.south);
\draw[edge] (gi.north)--node[val] {$vg_j$}(ai.south);
\node[anchor=north,inner sep=0pt,yshift=-2mm] (jr) at (gr.south) {$j=1,\ldots,k$};
\node[anchor=north west,inner sep=0pt,yshift=-1.5mm] at (jr.south west) {$\ell=1,\ldots,n_j$};
\end{tikzpicture}
\caption{Affine gadget, showing selected goods in decreasing bid order. The superscript $0$ identifies the floor block; its $n$ top supplier goods have the same bid. The other supplier and compensation goods are omitted.}
\label{fig:roi-affine-rewritten}
\end{figure}

Figure~\ref{fig:roi-affine-rewritten} shows how the input charges reach $A_{\mathrm{aux}}$ through its receiving goods. We use a common main-good value $m>0$. An ordinary pricing block, called the floor block and marked by superscript $0$, supplies the lower price paid by a main good. Its scale is $g_0$. Input branch $j$ has scale $g_j$ and $n_j$ receiving goods. These scales and multiplicities are chosen below.

\paragraph{Choosing the valuations and private adjustments.} Give $G_{\mathrm{out}}$ and $G_{\mathrm{aux}}$ the same value $m$, so equality of their bids will imply $\alpha_{\mathrm{out}}=\alpha_{\mathrm{aux}}$. We construct the floor block to provide the lower main-good price and compensate for its variation.

Give $A_{\mathrm{aux}}$ an ordinary signal pair $G_{\mathrm{sig}}^{0},G_{\mathrm{comp}}^{0}$ at scale $g_0$, with values $vg_0,eg_0$. Add $M=96$ compensation copies $G_{\mathrm{copy},\ell}^{0}$, for $\ell=1,\ldots,M$, each of value $eg_0$ and also owned by $A_{\mathrm{aux}}$. This fixed number is independent of both the source circuit and the design cap. It counts only the extra compensation copies; the full inventory also includes main, receiving, and signal-pair goods. Set

\[f_0=\frac m2,\qquad g_0=\frac{f_0}{MS}.\]
Here $f_0$ is a fixed payment amount, while the actual floor price may depend on the suppliers. Write $\alpha_{\mathrm{sup}}^{0},\alpha_{\mathrm{cmp}}^{0}$ for the supplier multipliers in this block. The highest foreign bid after the main goods is $f=M\alpha_{\mathrm{sup}}^{0}g_0$. The supplier identity gives

\[f+Mc\alpha_{\mathrm{cmp}}^{0}g_0=MSg_0=f_0.\]

For $\alpha_{\mathrm{aux}}\in[L,U]$, each extra copy pays $c\alpha_{\mathrm{cmp}}^{0}g_0$, so the copies' total payment together with the floor price is the fixed amount $f_0$. Below $L$, each copy pays at most this price. The ordinary pair itself contributes zero ROI surplus on $[L,U]$ and nonnegative surplus below $L$. The independent supplier bounds place every supplier bid in this block at most $2Mg_0=m/S<m$, and its core bids below $Tg_0<Mg_0$. Hence both main goods precede the whole floor block. The displayed identity also gives $0<f<f_0$.

For input branch $j$, use an ordinary signal pair of $A_{\mathrm{in},j}$ at scale $g_j$, and give $A_{\mathrm{aux}}$ goods $G_{\mathrm{recv},j,\ell}$ of value $rg_j$, for $\ell=1,\ldots,n_j$. Choose

\[n_jvg_j=md_j.\]

The inequalities $r>Uv$, $\alpha_{\mathrm{aux}}\ge1$, and $\alpha_{\mathrm{in},j}\le U$ put every receiving good before $G_{\mathrm{in},j}$. With only other receiving goods of $A_{\mathrm{aux}}$ intervening, the no-self-pricing rule makes each pay the signal bid $v\alpha_{\mathrm{in},j}g_j$. Their combined payment is thus $md_j\alpha_{\mathrm{in},j}$. For an explicit choice, order two inputs so that $d_1\ge d_2$, put $g_{\mathrm{prev}}=g_0$ before the first input branch, use $\delta=\delta(2)$, and choose successively

\begin{equation}\label{eq:roi-explicit-receiving-count}n_j=\left\lfloor\frac{98md_j}{v\delta g_{\mathrm{prev}}}\right\rfloor+1,\qquad g_j=\frac{md_j}{vn_j},\qquad g_{\mathrm{prev}}\gets g_j.\end{equation}
These choices give $98g_j<\delta g_{\mathrm{prev}}$, including when the quantity inside the floor is an integer. Lemma~\ref{lem:roi-common-supplier-multipliers} uses precisely these gaps to prove block separation. The counts depend only on the fixed arithmetic construction and design cap. Appendix~\ref{subsec:roi-inventory-certificate} bounds the complete inventory for all fixed $A>1$.

Finally, choose the private adjustments to set the payment targets. The goods of $A_{\mathrm{out}}$ outside its signal pairs have value $m$. Those of $A_{\mathrm{aux}}$ have total value $m+\sum_j rn_jg_j+Meg_0$, comprising its main good, receiving goods, and extra compensation copies. Set

\[\begin{aligned}\Delta_{\mathrm{out}}&=(L-1)m,\\\Delta_{\mathrm{aux}}&=m(C-L)+f_0-\left(m+\sum_{j=1}^{k}rn_jg_j+Meg_0\right).\end{aligned}\]

After accounting for the ordinary signal pairs and the private base payments, the payment targets are $Lm$ for $A_{\mathrm{out}}$ and $m(C-L)+f_0$ for $A_{\mathrm{aux}}$. Any private payment above its base amount will be included explicitly in the derivation. The private-scale conditions allow either sign of the adjustment while keeping every good value positive.

The next lemma extends the wire lower bound to the affine auxiliary.

\begin{lemma}[Lower bound on the affine auxiliary multiplier]\label{lem:roi-affine-aux-lower}
In every equilibrium of the constructed auction, an affine gadget satisfying \eqref{eq:roi-affine-promise} has $\alpha_{\mathrm{aux}}\ge L$.
\end{lemma}

\begin{proof}
Lemma~\ref{lem:roi-wire-lower-bound} gives $\alpha_{\mathrm{out}}\ge L$, and Lemma~\ref{lem:roi-core-upper-bound} places all core multipliers below $U$, so their ROI constraints are equalities. Suppose $\alpha_{\mathrm{aux}}<L$. Then $G_{\mathrm{aux}}$ follows $G_{\mathrm{out}}$ and pays $f$. The extra copies owned by $A_{\mathrm{aux}}$ jointly pay at most $Mc\alpha_{\mathrm{cmp}}^{0}g_0$. The ordinary pair of $A_{\mathrm{aux}}$ has nonnegative surplus, and its private pair contributes at least $\Delta_{\mathrm{aux}}$. The ROI equality of $A_{\mathrm{aux}}$ therefore requires

\[m(C-L)+f_0\le m\sum_{j=1}^{k}d_j\alpha_{\mathrm{in},j}+f+Mc\alpha_{\mathrm{cmp}}^{0}g_0=m\sum_{j=1}^{k}d_j\alpha_{\mathrm{in},j}+f_0.\]
This gives $H\le L$, and the range condition forces $H=L$. The endpoint condition then gives $\alpha_{\mathrm{out}}=L$. But $G_{\mathrm{out}}$ pays $m\alpha_{\mathrm{aux}}<Lm$, contradicting the output payment target. Hence $\alpha_{\mathrm{aux}}\ge L$.
\end{proof}

\paragraph{Deriving the equilibrium relation.}

We next use this lower bound to establish equal main bids. Let $s_{\mathrm{out}},s_{\mathrm{aux}}\ge0$ denote the private payments above their respective base amounts. Their nonnegativity follows from Lemmas~\ref{lem:roi-signal-pair-payments}, \ref{lem:roi-wire-lower-bound}, and~\ref{lem:roi-affine-aux-lower}. After cancelling ordinary pairs and subtracting receiving and extra-copy payments, the expected main payments are $Lm-s_{\mathrm{out}}$ for $G_{\mathrm{out}}$ and $(H-L)m+f-s_{\mathrm{aux}}$ for $G_{\mathrm{aux}}$. 

\begin{lemma}[Equal main bids in the affine ROI-GSP gadget]\label{lem:roi-affine-equal-bids}
For an affine gadget satisfying \eqref{eq:roi-affine-promise}, every equilibrium of the constructed auction has equal bids on $G_{\mathrm{out}}$ and $G_{\mathrm{aux}}$. Thus, $\alpha_{\mathrm{out}}=\alpha_{\mathrm{aux}}$.
\end{lemma}

\begin{proof}
Both multipliers are at least $L$ by Lemmas~\ref{lem:roi-wire-lower-bound} and~\ref{lem:roi-affine-aux-lower}. If $G_{\mathrm{aux}}$ bids strictly more than $G_{\mathrm{out}}$, it pays $m\alpha_{\mathrm{out}}\ge Lm$. Its required payment is smaller than $Lm$, because $s_{\mathrm{aux}}\ge0$ and $H-L+f/m<U-L+1/2\le3/4<L$. Conversely, if $G_{\mathrm{out}}$ bids strictly more, its price $m\alpha_{\mathrm{aux}}=Lm-s_{\mathrm{out}}$ forces $\alpha_{\mathrm{aux}}=L$ and $s_{\mathrm{out}}=0$. Lemma~\ref{lem:roi-signal-pair-payments} then gives $s_{\mathrm{aux}}=0$. The payment $f$ of $G_{\mathrm{aux}}$ forces $H=L$, and the independently checked endpoint condition gives $\alpha_{\mathrm{out}}=L$, contradicting the strict order. Thus the main bids and multipliers are equal.
\end{proof}

In either main order, the combined payment is $m\alpha_{\mathrm{out}}+f$. Adding the required payments gives

\[m\alpha_{\mathrm{out}}=mH-s_{\mathrm{out}}-s_{\mathrm{aux}}<mF.\]
Both multipliers lie in $[L,F)$, where the private payments equal their base amounts. Hence $s_{\mathrm{out}}=s_{\mathrm{aux}}=0$, and

\[\alpha_{\mathrm{out}}=\alpha_{\mathrm{aux}}=H=C-\sum_{j=1}^{k}d_j\alpha_{\mathrm{in},j}.\]

\paragraph{Completing the gadget.} We next complete each permitted input configuration.

\begin{lemma}[Exact realization of the affine ROI-GSP gadget]\label{lem:roi-affine-realization}
Provided \eqref{eq:roi-affine-promise} is verified independently, every equilibrium satisfies
\[
\alpha_{\mathrm{out}}=\alpha_{\mathrm{aux}}
=H=C-\sum_{j=1}^{k}d_j\alpha_{\mathrm{in},j}\in[L,F).
\]
Conversely, every permitted input assignment admits a local completion realizing this output and satisfying both core ROI equalities, with suppliers completed at their actual lower bids. The input bidders' upstream ROI conditions remain unchanged.
\end{lemma}

\begin{proof}

Set $\alpha_{\mathrm{out}}=\alpha_{\mathrm{aux}}=H$. After the supplier blocks have been completed at their actual lower bids, let $f$ be the resulting floor price. Since $0<f<f_0<Lm\le mH$, choose the probability that $G_{\mathrm{out}}$ occurs before $G_{\mathrm{aux}}$ as

\[q_* =\frac{Lm-f}{mH-f}\in[0,1].\]
Use the reverse main order with probability $1-q_*$. The expected payment for $G_{\mathrm{out}}$ is $q_*mH+(1-q_*)f=Lm$, while the expected payment for $G_{\mathrm{aux}}$ is $(1-q_*)mH+q_*f=(H-L)m+f$.

The receiving payments and extra-copy payments restore the full target $m(C-L)+f_0$ of $A_{\mathrm{aux}}$. Ordinary signal pairs contribute zero surplus, and private signal pairs contribute their prescribed adjustments because $H\in[L,F)$. Thus both core ROI equalities hold. The supplier completion depends on the numerical bids and actual lower prices, so it remains valid when this main-order probability is chosen. At $H=L$, the formula gives $q_*=1$ and the same verification applies. 
\end{proof}

\subsubsection{Nonlinear gadget}
\label{subsubsec:roi-nonlinear-rewritten}
For the nonlinear gadget, set $k=1$. We use a middle good $G_{\mathrm{mid}}$ to make the lower main-good price depend on the output multiplier. Using the relative width $s$ defined in Section~\ref{subsec:roi-common-rewritten}, set

\[P(t)=\frac{3t}{2}-\frac{t^2}{L}.\]

Our target is $P(\alpha_{\mathrm{out}})=\kappa\alpha_{\mathrm{in},1}$. Its output must lie in $(L,F)$ so that the private signal pairs provide their fixed adjustments and the output remains a valid input to subsequent constructions. The parameter choices give

\[\kappa L-P(U)=\frac{Ls^2}{2}>0,\qquad P(L)-\kappa U=\frac{Ls^2(2+s)}2>0.\]

The inner endpoint also has a strict margin for every fixed design cap, including caps arbitrarily close to one:

\[\kappa L-P(F)=Ls^2\left(\frac14-s+\frac{s^2}{4}\right)\ge\frac{Ls^2}{20}>0.\]
Here $0<s\le1/5$ gives $1/4-s\ge1/20$. Thus the use of $F$ does not rely on checking only the design $A=2$.

Since $P'(t)=3/2-2t/L<0$ on $[L,U]$, every $\xi\in[L,U]$ has a unique solution $R(\xi)\in(L,F)$ of $P(R(\xi))=\kappa\xi$. Thus the relation we will implement can also be written as $\alpha_{\mathrm{out}}=R(\alpha_{\mathrm{in},1})$. 

\begin{figure}[H]
\centering
\begin{tikzpicture}[
bidder/.style={ellipse,draw,line width=0.55pt,fill=blue!4,inner xsep=2.5pt,inner ysep=1.5pt},
good/.style={rectangle,draw,rounded corners=1pt,line width=0.4pt,fill=black!2,minimum height=6mm,inner sep=2pt},
edge/.style={->,>=stealth,line width=0.4pt},
val/.style={midway,fill=white,inner sep=1.2pt},
relation/.style={font=\Large,inner sep=0pt}
]
\node[good] (go) at (0,0) {$G_{\mathrm{out}}$};
\node[good] (ga) at (2.7,0) {$G_{\mathrm{aux}}$};
\node[good] (gm) at (5.4,0) {$G_{\mathrm{mid}}$};
\node[good] (gi) at (8.1,0) {$G_{\mathrm{in},1}$};
\node[relation] at (1.35,0) {$=$};
\node[relation] at (4.05,0) {$>$};
\node[relation] at (6.75,0) {$>$};
\node[bidder] (ao) at (0,1.6) {$A_{\mathrm{out}}$};
\node[bidder] (aa) at (2.7,1.6) {$A_{\mathrm{aux}}$};
\node[bidder] (am) at (5.4,1.6) {$A_{\mathrm{aux}}$};
\node[bidder] (ai) at (8.1,1.6) {$A_{\mathrm{in},1}$};
\draw[edge] (go.north)--node[val] {$m$}(ao.south);
\draw[edge] (ga.north)--node[val] {$m$}(aa.south);
\draw[edge] (gm.north)--node[val] {$m/2$}(am.south);
\draw[edge] (gi.north)--node[val] {$\kappa m$}(ai.south);
\end{tikzpicture}
\caption{Nonlinear gadget, showing only the core goods. The signal compensation good and the supplier goods are omitted.}
\label{fig:roi-nonlinear-rewritten}
\end{figure}
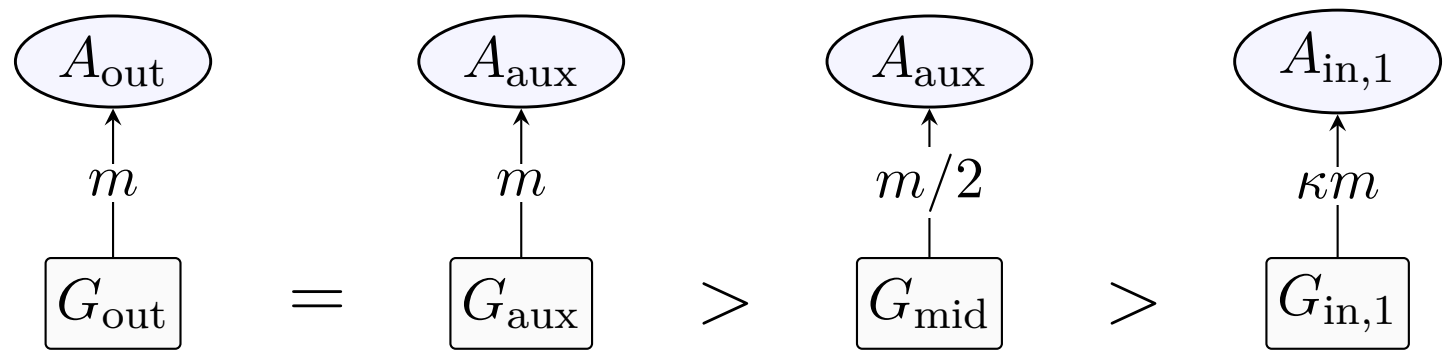

Figure~\ref{fig:roi-nonlinear-rewritten} shows the main goods, middle good, and input signal at common main scale $m>0$. Its compensation and supplier goods use the ordinary construction of Section~\ref{subsec:roi-suppliers-rewritten}.

\paragraph{Choosing the valuations and private adjustments.} Give $G_{\mathrm{out}}$ and $G_{\mathrm{aux}}$ value $m$, and give $A_{\mathrm{aux}}$ the middle good $G_{\mathrm{mid}}$ of value $m/2$, hence bid $m\alpha_{\mathrm{aux}}/2$. At a main tie this sets the lower output price. When $G_{\mathrm{out}}$ comes first, the no-self-pricing rule skips $G_{\mathrm{mid}}$ for $G_{\mathrm{aux}}$, allowing both auxiliary goods to pay the input bid.

Use an ordinary signal pair $G_{\mathrm{in},1},G_{\mathrm{comp},1}$ of $A_{\mathrm{in},1}$ at scale

\[g=\frac{\kappa m}{v}.\]
Then $G_{\mathrm{in},1}$ has value $vg=\kappa m$ and bid $\kappa m\alpha_{\mathrm{in},1}$. Place $G_{\mathrm{out}}$, $G_{\mathrm{aux}}$, and $G_{\mathrm{mid}}$ in the same ordinary block. Their normalized values are at most $v/\kappa<8/3=Q$, using $v<S<1011/1000$ and $\kappa\ge19/50$. The common lower-bound lemma, together with the core upper bound, puts all three goods before the signal, so the ordinary pair preserves the input bidder's ROI condition.

Choose the private adjustments so that the remaining payment target is $Lm$ for both $A_{\mathrm{out}}$ and $A_{\mathrm{aux}}$. The main good of $A_{\mathrm{out}}$ has value $m$, while the main and middle goods of $A_{\mathrm{aux}}$ have total value $3m/2$. Set

\[\Delta_{\mathrm{out}}=(L-1)m,\qquad \Delta_{\mathrm{aux}}=(L-3/2)m.\]

The private adjustment of $A_{\mathrm{aux}}$ removes the extra value contributed by its middle good from the required payment target. Although this adjustment is negative, the private-scale conditions keep its compensation value positive. Until the multipliers are proved to lie below $F$, any private payment above its base amount reduces the remaining target and must be kept in the calculation.

\paragraph{Deriving the equilibrium relation.} We first prove the order of the middle good and input signal using Lemmas~\ref{lem:roi-core-upper-bound} and~\ref{lem:roi-wire-lower-bound}, proved independently of the gate equations. These give $L\le\alpha_{\mathrm{out}},\alpha_{\mathrm{aux}}<U$ and $\alpha_{\mathrm{in},1}\in[L,U]$. After dividing bids by $m$, the choice of $\kappa$ gives

\[\kappa\alpha_{\mathrm{in},1}\le\kappa U<L/2\le\alpha_{\mathrm{aux}}/2<U/2<L\le\alpha_{\mathrm{out}}.\]
Hence $G_{\mathrm{mid}}$ follows both main goods and precedes $G_{\mathrm{in},1}$. Since the core multipliers lie below the actual cap, their ROI constraints are equalities. Since they are at least $L$, the private signal payments are at least their base amounts, so the remaining payment target of each of $A_{\mathrm{out}}$ and $A_{\mathrm{aux}}$ is at most $Lm$.
The next lemma uses these order and payment bounds to prove that the two main bids are equal.

\begin{lemma}[Equal main bids in the nonlinear ROI-GSP gadget]\label{lem:roi-nonlinear-equal-bids}
In every equilibrium of the constructed auction, the nonlinear gadget has equal bids on $G_{\mathrm{out}}$ and $G_{\mathrm{aux}}$. Thus, $\alpha_{\mathrm{out}}=\alpha_{\mathrm{aux}}$.
\end{lemma}

\begin{proof}
If $G_{\mathrm{out}}$ bids strictly more than $G_{\mathrm{aux}}$, its price $m\alpha_{\mathrm{aux}}\ge Lm$ and its remaining target at most $Lm$ force $\alpha_{\mathrm{aux}}=L$ and the private payment of $A_{\mathrm{out}}$ to equal its base amount. At multiplier $L$, Lemma~\ref{lem:roi-signal-pair-payments} also makes the private payment of $A_{\mathrm{aux}}$ equal its base amount. Its main and middle goods must therefore jointly pay $Lm$, but both pay the input bid, giving only $2\kappa m\alpha_{\mathrm{in},1}\le2\kappa mU<Lm$. Conversely, if $G_{\mathrm{aux}}$ bids strictly more, its two goods jointly pay $m\alpha_{\mathrm{out}}+\kappa m\alpha_{\mathrm{in},1}>Lm$, exceeding its remaining target. Both strict orders are impossible, proving equal main bids and multipliers.
\end{proof}

By Lemma~\ref{lem:roi-nonlinear-equal-bids}, the main goods have common bid $m\alpha_{\mathrm{out}}$.

We prove that the common multiplier is below $F$ before cancelling private excess payments. As in the affine gadget, let $s_{\mathrm{out}},s_{\mathrm{aux}}\ge0$ denote the private payments above their respective base amounts. To express the remaining targets after division by $m$, put $E_{\mathrm{out}}=L-s_{\mathrm{out}}/m$ and $E_{\mathrm{aux}}=L-s_{\mathrm{aux}}/m$. Let $q_*$ be the probability that $G_{\mathrm{out}}$ precedes $G_{\mathrm{aux}}$. The two normalized ROI equations are

\begin{equation}\label{eq:roi-nonlinear-payments}
\left\{\begin{aligned}&E_{\mathrm{out}}=\frac{\alpha_{\mathrm{out}}}{2}(1+q_*),\\&E_{\mathrm{aux}}=(1-q_*)\alpha_{\mathrm{out}}+(1+q_*)\kappa\alpha_{\mathrm{in},1}.\end{aligned}\right.
\end{equation}
The same probability must satisfy both equations. The first gives $E_{\mathrm{out}}>0$, and both targets are at most $L$. Eliminating $q_*$ therefore gives

\[\begin{aligned}\kappa\alpha_{\mathrm{in},1}&=\alpha_{\mathrm{out}}-\frac{\alpha_{\mathrm{out}}^2-E_{\mathrm{aux}}\alpha_{\mathrm{out}}/2}{E_{\mathrm{out}}}\\&\le\alpha_{\mathrm{out}}-\frac{\alpha_{\mathrm{out}}^2-L\alpha_{\mathrm{out}}/2}{L}=P(\alpha_{\mathrm{out}}).\end{aligned}\]
The inequality uses $0<E_{\mathrm{out}}\le L$, $E_{\mathrm{aux}}\le L$, and $\alpha_{\mathrm{out}}\ge L$, which make the numerator positive. If $\alpha_{\mathrm{out}}\ge F$, monotonicity of $P$ would give

\[\kappa\alpha_{\mathrm{in},1}\le P(\alpha_{\mathrm{out}})\le P(F)<\kappa L\le\kappa\alpha_{\mathrm{in},1},\]
a contradiction. Thus both main multipliers lie in $[L,F)$, where the private payments equal their base amounts. Hence $s_{\mathrm{out}}=s_{\mathrm{aux}}=0$ and $E_{\mathrm{out}}=E_{\mathrm{aux}}=L$. Eliminating $q_*$ from these exact targets now gives the desired relation

\begin{equation}\label{eq:roi-nonlinear-rewritten}\begin{aligned}P(\alpha_{\mathrm{out}})&=\frac{3\alpha_{\mathrm{out}}}{2}-\frac{\alpha_{\mathrm{out}}^2}{L}=\kappa\alpha_{\mathrm{in},1},\\\alpha_{\mathrm{aux}}&=\alpha_{\mathrm{out}}=R(\alpha_{\mathrm{in},1})\in(L,F).\end{aligned}\end{equation}
The second line follows from the uniqueness and range of $R$ proved above. 

\paragraph{Completing the gadget.} We next show that every allowed input has a local completion.

\begin{lemma}[Exact realization of the nonlinear ROI-GSP gadget]\label{lem:roi-nonlinear-realization}
Every equilibrium satisfies \eqref{eq:roi-nonlinear-rewritten}, with unique output $\alpha_{\mathrm{out}}=\alpha_{\mathrm{aux}}=R(\alpha_{\mathrm{in},1})\in(L,F)$ for each input in $[L,U]$. Conversely, every such input admits a local completion realizing this output and satisfying both core ROI equalities, with suppliers completed at their actual lower bids. The input bidder's upstream ROI condition remains unchanged.
\end{lemma}

\begin{proof}

Given $\alpha_{\mathrm{in},1}\in[L,U]$, set $\alpha_{\mathrm{out}}=\alpha_{\mathrm{aux}}=R(\alpha_{\mathrm{in},1})$ and choose the probability that $G_{\mathrm{out}}$ precedes $G_{\mathrm{aux}}$ as

\[q_*=\frac{2L}{\alpha_{\mathrm{out}}}-1.\]
Since $L<\alpha_{\mathrm{out}}<F<U<2L$, we have $0<q_*<1$. Equation~\eqref{eq:roi-nonlinear-rewritten} verifies~\eqref{eq:roi-nonlinear-payments} with $E_{\mathrm{out}}=E_{\mathrm{aux}}=L$, so the two remaining payment targets are met.
The ordinary input pair contributes zero ROI surplus, and both private signal pairs contribute their prescribed adjustments because the main multipliers lie in $(L,F)$. The two core ROI equalities therefore hold. Proposition~\ref{prop:roi-block-interface} completes the suppliers at the actual lower bid, independently of this main-order probability. 
\end{proof}

\subsection{Encoding and Exact Gate Implementation}
\label{subsec:roi-arithmetic-rewritten}
We implement the source gates using the inverse-feedback, square, and polarization methods of Theorem~\ref{thm:criterion}(i). Every affine call must satisfy \eqref{eq:roi-affine-promise}, which we check using subcircuit equations already valid for all inputs, independently of the new equation.

Encode a source value $x\in[0,1]$ by a wire multiplier $\alpha_X$ with

\[E=[\ell_0,\ell_1],\qquad \ell_0=L+h,\qquad \ell_1=L+2h,\qquad \alpha_X=\ell_0+hx.\]
Then $E\subset(L,F)$ and decoding is $x=(\alpha_X-\ell_0)/h$. Lemmas~\ref{lem:roi-core-upper-bound} and~\ref{lem:roi-wire-lower-bound} initially give only $\alpha_X\in[L,U)$, so we need a normalization map $\mathcal N:[L,U]\to E$ that is the identity on $E$. Internal wires may lie outside $E$. We use the closed interval $[L,U]$ for the checks below, which also covers all equilibrium inputs.

\paragraph{Restricted affine compilation.} The next lemma converts signed affine expressions into decreasing affine calls, subject to explicit range and endpoint checks.

\begin{lemma}[Restricted affine construction]\label{lem:roi-restricted-affine}
Let $H$ be a fixed rational affine expression in at most three distinct wire multipliers in $[L,U]$, possibly including its output $\alpha_Y$. If established subcircuit equations imply $L\le H<F$ and $H=L\Rightarrow\alpha_Y=L$ independently of $\alpha_Y=H$, fresh copies and at most three final binary gadgets from Lemma~\ref{lem:roi-affine-realization} enforce $\alpha_Y=H$. Every assignment satisfying those equations and $\alpha_Y=H$ extends to the construction. For fixed coefficients and design cap, its size is constant and input bidders' upstream ROI conditions are preserved.
\end{lemma}

\begin{proof}
Fix $\beta=7/8$, $\mu=(L+F)/2$, and $b_\nu=\mu+\beta\ell_1$. The decreasing affine copy

\[\nu(\alpha_X)=b_\nu-\beta\alpha_X,\qquad
\nu([L,U])=[L+2h(q-\beta),L+2h(q+\beta)]\subset(L,F)
\]
has strict range because $F-L=4hq$ and $q\ge9/10>\beta$. Lemma~\ref{lem:roi-affine-realization} supplies this unary call and its completion. A second copy also has strict range, since the first lies in $[L,U]$. The identities are

\[\alpha_X=\frac{b_\nu-\nu(\alpha_X)}\beta,\qquad
\alpha_X=\frac{\nu(\nu(\alpha_X))-b_\nu(1-\beta)}{\beta^2}.\]

Combine repeated variables and omit zero coefficients. Replace a positive term $b\alpha_X$ by $bb_\nu/\beta-(b/\beta)\nu(\alpha_X)$, using a fresh copy. Replace a negative self-input $-d\alpha_Y$ by $db_\nu(1-\beta)/\beta^2-(d/\beta^2)\nu(\nu(\alpha_Y))$. The resulting expression is decreasing affine, has no self-input, and equals $H$ using only the already established copy identities. Thus its range and endpoint checks hold before its output equation is imposed.

If there are at most two inputs, apply Lemma~\ref{lem:roi-affine-realization} directly. If there are three, write the expression as $C-d_1\alpha_1-d_2\alpha_2-d_3\alpha_3$ and put $D_*=d_1+d_2$. Introduce fresh wire multipliers

\[t=b_\nu-\frac{\beta}{D_*}(d_1\alpha_1+d_2\alpha_2),\qquad w=b_\nu-\beta t.\]
Both targets lie strictly in $(L,F)$: the weighted input mean belongs to $[L,U]$, so the first target is a $\nu$ image, and the second is another. Their identities turn the final target into

\[H=C+\frac{D_*b_\nu(1-\beta)}{\beta^2}-\frac{D_*}{\beta^2}w-d_3\alpha_3.\]

For the fixed arithmetic constructions below, choose the merged weights as in Table~\ref{tab:roi-binary-weights}, then order the two receiving branches by decreasing weight. The final expression is a two-input decreasing affine call with the original range and endpoint conditions. It therefore enforces $\alpha_Y=H$. 

Conversely, fix an assignment satisfying the previous relations and $\alpha_Y=H$. Assign all fresh copies their prescribed values and, when needed, assign $t,w$ by the displayed formulas. Complete each unary or binary gadget by Lemma~\ref{lem:roi-affine-realization}. Input preservation makes these choices compatible, including copies depending on the fixed candidate $\alpha_Y$. There are only constantly many new gadgets for fixed coefficients and design cap.
\end{proof}

\paragraph{Inverse feedback and squares.} We first show that the inverse feedback uses only values in the response range of Lemma~\ref{lem:roi-nonlinear-realization}. Recall that $P(R(\xi))=\kappa\xi$ and $-2\kappa<R'(\xi)<0$. Direct substitution gives

\[\begin{aligned}P(\ell_1+h)-\kappa L&=\frac{Ls(2-s)}{16}>0,\\P\bigl(\ell_1-(U-F)\bigr)-\kappa U&=\frac{Ls(-1+4s+4s^2-s^3)}4<0.\end{aligned}\]
The last sign follows from $0<s\le1/5$ and $4s+4s^2\le24/25<1$. Since $P$ is decreasing on $[L,U]$, these inequalities give

\[R(U)<\ell_1-(U-F)<\ell_1<\ell_1+h<R(L)<F.\]

For input $\alpha_X\in[L,U]$, use Lemma~\ref{lem:roi-restricted-affine} to create $\alpha_W=\ell_1+(\alpha_X-L)/4$, whose range is $[\ell_1,\ell_1+h]\subset(L,F)$. Create $\alpha_Y=R(\alpha_Z)$ with a nonlinear gadget, where $\alpha_Z$ is a fresh wire. This relation holds for every $\alpha_Z\in[L,U]$, before imposing the feedback equation. On these established relations the proposed right-hand side $H=\alpha_Z+\alpha_Y-\alpha_W$ is strictly increasing in $\alpha_Z$, since $1+R'>1-2\kappa>0$, and satisfies

\[H(L)\ge L+R(L)-(\ell_1+h)>L,\qquad H(U)\le U+R(U)-\ell_1<F.\]

The restricted affine construction can therefore impose $\alpha_Z=\alpha_Z+\alpha_Y-\alpha_W$. This gives $\alpha_Y=\alpha_W$ and $\kappa\alpha_Z=P(\alpha_W)$. Conversely, $\alpha_W\in(R(U),R(L))$ gives a unique $\alpha_Z\in(L,U)$ with $R(\alpha_Z)=\alpha_W$. At this value $\alpha_Z=H$, and the checked range of $H$ further gives $\alpha_Z\in(L,F)$. Thus every input has a completion of the feedback, including all its copies and the nonlinear auxiliary.

Next create a fresh output for the affine expression

\[\ell_0+\frac Lh\bigl(P(\ell_1)-\kappa\alpha_Z\bigr)+\frac{LP'(\ell_1)}{4h}(\alpha_X-L).\]
The established identity $\kappa\alpha_Z=P(\alpha_W)$, with $\alpha_W=\ell_1+(\alpha_X-L)/4$, makes this expression equal to

\[\mathcal Q(\alpha_X)=\ell_0+\frac{(\alpha_X-L)^2}{16h}\in E.\]
This strict range justifies Lemma~\ref{lem:roi-restricted-affine} before imposing the output equation. The resulting $\mathcal Q$ is forced on $[L,U]$, with a local completion for every input.

\paragraph{Square roots and normalization.} We next obtain a square root by feedback, making the only affine call in this construction that can reach $L$. For $0\le z\le9h^2$, encode $z$ by $\alpha_{Z_*}=\ell_0+z/(16h)$ and first create $\mathcal Q(\alpha_Y)$ using the preceding construction. Its identity holds for every candidate $\alpha_Y\in[L,U]$. The proposed affine right-hand side is

\[H=\alpha_Y+\frac83\bigl(\alpha_{Z_*}-\mathcal Q(\alpha_Y)\bigr)=\alpha_Y+\frac{z-(\alpha_Y-L)^2}{6h}.\]

Writing $\zeta=\alpha_Y-L\in[0,4h]$ gives

\[H-L=\zeta-\frac{\zeta^2}{6h}+\frac{z}{6h},\qquad \frac{\zeta}{3}+\frac{z}{6h}\le H-L\le3h<F-L.\]
Here the lower bound uses $\zeta\le4h$, and the upper bound follows by maximizing the concave quadratic at $\zeta=3h$ and using $z\le9h^2$. In particular, $H=L$ implies $\zeta=z=0$ and hence $\alpha_Y=L$, independently of the equation to be imposed. Both conditions in \eqref{eq:roi-affine-promise} now hold, so the restricted affine construction can impose $\alpha_Y=H$. This forces $(\alpha_Y-L)^2=z$ and, since $\alpha_Y\ge L$, uniquely gives $\alpha_Y=L+\sqrt z$. Conversely, this candidate lies in $[L,L+3h]\subset[L,F)$, satisfies the feedback, and extends through the square subcircuit and affine construction. This also covers $z=0$ and the endpoint output $L$.

For $j\in\{0,1\}$, first create $\mathcal Q(\alpha_X)$ and then the affine output

\[\alpha_{Z_j}=\mathcal Q(\alpha_X)+\frac{2(L-\ell_j)\alpha_X+\ell_j^2-L^2}{16h}=\ell_0+\frac{(\alpha_X-\ell_j)^2}{16h}.\]
The equality on the right uses the square identity already forced by the preceding subcircuit. Since $|\alpha_X-\ell_j|\le3h$, the affine target lies in $[\ell_0,\ell_0+9h/16]\subset E$, before its equation is imposed. Apply the square-root subcircuit to the encoded value $z=(\alpha_X-\ell_j)^2$, giving $\alpha_{Y_j}=L+|\alpha_X-\ell_j|$. A final constructed affine output is

\[\mathcal N(\alpha_X)=\frac{\ell_0+\ell_1+\alpha_{Y_0}-\alpha_{Y_1}}2.\]
The established absolute-value identities show that this target is $\ell_0$ below $\ell_0$, $\alpha_X$ on $E$, and $\ell_1$ above $\ell_1$. Its range is therefore $E\subset(L,F)$, as required before the final call. Every input in $[L,U]$ extends through both square-root subcircuits and this affine output. Thus $\mathcal N$ is implemented on the whole working interval and preserves every source encoding.

\paragraph{Wiring normalized inputs.} We now specify how the source arithmetic is connected. The circuit $\mathcal C'$ from Section~\ref{subsec:roi-suppliers-rewritten} contains the original source wires and the wires introduced by copying. For every input occurrence of every gate of $\mathcal C'$, insert a fresh implementation of $\mathcal N$ between that circuit wire and the gate's arithmetic implementation. Its output is the input $A_B$, or $A_C$ for a second input, used below. This also applies to every complement gate introduced by copying. For example, a complement equation with input wire $A_Y$ and output wire $A_X$ is implemented by

\begin{equation}\label{eq:roi-normalized-complement}\alpha_B=\mathcal N(\alpha_Y),\qquad \alpha_X=\ell_0+\ell_1-\alpha_B.\end{equation}

Identify each gate implementation's final output bidder with its output wire in $\mathcal C'$. In \eqref{eq:roi-normalized-complement}, $A_X$ is that output and $A_B$ is fresh to the input occurrence. Decode only original source wires. Insert normalization only at gate inputs of $\mathcal C'$, leaving internal feedback and $\nu$-copy equations unchanged.

\paragraph{The source gates.} The fresh input normalizers force $\alpha_B,\alpha_C\in E$ from raw circuit inputs anywhere in $[L,U]$. Write $\alpha_B=\ell_0+hx$ and $\alpha_C=\ell_0+hy$, where $x,y\in[0,1]$. This range follows from the full-interval implementation of $\mathcal N$, without assuming that the raw circuit wires are encoded. Complement is the constructed affine output $\ell_0+\ell_1-\alpha_B=\ell_0+h(1-x)$, whose target lies in $E$. For doubling, first construct $2\alpha_B-\ell_0\in[\ell_0,L+3h]\subset(L,F)$, then apply $\mathcal N$ with its output bidder identified with the circuit output. The result is $\ell_0+h\min\{2x,1\}$.

For the shifted product, create the affine mean $\alpha_D=(\alpha_B+\alpha_C)/2\in E$, followed by the three square outputs $\mathcal Q(\alpha_D)$, $\mathcal Q(\alpha_B)$, and $\mathcal Q(\alpha_C)$. Their established identities give the explicit final affine target

\[\begin{aligned}8\mathcal Q(\alpha_D)-2\mathcal Q(\alpha_B)-2\mathcal Q(\alpha_C)-3\ell_0&=\ell_0+\frac{(\alpha_B-L)(\alpha_C-L)}{4h}\\&=\ell_0+\frac h4(1+x)(1+y)\in[\ell_0+h/4,\ell_1].\end{aligned}\]
This range check precedes the final affine call; the restricted affine construction replaces its positive coefficient by fresh negative copies. We have therefore implemented $C(x)=1-x$, $D(x)=\min\{2x,1\}$, and $\phi(x,y)=(1+x)(1+y)/4$ with fixed-size rational constructions.

\paragraph{Closing the source encoding.} The input normalizers close the gap between the working interval and the source-encoding interval. The following lemma makes this simultaneous argument explicit, including the complement gates used for copying.

\begin{lemma}[Closure of the source encoding]\label{lem:roi-encoding-closure}
Under the input-normalization wiring above, every equilibrium places all circuit-wire multipliers of $\mathcal C'$ in $E$, where decoding gives a solution of $\mathcal C'$ and hence of the original source circuit. Conversely, every source solution extends through the copies and internal arithmetic within their permitted ranges, with the established local completions. Each source variable is decoded from one designated original-source multiplier.
\end{lemma}

\begin{proof}
The independently proved working bounds put every circuit and internal wire in $[L,U)$. The full-interval normalization forces each fresh gate input into $E$. The source-gate constructions above then put each circuit output in $E$, regardless of its raw inputs. Every wire of $\mathcal C'$ is such an output, so all circuit wires belong to $E$ simultaneously. Normalization is consequently the identity on them, and their equations are exactly the encoded equations of $\mathcal C'$. Lemma~\ref{lem:bounded-use} gives the original source solution and the stated decoder.

Conversely, extend a source solution through the complement copies and encode all circuit wires in $E$. Their normalizers act as the identity. For these fixed inputs and outputs, the preceding arithmetic constructions provide the internal values and local completions, including the inverse and square-root feedback variables. Fresh internal bidders make these choices compatible, and outgoing signal pairs preserve their senders' surplus by Lemma~\ref{lem:roi-wire-lower-bound}. Section~\ref{subsec:roi-completing-rewritten} completes the suppliers and global order distribution.
\end{proof}

\subsection{Completing the Proof}
\label{subsec:roi-completing-rewritten}
We prove global correspondence and check the size and cap restrictions.

\paragraph{Equilibrium correspondence and completion.} By Lemma~\ref{lem:roi-encoding-closure}, every equilibrium decodes to a source solution via $x=(\alpha_X-\ell_0)/h$ on the designated original-source coordinates. Conversely, fix the core multipliers extending a source solution. Apply Proposition~\ref{prop:roi-block-interface} in increasing price order, using each block's actual highest lower bid $p$. Fresh supplier identities make $p$ foreign to the bottom goods. After fixing supplier multipliers and affine floor prices, choose the main-order probabilities from Lemmas~\ref{lem:roi-affine-realization} and~\ref{lem:roi-nonlinear-realization} and uniform cyclic supplier-identity shifts. Each supplier then pays its value. For each core bidder, ordinary pairs have zero surplus, its private pair contributes $\Delta_i$, and its other goods have payment minus value $\Delta_i$, giving total ROI equality.

Block separation makes every combination of local tied orders valid. The lower price $p$ depends only on numerical bids in completed lower blocks. Supplier payments depend on local identity shifts, main payments on local main-order choices, and receiving payments and pair prices are fixed. Hence payments are sums of functions of individual local choices. With $b$ supplier blocks and $t$ main ties, Lemma~\ref{lem:local-coupling} preserves their marginals using at most $1+b(n-1)+t$ global orders, completing the equilibrium with polynomial support.

\paragraph{Size, caps, and complexity.} Each gate of $\mathcal C'$ has outdegree at most two and indegree at most two. We therefore add at most two input-normalization subcircuits per circuit gate. Each subcircuit has fixed size for fixed $A$ and reads its input only a fixed number of times; the number of ordinary signal uses per circuit wire remains bounded independently of source size. Internal wires also belong to fixed-size subcircuits. All added uses are ordinary signal pairs, whose conservation follows from Lemma~\ref{lem:roi-wire-lower-bound}, so they add no unaccounted sender surplus. Ownership, including receiving and compensation multiplicities, is thus bounded only by $A$. Each $A_{\mathrm{sup},b}$ owns three goods and each $A_{\mathrm{cmp},b}$ six. For fixed $A$, the complete bidder and good counts remain linear in source size. With the binary construction and \eqref{eq:roi-explicit-receiving-count}, Appendix~\ref{subsec:roi-inventory-certificate} gives fewer than $2^{43}\max\{1,(A-1)^{-5}\}$ goods per bidder, including full arithmetic expansion, input normalization, and private pairs. For design $A=2$ this is $2^{43}$. 

We check numerical encoding length separately from inventory. The geometric separating factor may be chosen as a fixed integer $\lambda>1$. With $N$ core bidders and constantly many blocks from the fixed constructions per bidder, ordinary scales are bounded by a fixed multiple of $\lambda^{O(N)}$ and have polynomial bit length. The receiving multiplicities are constants for fixed $A$ and do not depend on the circuit size. For the private scales, choose the fixed separating factor $D_0$ as an integer and put $X=\max_i(g_{\mathrm{base}}+1+W_i+|\Delta_i|)$. The recurrence in Section~\ref{subsec:roi-suppliers-rewritten} gives

\[\sum_{i=1}^{N}g_i\le X\bigl((1+D_0)^N-1\bigr).\]
All $W_i$ and $\Delta_i$ are rational combinations of ordinary values and fixed construction coefficients, so $X$ and a common denominator for them have polynomial bit length. Integer $D_0$ introduces no new denominators in the recurrence. The displayed bound gives polynomial numerator lengths as well. Adjusted private values remain positive by \eqref{eq:roi-private-scale-choice}. Thus the auction has polynomial description length for fixed design cap $A$. This hardness construction need not be polynomial in the binary length of a varying $A$; Appendix~\ref{sec:prescribed-membership} gives uniform membership for input caps.

All supplier and core multipliers are strictly below $U$, independently of the actual cap. The full equilibrium set is therefore unchanged for every $\widehat A\ge U$. For a prescribed cap, use design $A=\widehat A>U$. For the canonical cap, fix design $A=2$, giving $U=3/2$. Writing $v_G>0$ for the actual value of good $G$, the canonical cap satisfies

\[\widehat A_0=1+\frac{\sum_G v_G}{\min_G v_G}\ge2>U.\]

We now combine the established correspondence and instance bounds to prove Theorem~\ref{thm:roi-gsp-rewritten}.

\begin{proof}[Proof of Theorem~\ref{thm:roi-gsp-rewritten}]
Membership for prescribed caps follows from Lemma~\ref{lem:gsp-prescribed-cap-membership}; canonical-cap membership follows from \cite{HLN26}. Section~\ref{sec:complexity} gives the distance upper bound.
The equilibrium correspondence and polynomial-support completion were established above, using Lemma~\ref{lem:roi-encoding-closure}, Proposition~\ref{prop:roi-block-interface}, and Lemma~\ref{lem:local-coupling}.

Lemma~\ref{lem:roi-explicit-parameters} fixes the supplier parameters using only the design cap. Together with the counting and scale analysis above, this gives constant ownership for each fixed design and polynomial encoding length. The cap argument above applies to every prescribed cap and to the canonical cap. In the latter case, fixed design $A=2$ gives a universal ownership bound. Each source variable is decoded from one designated multiplier by $x=(\alpha_X-\ell_0)/h$, so the correspondence gives an SL reduction from Lemma~\ref{lem:source}. This proves $\FIXP$-hardness and, with membership, $\FIXP$-completeness.

A rational multiplier vector within $h\varepsilon$ in $\ell_\infty$ of an exact equilibrium decodes to a vector within $\varepsilon$ of a source solution. Since $h$ is a fixed positive rational for each design, the tolerance conversion preserves polynomial encoding length and proves distance hardness. The matching upper bound gives $\FIXPa$-completeness for multiplier-vector approximation in distance.
\end{proof}

\section{Second-Price Throttling}
\label{sec:throttle}

We now consider second-price throttling: separate auctions with fixed bids, independent participation, and one common tie priority. A bidder's conditional payment depends on others' participation probabilities. Its budget response turns this dependence into reciprocal arithmetic.

\begin{theorem}[Sparse Throttling Equilibria]
\label{thm:throttle}
Exact second-price throttling equilibrium computation is $\FIXP$-complete under separable linear reductions. Approximating the participation vector in distance to an exact equilibrium is $\FIXPa$-complete under the same restrictions. Hardness holds with at most four positive bids per bidder, three per good, a common tie priority across auctions, and positive bids and budgets in fixed finite rational sets.
\end{theorem}

Let $B_i>0$ be the budget of $A_i$, $\theta_i$ its participation probability, and $P_i$ its total expected payment conditional on participating. Its unconditional payment is $\theta_iP_i$, so the equilibrium budget response is

\begin{equation}\label{eq:throttle-budget-response-rewritten} \theta_i=\frac{B_i}{\max\{B_i,P_i\}}. \end{equation}

We implement Lemma~\ref{lem:source} through affine conditional payments. When $P_i>B_i$, the response gives $\theta_i=B_i/P_i$, supplying nonlinearity without a separate nonlinear auction gadget. We then add affine decoding and check that outgoing uses preserve the prescribed payment equations.

\subsection{A Common Auction Structure}
\label{subsec:throttle-common-rewritten}

We use the previous bidder roles and add dummy bidders. Lemma~\ref{lem:bounded-use} removes self-inputs, so each signal connects distinct bidders.

The bidder roles are:
\begin{itemize}
\item $A_{\mathrm{in},j}$: input bidder with participation probability $\theta_{\mathrm{in},j}$, for each input position $j$.
\item $A_{\mathrm{out}}$: output bidder with participation probability $\theta_{\mathrm{out}}$ and budget $B_{\mathrm{out}}=1$.
\item $A_{\mathrm{aux},\ell}$: auxiliary bidder with participation probability $\theta_{\mathrm{aux},\ell}$ and a budget specified in its gadget.
\item $A_{\mathrm{dum},\ell}$: dummy bidder with budget one, introduced for each absence-signal or constant-payment good.
\end{itemize}

Wire roles follow Section~\ref{sec:overview}. Auxiliary and dummy bidders are fresh and do not represent original source variables. A signal here is a dependence of one bidder's conditional payment on another's participation. Conditional on $A_i$ participating, a presence signal charges $a_{ij}>0$ when $A_j$ participates, contributing $a_{ij}\theta_j$. An absence signal charges $d_{ij}>0$ when $A_j$ is absent, contributing $d_{ij}(1-\theta_j)$. The latter also charges $A_j$ the constant $d_{ij}$ whenever it participates.

The goods are:
\begin{itemize}
\item $G_{\mathrm{pres},ij}$: presence-signal good for the contribution $a_{ij}\theta_j$ to $P_i$.
\item $G_{\mathrm{abs},ij}$: absence-signal good for the contribution $d_{ij}(1-\theta_j)$ to $P_i$.
\item $G_{\mathrm{const},i}$: constant-payment good providing a remaining positive constant contribution to $P_i$.
\end{itemize}

\begin{figure}[H]
\centering
\begin{tikzpicture}[
font=\normalsize,
bidder/.style={ellipse,draw,line width=0.55pt,fill=blue!4,inner xsep=3pt,inner ysep=2pt},
good/.style={rectangle,draw,rounded corners=1pt,line width=0.4pt,fill=black!2,minimum height=6mm,inner xsep=4pt,inner ysep=2pt},
edge/.style={->,>=stealth,line width=0.4pt},
bid/.style={midway,fill=white,inner sep=1.2pt},
panel/.style={anchor=north,inner sep=0pt}
]

\begin{scope}
\node[good] (gp) at (0,0) {$G_{\mathrm{pres},ij}$};
\node[bidder] (pi) at (-1.25,2) {$A_i$};
\node[bidder] (pj) at (1.25,2) {$A_j$};

\draw[edge] (gp.north)--node[bid] {$2a_{ij}$}(pi.south);
\draw[edge] (gp.north)--node[bid] {$a_{ij}$}(pj.south);

\node[panel] at (0,-0.65) {(a) Presence signal};
\end{scope}

\begin{scope}[xshift=5.2cm]
\node[good] (ga) at (0,0) {$G_{\mathrm{abs},ij}$};
\node[bidder] (ai) at (-1.7,2) {$A_i$};
\node[bidder] (aj) at (0,2) {$A_j$};
\node[bidder] (ad) at (1.7,2) {$A_{\mathrm{dum},\ell}$};

\draw[edge] (ga.north)--node[bid] {$d_{ij}$}(ai.south);
\draw[edge] (ga.north)--node[bid] {$2d_{ij}$}(aj.south);
\draw[edge] (ga.north)--node[bid] {$d_{ij}$}(ad.south);

\node[panel] at (0,-0.65) {(b) Absence signal};
\end{scope}

\begin{scope}[xshift=10.4cm]
\node[good] (gc) at (0,0) {$G_{\mathrm{const},i}$};
\node[bidder] (ci) at (-1.25,2) {$A_i$};
\node[bidder] (cd) at (1.25,2) {$A_{\mathrm{dum},\ell'}$};

\draw[edge] (gc.north)--node[bid] {$2z_i$}(ci.south);
\draw[edge] (gc.north)--node[bid] {$z_i$}(cd.south);

\node[panel] at (0,-0.65) {(c) Constant payment};
\end{scope}

\end{tikzpicture}
\caption{Auction constructions for presence signals, absence signals, and constant payments. The dummy bidders in parts~(b) and~(c) are distinct, and each has budget one.}
\label{fig:throttle-payment-patterns}
\end{figure}
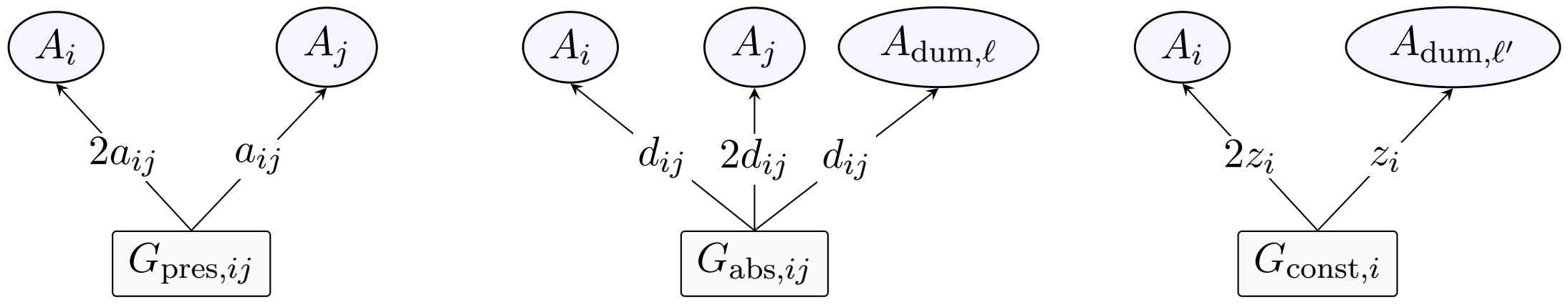

Each absence-signal or constant-payment good has a fresh dummy $A_{\mathrm{dum},\ell}$. Figure~\ref{fig:throttle-payment-patterns} gives the bids for $a_{ij},d_{ij},z_i>0$. Zero coefficients require no goods. Choose one common priority placing all bidders in the prescribed payment system before every dummy bidder, with arbitrary orders within these groups.

\paragraph{Presence signals.} In Figure~\ref{fig:throttle-payment-patterns}(a), participating $A_i$ always wins and pays $a_{ij}$ exactly when $A_j$ participates, giving conditional payment $a_{ij}\theta_j$. Bidder $A_j$ can win only at price zero, so its payment is unchanged.

\paragraph{Absence signals.} Consider $G_{\mathrm{abs},ij}$ in Figure~\ref{fig:throttle-payment-patterns}(b). Its dummy bidder $A_{\mathrm{dum},\ell}$ can win only when both $A_i$ and $A_j$ are absent, and then pays zero. Its conditional payment is therefore zero, and its positive budget forces $\theta_{\mathrm{dum},\ell}=1$. If $A_j$ is absent and $A_i$ participates, $A_i$ wins its tie with $A_{\mathrm{dum},\ell}$ and pays $d_{ij}$. If $A_j$ participates, its bid $2d_{ij}$ wins, and it pays $d_{ij}$ regardless of whether $A_i$ participates. Thus, this good contributes $d_{ij}(1-\theta_j)$ to $P_i$ and the constant $d_{ij}$ to $P_j$.

\paragraph{Constant payments and outgoing charges.} For each non-dummy bidder $A_i$, we realize the conditional-payment equation

\begin{equation}\label{eq:throttle-affine-system-rewritten} P_i=c_i+\sum_{j\ne i}\bigl(a_{ij}\theta_j+d_{ij}(1-\theta_j)\bigr),\qquad a_{ij},d_{ij}\ge0. \end{equation}

Each outgoing absence good $G_{\mathrm{abs},ji}$ charges $A_i$ the constant $d_{ji}$ conditional on participation. These charges total

\[ \tau_i=\sum_{j\ne i}d_{ji}. \]

We need $c_i\ge\tau_i$, since outgoing absence goods already charge $\tau_i$. If $c_i>\tau_i$, add $G_{\mathrm{const},i}$ from Figure~\ref{fig:throttle-payment-patterns}(c) with $z_i=c_i-\tau_i$. Its dummy always participates because it can win only at zero price. Thus $A_i$ pays exactly $z_i$ conditional on participating. If $c_i=\tau_i$, omit this good. These adjustments give exactly \eqref{eq:throttle-affine-system-rewritten}.

 Every auction equilibrium satisfies the prescribed conditional-payment equations because all dummy bidders participate fully. Conversely, any solution of those equations and the budget responses extends to an auction equilibrium by assigning participation probability one to every dummy bidder.

Finally, count one term for each positive-coefficient summand $a_{ij}\theta_j$ or $d_{ij}(1-\theta_j)$. Each uses one signal good. If $A_i$ occurs in $m_i$ terms, either providing a probability or receiving its payment effect, then $A_i$ bids positively on those $m_i$ goods and at most one constant-payment good. Hence $A_i$ bids positively on at most $m_i+1$ goods. Each dummy bids positively only on its own good; every good has at most three positive bids. We check $c_i\ge\tau_i$ and $m_i\le3$ for the complete construction. We next choose the arithmetic gadgets' payment coefficients and budgets so their budget responses enforce the required encoded input--output relations.

\subsection{Encoded Arithmetic Gadgets}
\label{subsec:throttle-gadgets-rewritten}
Write $q_i=1-\theta_i$ and $r_i=q_i/\theta_i$. Positive budgets give $\theta_i>0$ at equilibrium, and~\eqref{eq:throttle-response} is equivalent to
\begin{equation}\label{eq:throttle-odds-response-rewritten}
r_i=\left[\frac{P_i}{B_i}-1\right]_+,\qquad
\theta_i=\frac1{1+r_i},\qquad q_i=\frac{r_i}{1+r_i}.
\end{equation}
A wire representing $x\in[0,1]$ uses
\begin{equation}\label{eq:throttle-encoding-rewritten}
r_{\mathrm{out}}=\frac1{1+2x},\qquad
q_{\mathrm{out}}=\frac1{2(1+x)},\qquad
\theta_{\mathrm{out}}=\frac{1+2x}{2(1+x)}.
\end{equation}
Thus encoded wires have $r\in[1/3,1]$, $q\in[1/4,1/2]$, and $\theta\in[1/2,3/4]$. Set $\kappa=1/32$. Every output bidder has budget one. Table~\ref{tab:throttle-arithmetic} specifies the auxiliary budgets and conditional payments for
\[\begin{aligned}
&C(x)=1-x,\qquad D(x)=\min\{2x,1\},\qquad H(x)=x/2,\\
&I(x)=1/(1+x),\qquad G(x,y)=\min\{(1+x+y)/2,1\}.
\end{aligned}\]
Within a row, $P_\ell,B_\ell,\theta_\ell,q_\ell$ refer to its fresh auxiliary bidders. Section~\ref{subsec:throttle-completing-rewritten} verifies that constant terms cover outgoing absence charges, so the goods above realize these payment equations.

\begin{table}[H]
\centering
\begingroup
\renewcommand{\arraystretch}{1.35}
\begin{tabularx}{\textwidth}{
  @{}p{0.10\textwidth}p{0.21\textwidth}X@{}
}
\toprule
Gate & Auxiliary budgets & Conditional payments \\
\midrule
$C$ & $\kappa,\kappa^2$ &
$\begin{array}{@{}l@{\quad}l@{}}
P_1=\kappa(1/2+5q_{\mathrm{in},1}/2), &
P_2=9\kappa^2q_1, \\
P_{\mathrm{out}}=11/10+9\theta_2/10.
\end{array}$ \\

$D$ & $\kappa,1,\kappa$ &
$\begin{array}{@{}l@{\quad}l@{}}
P_1=\kappa(2/3+q_{\mathrm{in},1}), &
P_2=1/2+2\theta_1/3, \\
P_3=15\kappa q_2, &
P_{\mathrm{out}}=3/4+5\theta_3/4.
\end{array}$ \\

$H$ & $\kappa,\kappa$ &
$P_1=4\kappa q_{\mathrm{in},1},\quad
 P_2=2\kappa\theta_1,\quad
 P_{\mathrm{out}}=1+\theta_2.$ \\

$I$ & $\kappa$ &
$P_1=\kappa(1+4q_{\mathrm{in},1}),\quad
 P_{\mathrm{out}}=1+\theta_1.$ \\

$G$ & $\kappa$ for all five &
$\begin{array}{@{}l@{\quad}l@{}}
P_1=4\kappa q_{\mathrm{in},1}, &
P_2=4\kappa q_{\mathrm{in},2}, \\
P_3=2\kappa(\theta_1+\theta_2), &
P_4=3\kappa\theta_3, \\
P_5=3\kappa\theta_4, &
P_{\mathrm{out}}=1+\theta_5.
\end{array}$ \\
\bottomrule
\end{tabularx}
\endgroup
\caption{Payment equations for the five throttling gadgets.
The output budget is one in every row.}
\label{tab:throttle-arithmetic}
\end{table}

The next lemma verifies that these payment equations implement the stated operations and establishes their output ranges before assuming encoded inputs.

\begin{lemma}[Exact throttling arithmetic]\label{lem:throttle-encoded-arithmetic}
Table~\ref{tab:throttle-arithmetic} implements $C,D,H,I,G$ under encoding~\eqref{eq:throttle-encoding-rewritten}. Output odds are at most one for arbitrary input absence probabilities in $[0,1]$, and at least $1/3$ when all input absence probabilities are at most $1/2$. Successive budget responses uniquely determine the auxiliary and output probabilities for fixed inputs and extend every encoded input to the required encoded output.
\end{lemma}

We calculate the responses on arbitrary inputs before applying the encoding.

\paragraph{Complement.}
The responses give $r_1=[(5q_{\mathrm{in},1}-1)/2]_+$, $\theta_2=1/\max\{1,9q_1\}$, and $r_{\mathrm{out}}=1/10+9\theta_2/10$. Hence
\begin{equation}\label{eq:throttle-complement-response-rewritten}
r_{\mathrm{out}}=\begin{cases}
1,&0\le q_{\mathrm{in},1}\le1/4,\\
q_{\mathrm{in},1}/(5q_{\mathrm{in},1}-1),&1/4\le q_{\mathrm{in},1}\le1.
\end{cases}
\end{equation}
This lies in $[1/4,1]$, and in $[1/3,1]$ when $q_{\mathrm{in},1}\le1/2$. Substitution of an encoded input gives $r_{\mathrm{out}}=1/(3-2x)$, the encoding of $C(x)$.

\paragraph{Capped doubling.}
Successive responses give
\[r_1=[q_{\mathrm{in},1}-1/3]_+,\qquad
r_2=\left[\frac{1-3r_1}{6(1+r_1)}\right]_+,\qquad
\theta_3=\frac1{\max\{1,15q_2\}},\qquad
r_{\mathrm{out}}=\left[\frac{5\theta_3-1}{4}\right]_+.
\]
Since $r_2\le1/6$, we have $q_2\le1/7$ and $\theta_3\ge7/15$, so $r_{\mathrm{out}}\in[1/3,1]$. More explicitly,
\begin{equation}\label{eq:throttle-doubling-response-rewritten}
r_{\mathrm{out}}=\begin{cases}
1/3,&0\le q_{\mathrm{in},1}\le1/3,\\
q_{\mathrm{in},1}/(2-3q_{\mathrm{in},1}),&1/3\le q_{\mathrm{in},1}\le1/2,\\
1,&1/2\le q_{\mathrm{in},1}\le1.
\end{cases}
\end{equation}
On encoded inputs this is $1/(1+4x)$ for $x\le1/2$ and $1/3$ otherwise, exactly the encoding of $D(x)$.

\paragraph{Halving.}
The responses are $\theta_1=1/\max\{1,4q_{\mathrm{in},1}\}$, $\theta_2=1/\max\{1,2\theta_1\}$, and $r_{\mathrm{out}}=\theta_2$. Arbitrary inputs give $\theta_1\in[1/4,1]$ and $r_{\mathrm{out}}\in[1/2,1]$. For an encoded input, $\theta_1=(1+x)/2$, so $r_{\mathrm{out}}=1/(1+x)$, the encoding of $H(x)$.

\paragraph{Reciprocal.}
Here $r_1=4q_{\mathrm{in},1}$, so
\begin{equation}\label{eq:throttle-reciprocal-response-rewritten}
r_{\mathrm{out}}=\theta_1=\frac1{1+4q_{\mathrm{in},1}}.
\end{equation}
It belongs to $[1/5,1]$ on arbitrary inputs and to $[1/3,1]$ when $q_{\mathrm{in},1}\le1/2$. On an encoded input it equals $1/(1+2/(1+x))$, the encoding of $I(x)$.

\paragraph{Shifted average.}
The first two responses give $\theta_1,\theta_2\in[1/4,1]$, and thus $\theta_3=1/(2\theta_1+2\theta_2)\in[1/4,1]$. The last two give
\[\theta_4=\frac1{\max\{1,3\theta_3\}},\qquad
\theta_5=\frac1{\max\{1,3\theta_4\}},\qquad
r_{\mathrm{out}}=\theta_5.
\]
Splitting at $\theta_3=1/3$ yields
\begin{equation}\label{eq:throttle-average-response-rewritten}
r_{\mathrm{out}}=\max\{1/3,\theta_3\}\in[1/3,1].
\end{equation}
For encoded inputs, $\theta_1=(1+x)/2$ and $\theta_2=(1+y)/2$, so $\theta_3=1/(2+x+y)$. This is precisely the encoding of $G(x,y)$.

\begin{proof}[Proof of Lemma~\ref{lem:throttle-encoded-arithmetic}]
The five calculations above establish the encoded identities and the stated output ranges. In each row, every auxiliary response depends only on the inputs and earlier auxiliaries, so the budget responses uniquely determine all auxiliary and output probabilities for fixed inputs. They therefore also give the required extension for every encoded input assignment.
\end{proof}

\paragraph{Constants.}
For $z\in\{1/4,1/2\}$, an output bidder with budget one and $P_{\mathrm{out}}=1+1/(1+2z)$ has odds $1/(1+2z)$ and represents $z$. The constant-payment construction realizes these equations after the outgoing-charge adjustment.

\subsection{Encoding and Exact Gate Implementation}
\label{subsec:throttle-arithmetic-rewritten}

Complement and capped doubling are available. We now build the shifted product and then add affine decoding, which the nonlinear wire encoding does not itself provide.

\paragraph{Implementing the source functions.} The shifted-average gadget, together with complement and capped doubling, implements capped addition:

\[ \mathcal A(x,y)=C(D(C(G(x,y))))=\min\{x+y,1\}. \]

Indeed, if $x+y\le1$, the successive values are $G(x,y)=(1+x+y)/2$, $C(G(x,y))=(1-x-y)/2$, and $D(C(G(x,y)))=1-x-y$. The final complement gives $x+y$. If $x+y\ge1$, then $G(x,y)=1$, and the remaining operations give one. Every intermediate value belongs to $[0,1]$.

These operations meet Theorem~\ref{thm:criterion}(ii), giving a fixed-size rational implementation of $\phi(x,y)=(1+x)(1+y)/4$ with all intermediate arguments in $[0,1]$. Lemma~\ref{lem:throttle-encoded-arithmetic} supplies the local completions for this implementation.

\paragraph{Obtaining an affine decoding rule.}  For each original source variable $x$, let $A_W$ denote its wire bidder and introduce a fresh decoding bidder $A_{\mathrm{dec}}$ with

\[ B_{\mathrm{dec}}=\frac{\kappa}{4},\qquad P_{\mathrm{dec}}=\kappa q_W. \]
For an encoded wire, $q_W\in[1/4,1/2]$, so $P_{\mathrm{dec}}\ge B_{\mathrm{dec}}$. The budget response therefore gives

\begin{equation}\label{eq:throttle-decoding-rewritten} \theta_{\mathrm{dec}}=\frac1{4q_W}=\frac{1+x}{2},\qquad x=2\theta_{\mathrm{dec}}-1. \end{equation}
The payment $P_{\mathrm{dec}}=\kappa q_W$ is realized by an absence-signal good on which $A_W$ has bid $2\kappa$, $A_{\mathrm{dec}}$ has bid $\kappa$, and a fresh dummy bidder has bid $\kappa$. This good also charges $A_W$ the constant $\kappa$ whenever it participates. We add the decoding bidders after copying and include this charge in the complete payment accounting below.

\subsection{Completing the Proof}
\label{subsec:throttle-completing-rewritten}

Expand every source gate into its fixed arithmetic implementation from Section~\ref{subsec:throttle-arithmetic-rewritten}, then apply Lemma~\ref{lem:bounded-use} to the expanded circuit. The resulting complement copies preserve the solution set, each original wire of the expanded circuit has at most one outgoing circuit use, each new copy wire has at most two, and no gate uses its own output as an input. Add a decoding bidder only for each designated original source variable, after this copying step.

Create one wire bidder for each wire and fresh auxiliary bidders for each encoded gadget. Realize their prescribed conditional payments using the goods of Section~\ref{subsec:throttle-common-rewritten}, with a fresh dummy bidder for each absence-signal or constant-payment good. Choose a common priority that places all wire, auxiliary, and decoding bidders before all dummy bidders.

\paragraph{Covering outgoing absence charges.} We first verify the parameter condition $c_i\ge\tau_i$. Every wire bidder has constant term at least $3/4$, and each circuit input occurrence imposes an outgoing absence charge of at most $4\kappa$. The copying construction in Lemma~\ref{lem:bounded-use} gives a designated original source wire at most one circuit use; it also has one decoding use, the latter imposing charge $\kappa$. Every other wire has at most two circuit uses. Thus the outgoing absence charges of any wire sum to at most $8\kappa=1/4<3/4$. We next check the auxiliary bidders that send absence signals. In a complement gadget, $A_{\mathrm{aux},1}$ incurs charge $9\kappa^2$, covered by its constant term $\kappa/2$. In a capped-doubling gadget, $A_{\mathrm{aux},2}$ incurs charge $15\kappa$, covered by its constant term $1/2$. These inequalities hold because $\kappa=1/32$. Every other auxiliary sends only presence signals, which impose no payment on the bidder providing the signal. Each decoding bidder has no outgoing use and has constant term zero. Hence $c_i\ge\tau_i$ holds for every bidder in the prescribed system, and the realization in Section~\ref{subsec:throttle-common-rewritten} applies.

\paragraph{Establishing the encoding interval.} We prove the wire ranges without assuming that any input is already encoded. For arbitrary input absence probabilities in $[0,1]$, Lemma~\ref{lem:throttle-encoded-arithmetic} gives output odds at most one for every gadget. Constant outputs satisfy the same bound. Since every wire is the output of one gadget, all wires simultaneously satisfy $r_i\le1$ and therefore $q_i\le1/2$. We then obtain the lower bound on every wire's odds. The capped-doubling, halving, and shifted-average gadgets have output odds at least $1/3$ for arbitrary inputs. The complement and reciprocal gadgets have the same lower bound whenever their input absence probabilities are at most $1/2$, which has just been proved for all wires. Constant outputs also satisfy this bound. Thus every wire has $r_i\in[1/3,1]$ and $q_i\in[1/4,1/2]$. These bounds hold simultaneously on the entire circuit, including directed cycles, and justify every encoded gate identity.

\paragraph{Equilibrium correspondence and completion.} We first decode an arbitrary auction equilibrium. Every dummy bidder has zero conditional payment and therefore participates with probability one. The auction thus realizes the prescribed conditional-payment equations exactly. The wire bounds just proved allow the encoded identities of Lemma~\ref{lem:throttle-encoded-arithmetic} to be applied simultaneously. The represented values therefore satisfy the arithmetic circuit and all complement-copy equations. Removing the copy wires and using the arithmetic implementations gives a solution of the original source circuit. Each source value is read from its decoding bidder by \eqref{eq:throttle-decoding-rewritten}.

Conversely, extend any source solution through the arithmetic and complement copies. Set each wire's encoded participation probability, choose fresh auxiliary probabilities successively from their budget responses, and set decoding probabilities by \eqref{eq:throttle-decoding-rewritten}. Set all dummy probabilities to one. The completions in Lemma~\ref{lem:throttle-encoded-arithmetic} verify every prescribed response. Constant-payment adjustments already cover outgoing absence charges, so realized and prescribed conditional payments agree. All budget responses hold, giving an equilibrium. Independent participation and the fixed common priority determine the outcomes; no further allocation or tie probabilities are needed.

\paragraph{Sparsity and complexity.} Each wire bidder occurs in at most one incoming payment term and two outgoing uses, including decoding when present. The shifted-average auxiliary $A_{\mathrm{aux},3}$ has two incoming terms and one outgoing use. Other auxiliaries have at most one incoming term and one outgoing use; decoding bidders have one incoming term and no outgoing use. Thus each system bidder occurs in at most three terms and bids positively on at most four goods, including at most one constant-payment good. Each dummy bids on one good; each good has at most three positive bids.
We verify the finite numerical sets and the reduction size. Budgets belong to $\{1,\kappa,\kappa^2,\kappa/4\}$. All conditional-payment constants and coefficients belong to fixed finite rational sets determined by the gadget formulas. Each constant adjustment $c_i-\tau_i$ subtracts at most two outgoing absence charges from one prescribed constant. Hence the positive adjustments, and therefore all realized positive bids, belong to a fixed finite rational set. Arithmetic expansion, copying, decoding, and auction realization increase size by only a constant factor. The auction has size linear in the source circuit and polynomial encoding length. We now combine the established correspondence and instance bounds to prove Theorem~\ref{thm:throttle}.

\begin{proof}[Proof of Theorem~\ref{thm:throttle}]
The response in \eqref{eq:throttle-budget-response-rewritten} gives membership in $\FIXP$; see Section~\ref{sec:complexity} and \cite{CKK21}. Each conditional expected payment has a polynomial-size arithmetic circuit obtained by summing over possible second-highest participating bids. Since $B_i>0$, every denominator in the response is positive throughout the domain. This representation also gives the distance upper bound.

The charge checks above let the auction realize the systems of Lemma~\ref{lem:throttle-encoded-arithmetic}. Its arbitrary-input bounds establish the encoding interval, and its identities and completions, with Section~\ref{subsec:throttle-arithmetic-rewritten}, give both global solution directions as proved above. The counts verify four goods per bidder, three positive bids per good, a common priority, fixed finite rational data sets, and polynomial size.

The decoder $x=2\theta_{\mathrm{dec}}-1$ reads each source variable affinely from one participation probability. Thus the correspondence is an SL reduction from Lemma~\ref{lem:source}, proving $\FIXP$-hardness and, with membership, $\FIXP$-completeness. A participation vector within $\varepsilon/2$ in maximum norm of an exact equilibrium decodes to a vector within $\varepsilon$ of a source solution. This constant tolerance conversion gives distance hardness; the matching upper bound proves $\FIXPa$-completeness.
\end{proof}

\section{Conclusion}\label{sec:conclusion}
This work shows that simple scalar bidding controls can lead to the full complexity of algebraic fixed points. Exact equilibrium computation is $\FIXP$-complete in all four models, and approximation in distance is $\FIXPa$-complete. The sparse constructions for ordinary ROI pacing and throttling show that this difficulty persists with few goods per bidder and fixed finite rational data. For GSP, the results show that bounding each advertiser's inventory alone does not make exact computation tractable.

A natural algorithmic question is whether exact budget- and ROI-constrained GSP equilibria can be computed in polynomial time when each advertiser owns at most two products, while the numbers of advertisers and display slots remain unrestricted. Our hardness results use a constant inventory bound but leave this two-product case unresolved. An efficient algorithm for this case would identify a tractable restriction on individual catalogs and help locate the inventory threshold at which exact computation becomes hard.

\Needspace{10\baselineskip}
\section*{AI Methodology}
OpenAI Codex assisted with the development and review of the auction constructions and proofs, literature comparison, figures, and consistency of notation and exposition. ChatGPT assisted with exposition, translation, bibliography formatting, and LaTeX consistency. The human authors are responsible for the results' correctness and originality, reference accuracy, and final manuscript.

\appendix \crefalias{section}{appendix} \crefalias{subsection}{appendix}
\addtocontents{toc}{\protect\setcounter{tocdepth}{1}}
\section{ROI-GSP Membership with a Prescribed Cap}
\label{sec:prescribed-membership}
We adapt the group-allocation and payment circuits in Sections~4.1--4.2 of~\cite{HLN26} to a prescribed rational cap. Their ordering construction is unchanged; we verify the cap-dependent bounds and multiplier update.

\begin{lemma}[ROI-GSP membership with a prescribed cap]
\label{lem:gsp-prescribed-cap-membership}
For every rational cap $A>1$, ROI-GSP equilibria exist and can be computed in $\FIXP$ with polynomial support over legal orders. Circuit size and constant bit lengths are polynomial in the auction encoding length and the binary length of $A$, including when $A$ is an input. At $A=1$, an equilibrium is computable in polynomial time.
\end{lemma}

\begin{proof}
Let $m$ be the number of goods, $T\le m$ the number of equal-value groups used in~\cite{HLN26}, $v^*$ the largest value, and $c_\Sigma=\max\{1,\sum_k c_k\}$. Slots after $m$ can be omitted. If $T\le1$, all goods have the same owner and zero prices, so set all multipliers to $A$. Assume $T\ge2$.

Use the group-insertion and comparison construction of~\cite{HLN26}. Each Heaviside pseudogate with input $x$ is represented by a coordinate $\eta\in[0,1]$ with the continuous update $\eta'=\operatorname{clip}_{[0,1]}(\eta+x)$. At a fixed point it has exactly the required comparison relation. The insertion probabilities are nonnegative and normalized throughout the domain. Sorting their cumulative probabilities and using the intervals between consecutive breakpoints gives $Q=\sum_{t=2}^T(t+1)-1=O(T^2)$ order coordinates. Lemmas~4.2--4.4 of~\cite{HLN26} show that at a fixed point the positive-weight orders are legal and give the exact payment and value formulas. These statements depend on bid comparisons, not on the particular cap.

For completeness, all intermediate and output coordinates have finite rational bounds for the supplied cap. Group bids lie in $[0,Av^*]$. Even when comparison selectors are fractional, group-position expressions are bounded by $2T^2$, item positions by $m$, price expressions by $TAv^*$, and CTR expressions by $c_\Sigma$. The latter bounds follow by allowing every unit-bounded selector in the corresponding sums to be nonzero. The order weights sum to one, so
\[0\le P_i\le mTAv^*c_\Sigma,\qquad 0\le V_i\le mv^*c_\Sigma.\]
The construction uses finite endpoint comparisons and the finite sentinel $T$ for a missing foreign group, as in the cited payment circuit.

Take multiplier coordinates in $[1,A]$, comparison and order-weight coordinates in $[0,1]$, and item-position coordinates in $[0,m]$. Update comparisons by their clipped rules, assign the computed order weights and positions to the corresponding output coordinates, and update multipliers by
\[\alpha_i'=\operatorname{clip}_{[1,A]}(\alpha_i+V_i-P_i).\]
The bounds above make this a continuous algebraic self-map of a compact rational box. Affine rescaling gives a polynomial-size circuit on a unit cube. Brouwer's theorem gives a fixed point, and the cited allocation lemmas identify its payment and value coordinates with a legal order distribution.

At a fixed point, $P_i>V_i$ would decrease $\alpha_i$ unless $\alpha_i=1$. At that endpoint, every good's price is at most its own unscaled value in every legal order, so $P_i\le V_i$. Thus ROI feasibility always holds. Similarly, $P_i<V_i$ would increase $\alpha_i$ unless $\alpha_i=A$. These are exactly the required equilibrium conditions.

The ordering circuits have polynomial size, and replacing the canonical cap by $A$ changes only bounds involving $Av^*$, $mTAv^*c_\Sigma$, and $[1,A]$. Their bit lengths are polynomial in the input and the binary length of $A$. Multipliers and the polynomial-support distribution are represented by affine output coordinates, proving uniform $\FIXP$ membership. At $A=1$, sort the unscaled bids and use any legal deterministic order; prices are at most values and every multiplier is at the cap.
\end{proof}

\section{Shared Arithmetic and Tie Coupling}
\label{sec:arithmetic}
We state the ordering lemma used by both GSP reductions, prove the bounded affine construction, and prove Theorem~\ref{thm:criterion}. Auction-specific range conditions are verified in the corresponding reductions.

\subsection{Coupling Local Choices}
\label{sec:copy-proof}
\label{sec:framework}

\begin{lemma}[Coupling Local Choices, {\cite{HLN26}}]
\label{lem:local-coupling}
Consider $m$ local choice locations, with $k_j\ge1$ choices at location $j$ for $j=1,\ldots,m$. Suppose every tuple of local choices can be combined into a valid global order, and each bidder's payment is a sum of functions of the individual local choices. Any prescribed local marginal distributions can then be realized by at most $1+\sum_{j=1}^{m}(k_j-1)$ global orders, with the same expected payments.
\end{lemma}

This is the quantile coupling in Lemma~4.3 of~\cite{HLN26}, applied to the local choice distributions; the support bound counts their distinct cumulative-probability intervals.

\subsection{Bounded Affine Arithmetic}
\label{sec:affine-compiler}
The next lemma builds signed clipped affine expressions from decreasing affine gates with an upper cap. Auction constructions with additional range conditions must check them separately.

\begin{lemma}[Constructing Clipped Affine Expressions]
\label{lem:clipped-affine}
Suppose gates with at most two inputs implement

\[y=\min\left\{1,c-\sum_{j=1}^{k}a_jx_j\right\},\qquad k\in\{0,1,2\},\qquad a_j\ge0,\qquad c\ge\sum_{j=1}^{k}a_j,\]
where the inputs belong to $[0,1]$ and the constants are rational. These gates implement any fixed rational signed affine expression followed by clipping to $[0,1]$. The implementation has fixed size, all intermediate values lie in $[0,1]$, and every input assignment uniquely determines its internal values.
\end{lemma}

\begin{proof}
We first construct the elementary operations. The unary gate $C(x)=1-x$ implements complement. Complementing the inputs gives capped addition and capped scaling:

\[\mathcal A(x,y)=\min\{1,2-C(x)-C(y)\}=\min\{x+y,1\},\qquad \min\{1,k-kC(x)\}=\min\{kx,1\}\quad(k\ge0).\]
Each displayed expression uses at most two inputs and satisfies the required inequality between its constant and the sum of its nonnegative coefficients. Constants in $[0,1]$ use zero-input gates. Complement and capped addition also give nonnegative subtraction:

\[\mathcal T(x,y)=C(\mathcal A(C(x),y))=[x-y]_+.\]

We now construct a general affine expression. Let $f$ be the desired fixed signed affine expression. For each positive term $bx$, use the identity $bx=b-bC(x)$. Collecting constants then writes

\[f=c_0-\sum_j a_jz_j,\qquad a_j\ge0,\qquad z_j\in[0,1],\]
where each $z_j$ is an original input or its complement. If $c_0\le0$, then $f\le0$ throughout the domain, so its clipped value is the constant zero. Otherwise choose

\[M=\max\left\{1,c_0,\sum_j a_j\right\},\qquad q=\sum_j\frac{a_j}{M}z_j.\]
Each weighted term is obtained by capped scaling, and every partial sum is at most one. Binary capped additions therefore compute $q$ exactly. Since $c_0/M\in[0,1]$, nonnegative subtraction followed by capped scaling gives

\[\clip(f)=\min\left\{1,M\left[\frac{c_0}{M}-q\right]_+\right\}.\]
All intermediate values lie in $[0,1]$, and each input has a unique extension. The size and rational constants depend only on the fixed expression.
\end{proof}

\subsection{Nonlinear Arithmetic}
\label{sec:reciprocal}
We prove the arithmetic criterion stated in Section~\ref{sec:overview}.

\begin{proof}[Proof of Theorem~\ref{thm:criterion}]
We reduce the source product to squaring and construct a square under each alternative.
Once $\Sq(x)=x^2$ is available on $[0,1]$, polarization gives

\[xy=2\left(\Sq\left(\frac{x+y}{2}\right)-\Sq\left(\frac x2\right)-\Sq\left(\frac y2\right)\right).\]

The mean is obtained by adding the two halved inputs, whose sum is at most one. The two successive differences in the displayed formula are nonnegative, and the final difference is $xy/2\in[0,1/2]$. Nonnegative subtraction and capped doubling therefore implement the identity exactly. Applying this product to $(1+x)/2$ and $(1+y)/2$ gives $\phi(x,y)=(1+x)(1+y)/4$. Complement and capped doubling already provide the other two source functions.

Under alternative (i), write $p=h^{-1}$. We first realize $p(t)$ for a desired input $t\in J$. Set $\eta=1$ when $h$ is decreasing and $\eta=-1$ when $h$ is increasing, and impose the simultaneous equations

\[\left\{\begin{aligned}&y=h(u),\\&u=\mathrm{clip}_{I_0}(u+\eta(y-t)).\end{aligned}\right.\]

Clipping to $I_0$ is implemented by an affine encoding of that interval into $[0,1]$. At the lower endpoint of $I_0$, strict containment of $J$ inside $h(I_0)$ gives $\eta(h(u)-t)>0$, so the update cannot return that endpoint. At the upper endpoint it gives $\eta(h(u)-t)<0$, excluding that endpoint as well. Hence $u$ is interior. Clipping returns an interior value only when its argument equals that value, so the second equation forces $y=t$. The first then gives $u=p(t)$. Conversely, $u=p(t)$ and $y=t$ satisfy both equations.

To extract a square under alternative (i), we use the assumed form of $p$. Polynomial division gives either a nonaffine quadratic polynomial or an affine function plus a nonzero reciprocal of an affine function. In the polynomial case, write $p(t)=at^2+bt+c$ with $a\ne0$. Choose rational $t_0$ and $\epsilon>0$ such that $t_0+\epsilon[0,1]\subset J$. Then

\[p(t_0+\epsilon x)=p(t_0)+\epsilon(2at_0+b)x+a\epsilon^2x^2.\]
Subtracting the constant and linear terms and dividing by the nonzero rational $a\epsilon^2$ gives $x^2$. These are signed affine operations on bounded encoded variables, and the final expression lies in $[0,1]$.

In the remaining case, subtract the affine part of $p$. Its denominator has constant nonzero sign on $J$, so affine changes of argument and output give a reciprocal operation $q(w)=1/w$ on a positive rational interval $K$. Choose rational $a,d,z_0>0$ such that $a,a+d,z_0$ lie inside $K$, and set $k=z_0a(a+d)/d$. At $w=a$, the argument $k(q(w)-q(w+d))$ equals $z_0$. By continuity, choose a rational $\epsilon>0$ small enough that, for every $x\in[0,1]$, all three arguments $w=a+\epsilon x$, $w+d$, and $k(q(w)-q(w+d))$ remain in $K$. The reciprocal identities then give

\[dk\,q\bigl(k(q(w)-q(w+d))\bigr)=w(w+d)=a(a+d)+\epsilon(2a+d)x+\epsilon^2x^2.\]
Subtracting the constant and linear terms and dividing by $\epsilon^2$ again gives $x^2$. The compact intervals bound every denominator away from zero. All intermediate quantities have fixed bounded ranges and can be represented by affine encodings into $[0,1]$. The signed clipped affine operations carry out the required transformations on these encodings without changing valid represented values.

Under alternative (ii), we first obtain the elementary operations needed to extract a square. Complement and capped addition give $\mathcal T(x,y)=[x-y]_+$ as above. The constant one is $\mathcal A(x,C(x))$, and halving it gives $1/2$. Put $z=x/2$ and use the reciprocal operation to form

\begin{equation}\label{eq:common-square}\begin{aligned}d&=I(z)-\frac12I(z/2)=\frac1{(1+z)(2+z)},\\r&=I(4d-1)=\frac12+\frac{3x}{8}+\frac{x^2}{16}.\end{aligned}\end{equation}

For $x\in[0,1]$, we have $z\in[0,1/2]$ and $d\in[4/15,1/2]$, so the subtraction defining $d$ is nonnegative. The argument of the last reciprocal is implemented by

\[C(D(C(D(d))))=1-2(1-2d)=4d-1\in[1/15,1],\]
where both doublings occur before their upper cap is exceeded. Thus all three reciprocal calls use arguments in $[0,1]$. The term $3x/8$ is obtained by adding $x/4$ and $x/8$. Subtracting $1/2$ and then $3x/8$ from $r$ gives $x^2/16$, with both subtractions nonnegative. Four capped doublings recover $x^2$ exactly.

Both alternatives therefore implement a square and hence the source product. The constructions use fixed rational constants and a fixed number of gates. Every source input extends through them, including the uniquely determined inverse-feedback variables in (i), and the identities above force the required output in every solution.
\end{proof}

\section{Additional Proofs}
\label{sec:roi-parameters}
This appendix proves Lemma~\ref{lem:roi-explicit-parameters} and the order-independent identical-supplier bounds used in Section~\ref{subsec:roi-suppliers-rewritten}. All notation is from that section. Section~\ref{subsec:roi-inventory-certificate} gives an inventory certificate for every fixed design, including a universal bound at $A=2$.

\subsection{Supplier Parameters}
\label{subsec:roi-parameter-existence}
The construction fixes the supplier values and chooses the margins and supplier count directly from $h$. We prove Lemma~\ref{lem:roi-explicit-parameters} for both thresholds $\theta\in\{\vartheta,2\}$.

\begin{lemma}[Explicit supplier parameters]\label{lem:roi-explicit-parameters}
For every fixed rational design cap $A>1$, the parameters defined in Section~\ref{subsec:roi-suppliers-rewritten} satisfy \eqref{eq:small-cap-supplier-margins} and, for both $\theta\in\{\vartheta,2\}$,

\[\left\{\begin{aligned}&1<v<S<1+\sigma,\qquad c<\sigma/6,\qquad q+\rho<1,\\&\frac n{n-1}<1+\sigma<U,\qquad u(1+\sigma)<\frac14,\qquad Bc(q+\rho)<\frac14,\\&a>\frac{2\theta}{1-q},\qquad a(\vartheta-1)>2B,\qquad S+\gamma<2,\\&T>BQ+2,\qquad G>UQ,\qquad T>U(Q+1)+1,\\&T>G>u>c>2\delta>0,\quad N>2T,\quad c(q-\rho)>\delta,\\&r>Uv,\quad v/\kappa<Q,\quad M(1+\sigma)<98,\quad S<1011/1000.\end{aligned}\right.\]
For $A=2$, the construction uses $n=66$ suppliers in each group and $M=96$ floor compensation copies.
\end{lemma}

\begin{proof}[Proof of Lemma~\ref{lem:roi-explicit-parameters}]
We have $0<h\le1/16$, $L\le5/4$, $U\le3/2$, $q\ge9/10$, $\sigma\le1/64$, and $F-L=4hq$. To check that the interval defining $\vartheta$ is nonempty, note that $2/L-F\ge1/10$. Thus

\[\begin{aligned}
q\bigl(U-(1+\sigma)F\bigr)&=4qh^2(2/L-F)\ge\frac9{25}h^2,\\
\rho\bigl(U+(1+\sigma)F\bigr)&<4\rho=\frac{h^4}{4}\le\frac{h^2}{1024}.
\end{aligned}\]
Thus $(q+\rho)F<(q-\rho)U/(1+\sigma)$, and their midpoint gives the last two inequalities in \eqref{eq:small-cap-supplier-margins}. Also $qL=1+2h$ and $\rho L<h^2$ give

\[(q-\rho)L-(1+\sigma)>2h-5h^2>0.\]
These comparisons imply $1<qL<qF<\vartheta<qU<3/2$, and $q+\rho<1$ follows from $\rho<2h/L=1-q$.
The choice of $n$ gives $n\ge66$, $n/(n-1)<1+\sigma<U$, and $a>145/\sigma$. Since $K-a=M+1=97$ and $1<\theta\le2$,

\[S-1=\frac{97}{a}<\frac{97\sigma}{145}<\frac{11}{1000},\qquad
0<c<\frac{(47/2)\sigma}{145-2\sigma}<\frac\sigma6.\]
Moreover $97(a-\theta)-qTa>(97-T)a-194>0$, so $v-1=97/a-qT/(a-\theta)>0$. Hence $1<v<S<1+\sigma$. The bounds $u\le2c<\sigma/3$ and $B=401/50$ give $u(1+\sigma)<1/4$ and $Bc(q+\rho)<1/4$. The definition $\delta=49c/100$ yields $2\delta<c$, while $q-\rho>49/100$ gives $c(q-\rho)>\delta$.

For the remaining size comparisons, $1-q=2h/L$ and $\vartheta-1>2h$ imply

\[\frac{2\theta}{1-q}\le\frac{5}{2h}<\frac{145}{4h^2}<a,\qquad
a(\vartheta-1)>\frac{145}{2h}\ge1160>2B.\]
Also $\gamma<c(\vartheta)/4<\sigma/24$, so $S+\gamma<2$. Direct substitution gives $N>2T$, $T>BQ+2$, $G>UQ$, $T>U(Q+1)+1$, and $M(1+\sigma)\le97.5<98$. Together with $u<\sigma/3$ this gives $T>G>u>c>2\delta>0$. Finally $v<S<1011/1000$, $\kappa\ge19/50$, and $U\le3/2$ imply $v/\kappa<8/3=Q$ and $Uv<8/5=r$. At design $A=2$, $\sigma=1/64$ and $n=66$, while $M=96$ for every design.
\end{proof}

\subsection{Proof of Lemma~\ref{lem:identical-suppliers}}
\label{sec:identical-suppliers}

\begin{proof}[Proof of Lemma~\ref{lem:identical-suppliers}]
We compare each supplier's prices with bids submitted on foreign copies of its own goods. Let $V>0$ denote the total value of one supplier's catalog. Full display and unit click-through rates make this the allocated value of every supplier in the group.

First suppose that $A_b$ and $A_{b'}$ have $\alpha_b<\alpha_{b'}$. For every good of $A_{b'}$ with value $w$, the corresponding foreign good of $A_b$ bids $\alpha_bw$ and appears later. The price of the good of $A_{b'}$ is therefore at least $\alpha_bw$. Summing over its catalog gives $P_{b'}\ge\alpha_bV$, while ROI feasibility gives $P_{b'}\le V$. Hence $\alpha_b\le1$, and the multiplier domain implies $\alpha_b=1$. In particular, an unequal group has minimum multiplier one.

We now exclude a multiplier above two in an unequal group. Suppose $\alpha_b>2$ and choose a supplier with multiplier one. The catalog of $A_b$ contains goods of values $w$ and $2w$ for some $w>0$. Its good of value $w$ bids more than $2w$, so the foreign copy of value $2w$ owned by the multiplier-one supplier appears later and forces its price to be at least $2w$. Every other good of $A_b$ also has a later foreign copy bidding its unscaled value, so its price is at least that value. Thus $P_b\ge V+w>V$, contradicting ROI feasibility.

Now suppose that all suppliers share $\alpha_{\mathrm{grp}}$. At each distinct value, select the owner of the last good at that value. Every good of every other supplier at that value has a foreign equal-bid good after it and therefore pays the common bid. Since the catalogs are identical, the group thus pays at least $(n-1)\alpha_{\mathrm{grp}}V$ in total. Its total allocated value is $nV$, so ROI feasibility gives

\[(n-1)\alpha_{\mathrm{grp}}V\le nV,\qquad \alpha_{\mathrm{grp}}\le\frac n{n-1}\le2.\]
These orderwise comparisons bound all supplier multipliers by two under any valid-order distribution, independently of block separation and gate equations.
\end{proof}

\subsection{Proof of the Inventory Bound in Theorem~\ref{thm:roi-gsp-rewritten}}\label{subsec:roi-inventory-certificate}
We count the expanded binary gate constructions for every fixed rational design $A>1$. The resulting bound is less than $2^{43}\max\{1,(A-1)^{-5}\}$ goods per bidder; design $A=2$ gives the canonical-cap bound $2^{43}$. Use fresh internal subcircuits and auxiliary copies, except for the circuit-wire identifications required in Section~\ref{subsec:roi-arithmetic-rewritten}. 

\paragraph{Sizes of the expanded gate constructions.} Count a signed affine call before the restricted construction as one call. A square subcircuit $\mathcal Q$ uses three such calls and one nonlinear gadget: the inverse target, inverse feedback, square output, and nonlinear response. A square-root subcircuit uses one square subcircuit and one additional affine call. For normalization, use fresh square and root subcircuits for each of its two distance terms, followed by one affine call. The following counts include the fresh input normalizers of each source gate.

\begin{table}[H]
\centering
\sbox{\TableBody}{%
\begin{tabular}{lrr}
\toprule
Subcircuit & Signed affine operations & Nonlinear gadgets\\
\midrule
Square $\mathcal Q$ & $3$ & $1$\\
Square root & $4$ & $1$\\
Normalization $\mathcal N$ & $17$ & $4$\\
Normalized complement & $18$ & $4$\\
Normalized capped doubling & $35$ & $8$\\
Normalized shifted product & $45$ & $11$\\
\bottomrule
\end{tabular}%
}
\usebox{\TableBody}
\caption{Gate counts before the restricted affine construction, with no sharing of the two distance-square subcircuits.}
\label{tab:roi-template-counts}
\end{table}

Every signed affine call has at most three distinct variable terms. The restricted construction uses at most two copies per term and one final decreasing call, at most seven calls; use the larger bound nine. Replacing each three-input call by the binary construction uses at most three basic gadgets. Thus each source-gate construction, including complement copying and input normalization, has fewer than $3(9\cdot45+11)<1536$ basic gadgets and fewer than $2\cdot1536$ input occurrences. An internal wire belongs to one source-gate construction; a circuit wire belongs to its defining construction and at most two constructions using it. Hence every wire has fewer than $3\cdot2\cdot1536=9216$ ordinary signal uses. 

\paragraph{Bounding receiving multiplicities.} The fixed gate constructions have no negative self-input after combining terms: only inverse and root feedback reuse their output, with a positive coefficient. All other internal subcircuits and their outputs are fresh. With $\beta=7/8$, the three-input decreasing calls and a choice of merged weights are

\begin{table}[H]
\centering
\begingroup
\sbox{\TableBody}{%
\begin{tabular}{@{}lcc@{}}
\toprule
Call & Three weights & Weights merged in binary splitting\\
\midrule
Inverse feedback & $8/7,8/7,1$ & $8/7,1$\\
Root feedback & $8/7,64/21,8/3$ & $8/7,8/3$\\
Product output & $64/7,2,2$ & $2,2$\\
\bottomrule
\end{tabular}%
}
\usebox{\TableBody}
\caption{The three-input calls before binary splitting. All other basic calls have at most two inputs.}
\label{tab:roi-binary-weights}
\endgroup
\end{table}

The respective final binary weight pairs are

\[(960/343,8/7),\qquad (5120/1029,64/21),\qquad (256/49,64/7).\]
The weighted-mean copies have weights at least $1/8$. The only other coefficients that grow as the cap approaches one are the square-output weights

\[\frac{L\kappa}{h}=\frac2s-2-2s,\qquad
-\frac{LP'(\ell_1)}{4h}=\frac1{2s}+1.\]
For $0<s\le1/5$, the first is larger. Order the receiving branches by decreasing weight. The listed calls and the remaining affine expressions in Section~\ref{subsec:roi-arithmetic-rewritten} then give

\[d_1,d_2\ge\frac18,\qquad
d_1\le\frac2s=\frac L{2h}\le\frac5{8h},\qquad
d_2\le\max\left\{\frac{256}{49},\frac1{2s}+1\right\}\le\frac{21}{64h}.\]

The second weight is present only for binary calls. This estimate includes negative copies, the weighted-mean copies introduced by splitting, distance squares, normalization, complement, doubling, and mean calls.

We bound the ordinary block's smallest value separately. Since $n=\lceil(4h^2)^{-1}\rceil+2$,

\[a-2<\frac{145}{4h^2}+336<\frac{38}{h^2},\qquad
\delta(2)=\frac{(49/100)(47/2)}{a-2}>\frac3{10}h^2.\]
Also $1<v<2$ and $g_0/m=1/(192S)>1/195$. Normalize all local scales by $m$. Set $C_*=980/(3h^2)$, so that $98/\delta<C_*$. Equation~\ref{eq:roi-explicit-receiving-count} gives $n_j\le98d_j/(v\delta g_{\mathrm{prev}})+2$, and therefore

\[\begin{aligned}
n_1&<C_*\frac5{8h}\,195+2=\frac{79625}{2}h^{-3}+2,\\
\frac1{g_1}=\frac{vn_1}{d_1}&<\frac{98}{\delta g_0}+\frac{2v}{d_1}<63700h^{-2}+32,\\
n_2&<C_*\frac{21}{64h}(63700h^{-2}+32)+2
=\frac{27311375}{4}h^{-5}+3430h^{-3}+2.
\end{aligned}\]
These inequalities use at most two local branches and the price induction of Lemma~\ref{lem:roi-common-supplier-multipliers}.

\paragraph{Counting goods per owner.} A wire bidder owns its main good, a private pair, and two goods per ordinary signal use, hence fewer than $3+2\cdot9216=18435$ goods. An affine auxiliary owns its main good, private pair, ordinary floor pair, $96$ compensation copies, and at most two groups of receiving goods. Its inventory is at most

\[\begin{aligned}
101+n_1+n_2
&<\frac{27311375}{4}h^{-5}+\frac{86485}{2}h^{-3}+105\\
&<7{,}000{,}000h^{-5}\\
&=7{,}340{,}032{,}000{,}000\bigl(\min\{A-1,1\}\bigr)^{-5}\\
&<2^{43}\max\{1,(A-1)^{-5}\}.
\end{aligned}\]
The second inequality uses $h\le1/16$. Nonlinear auxiliaries own four goods; first- and second-group suppliers own three and six, respectively. All are below the displayed bound. For design $A=2$, each group has $66$ suppliers and the inventory is less than $2^{43}$. Receiving goods dominate this sufficient bound. For fixed $A$, every multiplicity is constant; geometric and private scales affect rational encoding lengths but not ownership. The hardness construction is for fixed $A$; uniform membership for input caps is proved in Appendix~\ref{sec:prescribed-membership}.

\begingroup
\interlinepenalty=10000
\printbibliography[title={References}]

@inproceedings{ABM19,
  author = {Gagan Aggarwal and Ashwinkumar Badanidiyuru and Aranyak Mehta},
  title = {{Autobidding with Constraints}},
  booktitle = {Proceedings of WINE 2019},
  date = {2019},
  year = {2019},
  url = {https://research.google/pubs/autobidding-with-constraints/},

}

@misc{AFZ24,
  author = {Gagan Aggarwal and Giannis Fikioris and Mingfei Zhao},
  title = {{No-Regret Algorithms in Non-Truthful Auctions with Budget and {ROI} Constraints}},
  date = {2024},
  year = {2024},
  eprint = {2404.09832},
  howpublished = {arXiv preprint \href{https://arxiv.org/abs/2404.09832}{arXiv:2404.09832}},
  eprinttype = {arxiv},

}

@inproceedings{CKK21,
  author = {Xi Chen and Christian Kroer and Rachitesh Kumar},
  title = {{Throttling Equilibria in Auction Markets}},
  booktitle = {Web and Internet Economics (WINE 2021)},
  series = {Lecture Notes in Computer Science},
  volume = {13112},
  pages = {551},
  publisher = {Springer},
  date = {2022},
  year = {2022},
  eprint = {2107.10923v3},
  eprinttype = {arxiv},
  note = {Extended abstract. References use the full version of September 30, 2026, \href{https://arxiv.org/abs/2107.10923v3}{arXiv:2107.10923v3}},

}

@article{CKK24,
  author = {Xi Chen and Christian Kroer and Rachitesh Kumar},
  title = {{The Complexity of Pacing for Second-Price Auctions}},
  journaltitle = {Mathematics of Operations Research},
  journal = {Mathematics of Operations Research},
  volume = {49},
  number = {4},
  pages = {2109--2135},
  date = {2024},
  year = {2024},
  doi = {10.1287/moor.2022.0009},

}

@article{CKP22,
  author = {Vincent Conitzer and Christian Kroer and Debmalya Panigrahi and Okke Schrijvers and Nicolas E. Stier-Moses and Eric Sodomka and Christopher A. Wilkens},
  title = {{Pacing Equilibrium in First Price Auction Markets}},
  journaltitle = {Management Science},
  journal = {Management Science},
  volume = {68},
  number = {12},
  pages = {8515--8535},
  date = {2022},
  year = {2022},
  doi = {10.1287/mnsc.2022.4310},

}

@article{CKSM22,
  author = {Vincent Conitzer and Christian Kroer and Eric Sodomka and Nicolas E. Stier-Moses},
  title = {{Multiplicative Pacing Equilibria in Auction Markets}},
  journaltitle = {Operations Research},
  journal = {Operations Research},
  volume = {70},
  number = {2},
  pages = {963--989},
  date = {2022},
  year = {2022},
  doi = {10.1287/opre.2021.2167},

}

@inproceedings{CL25,
  author = {Xi Chen and Yuhao Li},
  title = {{Constant Inapproximability of Pacing Equilibria in Second-Price Auctions}},
  booktitle = {Web and Internet Economics (WINE 2025)},
  series = {Lecture Notes in Computer Science},
  volume = {16266},
  pages = {294--306},
  publisher = {Springer},
  date = {2026},
  year = {2026},
  doi = {10.1007/978-3-032-18660-7_15},
  eprint = {2501.15295v2},
  eprinttype = {arxiv},

}

@misc{CMY26,
  author = {Li Chen and Jamie Morgenstern and Yuanyuan Yang},
  title = {{Beyond the {PPAD} Hardness of Auto-bidding Auctions}},
  date = {2026},
  year = {2026},
  eprint = {2608.01889},
  howpublished = {arXiv preprint \href{https://arxiv.org/abs/2608.01889}{arXiv:2608.01889}},
  eprinttype = {arxiv},

}

@misc{DLPST26,
  author = {Paul D{\"u}tting and Yuhao Li and Paes Leme, Renato and Kelly Spendlove and Yifeng Teng},
  title = {{Autobidding Equilibria in Sponsored Shopping}},
  date = {2026},
  year = {2026},
  eprint = {2602.21966v2},
  howpublished = {arXiv preprint \href{https://arxiv.org/abs/2602.21966v2}{arXiv:2602.21966v2}},
  eprinttype = {arxiv},

}

@misc{DLTZ25,
  author = {Yuan Deng and Yilin Li and Wei Tang and Hanrui Zhang},
  title = {{No-Regret Online Autobidding Algorithms in First-Price Auctions}},
  date = {2025},
  year = {2025},
  eprint = {2510.16869},
  howpublished = {arXiv preprint \href{https://arxiv.org/abs/2510.16869}{arXiv:2510.16869}},
  eprinttype = {arxiv},

}

@inproceedings{DMMZ21,
  author = {Yuan Deng and Jieming Mao and Vahab Mirrokni and Song Zuo},
  title = {{Towards Efficient Auctions in an Auto-bidding World}},
  booktitle = {Proceedings of The Web Conference 2021},
  pages = {3965--3973},
  date = {2021},
  year = {2021},
  eprint = {2103.13356},
  eprinttype = {arxiv},

  doi = {10.1145/3442381.3450052}
}

@article{EOS07,
  author = {Benjamin Edelman and Michael Ostrovsky and Michael Schwarz},
  title = {{Internet Advertising and the Generalized Second-Price Auction: Selling Billions of Dollars Worth of Keywords}},
  journaltitle = {American Economic Review},
  journal = {American Economic Review},
  volume = {97},
  number = {1},
  pages = {242--259},
  date = {2007},
  year = {2007},
  doi = {10.1257/aer.97.1.242},

}

@article{EY10,
  author = {Kousha Etessami and Mihalis Yannakakis},
  title = {{On the Complexity of {Nash} Equilibria and Other Fixed Points}},
  journaltitle = {SIAM Journal on Computing},
  journal = {SIAM Journal on Computing},
  volume = {39},
  number = {6},
  pages = {2531--2597},
  date = {2010},
  year = {2010},
  doi = {10.1137/080720826},

}

@article{FGHLP23,
  author = {Aris Filos-Ratsikas and Yiannis Giannakopoulos and Alexandros Hollender and Philip Lazos and Diogo Po{\c{c}}as},
  title = {{On the Complexity of Equilibrium Computation in First-Price Auctions}},
  journaltitle = {SIAM Journal on Computing},
  journal = {SIAM Journal on Computing},
  volume = {52},
  number = {1},
  pages = {80--131},
  date = {2023},
  year = {2023},
  doi = {10.1137/21M1435823},
  eprint = {2103.03238v3},
  eprinttype = {arxiv},

}

@inproceedings{FHHH24,
  author = {Aris Filos-Ratsikas and Kristoffer Arnsfelt Hansen and Kasper H{\o}gh and Alexandros Hollender},
  title = {{{PPAD}-Membership for Problems with Exact Rational Solutions: A General Approach via Convex Optimization}},
  booktitle = {Proceedings of STOC 2024},
  pages = {1204--1215},
  date = {2024},
  year = {2024},
  eprint = {2312.01237v1},
  eprinttype = {arxiv},
  note = {Section references use the full version, \href{https://arxiv.org/abs/2312.01237v1}{arXiv:2312.01237v1}},

}

@misc{HLN26,
  author = {Kristoffer Arnsfelt Hansen and Yuhao Li and Xinhao Nie},
  title = {{Complexity of Autobidding Equilibria in {GSP} Sponsored Shopping}},
  date = {2026-10-09},
  year = {2026},
  eprint = {2610.13716v1},
  howpublished = {arXiv preprint \href{https://arxiv.org/abs/2610.13716v1}{arXiv:2610.13716v1}},
  eprinttype = {arxiv},

}

@inproceedings{LT24,
  author = {Juncheng Li and Pingzhong Tang},
  title = {{Vulnerabilities of Single-Round Incentive Compatibility in Auto-bidding: Theory and Evidence from {ROI}-Constrained Online Advertising Markets}},
  booktitle = {Proceedings of IJCAI 2024},
  date = {2024},
  year = {2024},
  eprint = {2210.06107v6},
  eprinttype = {arxiv},

}

@misc{Liu26,
  author = {Zhengyang Liu},
  title = {{Tight Inapproximability of Pacing and Throttling Equilibria in Second-Price Auctions}},
  date = {2026},
  year = {2026},
  eprint = {2608.16682},
  howpublished = {arXiv preprint \href{https://arxiv.org/abs/2608.16682}{arXiv:2608.16682}},
  eprinttype = {arxiv},

}

@inproceedings{Mehta22,
  author = {Aranyak Mehta},
  title = {{Auction Design in an Auto-bidding Setting: Randomization Improves Efficiency beyond {VCG}}},
  booktitle = {Proceedings of The Web Conference 2022},
  date = {2022},
  year = {2022},
  eprint = {2204.10956},
  eprinttype = {arxiv},

  doi = {10.1145/3485447.3512062}
}

@inproceedings{PLPSZ24,
  author = {Paes Leme, Renato and Georgios Piliouras and Jon Schneider and Kelly Spendlove and Song Zuo},
  title = {{Complex Dynamics in Autobidding Systems}},
  booktitle = {Proceedings of EC 2024},
  date = {2024},
  year = {2024},
  eprint = {2406.19350},
  eprinttype = {arxiv},

}

@article{Survey24,
  author = {Gagan Aggarwal and Ashwinkumar Badanidiyuru and Santiago R. Balseiro and Kshipra Bhawalkar and Yuan Deng and Zhe Feng and Gagan Goel and Christopher Liaw and Haihao Lu and Mohammad Mahdian and Jieming Mao and Aranyak Mehta and Vahab Mirrokni and Paes Leme, Renato and Andres Perlroth and Georgios Piliouras and Jon Schneider and Ariel Schvartzman and Balasubramanian Sivan and Kelly Spendlove and Yifeng Teng and Di Wang and Hanrui Zhang and Mingfei Zhao and Wennan Zhu and Song Zuo},
  title = {{Auto-bidding and Auctions in Online Advertising: A Survey}},
  journaltitle = {ACM SIGecom Exchanges},
  journal = {ACM SIGecom Exchanges},
  volume = {22},
  number = {1},
  pages = {159--183},
  date = {2024},
  year = {2024},
  eprint = {2408.07685},
  eprinttype = {arxiv},

}

@article{Varian07,
  author = {Hal R. Varian},
  title = {{Position Auctions}},
  journaltitle = {International Journal of Industrial Organization},
  journal = {International Journal of Industrial Organization},
  volume = {25},
  number = {6},
  pages = {1163--1178},
  date = {2007},
  year = {2007},
  doi = {10.1016/j.ijindorg.2006.10.002},

}

@inproceedings{LPSZ24,
  author = {Brendan Lucier and Sarath Pattathil and Aleksandrs Slivkins and Mengxiao Zhang},
  title = {{Autobidders with Budget and ROI Constraints: Efficiency, Regret, and Pacing Dynamics}},
  booktitle = {Proceedings of the 37th Conference on Learning Theory},
  series = {Proceedings of Machine Learning Research},
  volume = {247},
  pages = {3642--3643},
  date = {2024},
  year = {2024},
  publisher = {PMLR},
  url = {https://proceedings.mlr.press/v247/lucier24a.html},
  eprint = {2301.13306},
  eprinttype = {arxiv},
}

@article{BG19,
  author = {Santiago R. Balseiro and Yonatan Gur},
  title = {{Learning in Repeated Auctions with Budgets: Regret Minimization and Equilibrium}},
  journaltitle = {Management Science},
  journal = {Management Science},
  volume = {65},
  number = {9},
  pages = {3952--3968},
  date = {2019},
  year = {2019},
  doi = {10.1287/mnsc.2018.3174},
}

@inproceedings{GLLLS23,
  author = {Jason Gaitonde and Yingkai Li and Bar Light and Brendan Lucier and Aleksandrs Slivkins},
  title = {{Budget Pacing in Repeated Auctions: Regret and Efficiency Without Convergence}},
  booktitle = {14th Innovations in Theoretical Computer Science Conference (ITCS 2023)},
  series = {Leibniz International Proceedings in Informatics (LIPIcs)},
  volume = {251},
  pages = {52:1--52:1},
  publisher = {Schloss Dagstuhl -- Leibniz-Zentrum f{\"u}r Informatik},
  date = {2023},
  year = {2023},
  doi = {10.4230/LIPIcs.ITCS.2023.52},
  eprint = {2205.08674},
  eprinttype = {arxiv},
  note = {Extended abstract},
}

@article{CWW24,
  author = {Zhaohua Chen and Chang Wang and Qian Wang and Yuqi Pan and Zhuming Shi and Zheng Cai and Yukun Ren and Zhihua Zhu and Xiaotie Deng},
  title = {{Dynamic Budget Throttling in Repeated Second-Price Auctions}},
  journaltitle = {Proceedings of the AAAI Conference on Artificial Intelligence},
  journal = {Proceedings of the AAAI Conference on Artificial Intelligence},
  volume = {38},
  number = {9},
  pages = {9598--9606},
  date = {2024},
  year = {2024},
  doi = {10.1609/aaai.v38i9.28816},
}

@article{DFHM24,
  author = {Argyrios Deligkas and John Fearnley and Alexandros Hollender and Themistoklis Melissourgos},
  title = {{Pure-Circuit: Tight Inapproximability for {PPAD}}},
  journaltitle = {Journal of the ACM},
  journal = {Journal of the ACM},
  volume = {71},
  number = {5},
  eid = {31},
  date = {2024},
  year = {2024},
  doi = {10.1145/3678166},
  eprint = {2209.15149},
  eprinttype = {arxiv},
}

@inproceedings{BDMMZ21,
  author = {Santiago Balseiro and Yuan Deng and Jieming Mao and Vahab Mirrokni and Song Zuo},
  title = {{Robust Auction Design in the Auto-bidding World}},
  booktitle = {Advances in Neural Information Processing Systems},
  volume = {34},
  date = {2021},
  year = {2021},
  url = {https://proceedings.neurips.cc/paper_files/paper/2021/hash/948f847055c6bf156997ce9fb59919be-Abstract.html},
}

@misc{HLYW26,
  author = {Yiyang Huang and Zhengyang Liu and Yonglei Yan and Zihe Wang},
  title = {{GSP} Autobidding Equilibria with Few Advertisers or Slots},
  date = {2026-10-21},
  year = {2026},
  eprint = {2610.29310v1},
  eprinttype = {arxiv},
  howpublished = {arXiv preprint \href{https://arxiv.org/abs/2610.29310v1}{arXiv:2610.29310v1}},
}

@misc{AK26,
  author = {Salam Afiouni and Christian Kroer},
  title = {Pacing Equilibria in Abstract Mechanisms},
  year = {2026},
  eprint = {2609.27285},
  archivePrefix = {arXiv},
  primaryClass = {cs.GT},
  url = {https://arxiv.org/abs/2609.27285}
}

@misc{AGPS26,
  author = {Ioannis Anagnostides and Ian Gemp and Georgios Piliouras and Kelly Spendlove},
  title = {Tight Inapproximability for Welfare-Maximizing Autobidding Equilibria},
  year = {2026},
  eprint = {2602.09110},
  archivePrefix = {arXiv},
  primaryClass = {cs.GT},
  url = {https://arxiv.org/abs/2602.09110}
}

@misc{DMMMZZ23,
  author = {Yuan Deng and Mohammad Mahdian and Jieming Mao and Vahab Mirrokni and Hanrui Zhang and Song Zuo},
  title = {Efficiency of the Generalized Second-Price Auction for Value Maximizers},
  year = {2023},
  eprint = {2310.03105},
  archivePrefix = {arXiv},
  primaryClass = {cs.GT},
  url = {https://arxiv.org/abs/2310.03105}
}
\endgroup
\end{document}